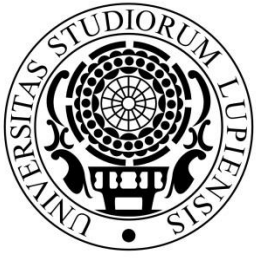

**UNIVERSITA' DEL SALENTO**

DIPARTIMENTO DI MATEMATICA E FISICA
"ENNIO DE GIORGI"

---

Dottorato di Ricerca in Fisica e Nanoscienze

PHD THESIS

# Study of the fluvial activity on Mars through mapping, sediment transport modelling and spectroscopic analyses

Supervisors:

Prof. Vincenzo OROFINO

Prof. Sergio FONTI

PhD student:

Giulia ALEMANNO

---

XXX ciclo

*"...keep Ithaka always in your mind"*

*Costantino Kavafis*

# Table of Contents

# *Introduction*

*Mars is the only planet, other than Earth, for which there are several evidences that liquid water was abundant at the surface. Among them, one of the best evidence is that the ancient terrains of Mars are covered with fluvial and lacustrine features such as valley networks, longitudinal valleys and basin lakes. In addition, spectroscopic data have shown the widespread presence of aqueous alteration minerals on the Martian surface, including clays composed by phyllosilicates and hydrated salts. However, on the present Mars, the liquid water cannot be stable at its surface for long period of time owing to its atmospheric and superficial conditions. In this view, how do we explain the presence of these valleys on the surface of the planet?*

*There are no doubts that these structures were carved by liquid water, but several questions still remain about the mechanisms of their formation and the climatic conditions under which this formation occurred. In this context, the study of these structures is a key in the investigation of the past climatic conditions of the planet which can have, in turn, important astrobiological implications.*

*Several hypotheses have been done on the possible origin of these structures and the paleoclimate of the planet. However, the increasing quantity of data available requires more and more detailed analysis and effort for the interpretation of them. For this reason, we decided to study in detail these fluvial structures from a geological and spectroscopic point of view.*

*In the **first chapter**, evidences of liquid water flow on the surface of the planet are described. Particular attention is focused on the observation of fluvial systems, such as valley networks in association with open- and /or closed-basin lakes. In addition, a wide interest is given to the detection of aqueous alteration minerals, such as clay minerals and hydrated salts. Finally, the paleoclimatic and astrobiological implications of these observations are briefly discussed.*

*On the basis of these observations we decided to provide an update of the previous manual maps of Martian valleys using a new mosaic of the Martian surface*

*and data at higher resolution. The data used and the methodology applied for the production of this global map are described in detail in* **Chapter II** *along with some analysis I performed on the valleys. I produced a more detailed global map of Martian valleys classified on the basis of their morphology. The geographic distribution of these valleys along with rough estimations of their ages allowed us to draw some conclusions about the climatic conditions at the time of the formation of these structures.*

*In addition, as reported in* **Chapter III**, *I extracted a subset of 63 valleys (containing Martian valley networks with a main branch longer than 150 km and a total length greater than 600 km) and estimated their formation timescales by means of a model based on the evaluation of water and sediment discharges. The adopted methodology is based on the application of sediment transport models adapted to the Martian cases on the basis of an accurate analysis of terrestrial rivers.*

*In addition, I estimated the total eroded volume produced by all the mapped Martian valley networks using the data extracted from the global map and the volume of the 63 valleys of our sample.*

*Finally, in* **Chapter IV**, *it has been reported the spectroscopic part of my PhD work. I analyzed some of these structures and associated open and closed-basin lakes using orbital spectroscopic data in search for aqueous alteration minerals. The description of the data used and of the methodology applied for the spectral analysis is thoroughly reported. The attention is especially focused on one closed-basin lake and three open ones where interesting spectral observations were found. The spectral behavior of several areas in these structures has, in fact, suggested the presence of Fe/Mg phyllosilicates sometimes mixed together with unaltered mafic material.*

# CHAPTER I

# The Martian surface

## *1.1 Mars today*

Among all the planets of the Solar System, except for Earth, Mars is the one that has the greatest potential to host living organisms, extinct or current. In fact, there are several similarities with our planet, such as the rotation period around its axis of 24h 37m 22.6s, very close to the Earth's value of 23h 56m 04s. Another analogy between Mars and Earth is the axial tilt of the two planets. For the Red Planet, the rotation axis has a tilt of 25.19°, slightly higher than the Earth's value of 23.45°. This fact allows the presence on Mars of a seasonal cycle similar to that of the Earth, although the Martian seasons have a duration that is twice compared to ours due to the greater orbital period of the planet. However, the data of the first spacecraft missions to Mars revealed a desert world full of craters such as the Moon, thus breaking down the hopes of finding past or present forms of life on the Red Planet.

A new important discovery occurred in 1971, when the Mariner 9 mission observed, for the first time, on the surface of the planet, fluvial structures and channels carved by liquid water flows (Masursky, 1973). In particular, the images showed huge channels carved by catastrophic floods known as outflow channels and large-scale valley networks that reminded one of the typical morphology of terrestrial fluvial systems (Craddock and Howard, 2002; Irwin et al., 2005a; Carr, 2006; Ansan and Mangold, 2013). This discovery is one of the most important steps in the exploration of the Solar System because it indicates the presence of abundant liquid water (even though in the past) on the surface of a planet. However, the data sent by the spacecraft show that today Mars is a cold and arid world. Its atmosphere is very thin and consists mainly of carbon dioxide. The pressure exerted by the Martian atmosphere on the planet surface is less than 1/100 respect to the Earth's value. Even the partial water vapor pressure, equal to about 0.002 mbar, is far lower than that of the Earth. Although the maximum daily temperature can reach 300K (27° C) during summer at medium latitudes, the average solar temperatures on the Red Planet range from 150K (-123° C) to 240K (-33° C) in the warmer regions of the southern hemisphere of the planet (Kieffer et al., 1977). Surface temperatures depend on latitude, season, albedo and thermal inertia of the surface. These temperatures are also related to the slope and the configuration of the terrain. It is clear that, under these conditions, liquid water currently cannot exist on the surface of the planet. Liquid water is stable when temperatures exceed 273K and partial water vapor pressure reaches 6.1 mbar (Carr, 2006). These conditions are very unlikely on the surface of the planet in this Era. How do we explain the presence of valleys on Mars? The hypothesis that in the past Mars experienced climatic conditions different than the present has been advanced. Before analyzing the possible hypothesized climatic changes let's see in detail some features observed on the surface of the Red Planet that suggest the presence of aqueous environments.

## *1.2*  *Signs of liquid water on the Martian Surface*

Understanding the history of water on Mars is of fundamental importance to disclose the evolution of its surface and atmosphere and its potential habitability. Thanks to the increasing quantity of data sent to Earth from the spacecraft in orbit around Mars, a considerable variety of evidences of past water flow on the surface of the planet have been discovered. In the next subsections, we will talk about those evidences, dividing them in geological and spectroscopic, even though, as we will see, all of them are strictly related one to another.

### *1.2.1 Geological evidences: fluvial systems and open/closed-basin lakes*

As already mentioned, the presence of fluvial valleys on the Martian surface (in particular, of valley networks), discovered during the Mariner 9 mission (Masursky, 1973), is probably the most compelling evidence that Mars was once capable of sustaining liquid water at or near the surface. The ancient terrains of Mars are covered with fluvial and lacustrine features classified based on their morphology, as we will see in detail in Chapter II, such as valley networks, longitudinal valleys, outflow channels, valleys on volcanoes. Among them, the valley networks (see **Fig. 1.1**), which are common in the southern highlands of the planet (e.g. Masursky et al., 1977; Hynek et al., 2010), have a dendritic pattern that resembles that of terrestrial river valleys, suggesting a formation by fluvial processes (Masursky, 1973; Craddock and Howard, 2002; Irwin et al., 2005a; Carr, 2006; Ansan and Mangold, 2013).

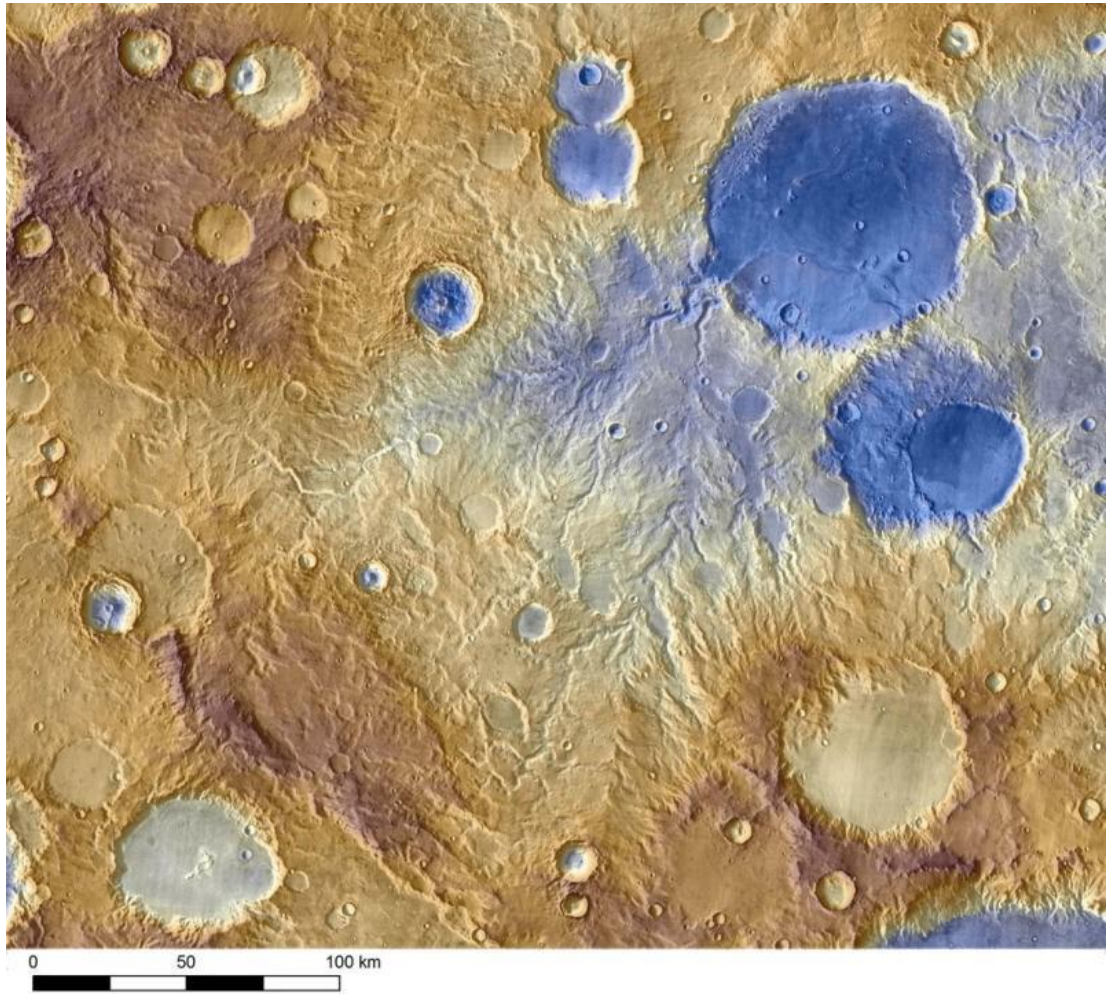

**Fig. 1.1.** False color image obtained by the Mars Odyssey spacecraft showing several valley networks on the Martian surface that appear to be formed by rainfalls resulting from rains or melting snow. The color scale ranges from blue for lowest areas to brown for the highest ones. Image credits: NASA

However, the exact nature of these processes is still not well understood. Two scenarios are mainly proposed to explain their origin: runoff produced by the flowing of liquid water or groundwater sapping subsequent to the collapse of the terrain. Some valley networks show features suggestive of groundwater sapping, a process which is not very common on Earth (Higgins, 1984). Such features are: the short, stubby tributaries with abrupt alcove-like headward terminations (Sharp and Malin, 1975; Pieri, 1980, Baker and Partridge, 1986); the constant valley width downstream (Goldspiel et al., 1993); the U-shaped cross sections (Carr, 1995); the relatively flat longitudinal profiles (Aharonson et al., 2002); and the low sinuosity, coupled with generally low drainage densities (Carr, 1995; Carr and Chuang, 1997; Grant, 2000).

On the contrary, other Martian valley networks show features that on Earth are the results of water runoff processes (Knighton, 1984) due to precipitation (snow or rain). Indeed, runoff is suggested by: the branching, dendritic patterns, with sources near drainage divides (Irwin and Howard, 2002; Hynek and Phillips, 2003); the V-shaped cross sections (Williams and Phillips, 2001; Kereszturi, 2005); the progressive widening downstream (Penido et al., 2013); the gradual deepening of branches with increasing Strahler order (Ansan and Mangold, 2013; for the definition of Strahler order, see Strahler, 1952); and the rather frequent presence of interior channels (Irwin et al., 2005a; Kereszturi, 2005). In some cases, these drainage systems are quite mature, with well-developed tributaries that reach the seventh Strahler's order (Hynek and Phillips, 2003; Ansan and Mangold, 2013). A typical example is the *Warrego Valles* system (see **Fig. 1.2**), studied by Ansan and Mangold (2006).

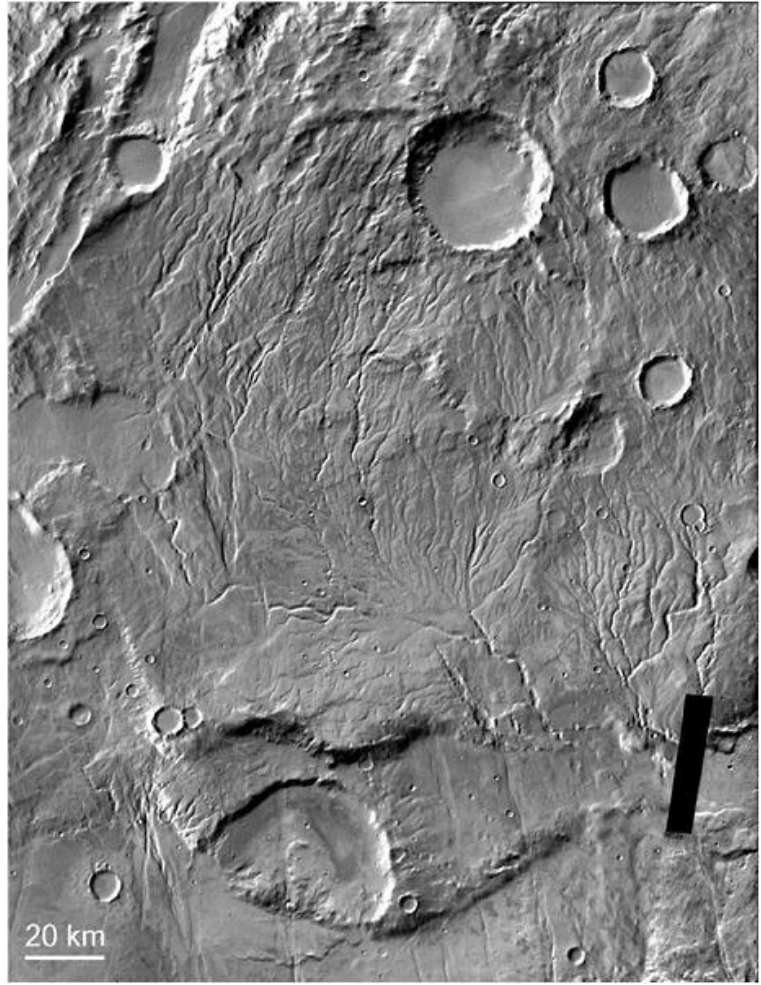

**Fig. 1.2.** *Warrego Valles* system located at 42.2°S, 93.0°W. The drainage density of this Noachian terrain is comparable to terrestrial values and implies precipitation and surface runoff. Image from the Thermal Emission Imaging System on the Mars Odyssey spacecraft. Image credits: NASA/JPL/Arizona State University.

Many networks, showing both kinds of cross-sections (V-shaped in the upper reaches, while U-shaped in the lower ones), seems to have mixed characteristics suggesting an origin due to both runoff and groundwater sapping (Kereszturi, 2005; Hoke and Hynek, 2009). In any case, as suggested by some authors (Craddock and Howard, 2002; Lamb et al., 2006), surface runoff seems to be necessary to carry away the debris generated by potential groundwater sapping.

Moreover, the present morphology of the valleys could be not pristine since the later structures may have been modified by mass wasting processes, and in particular by the failure of the side walls of the valley, likely mimicking the typical morphology of groundwater sapping valleys (Gulick and Baker, 1990). Erosion, faulting and dust cover could have further modified these valley systems. Therefore, in most cases, only the remnants of the original drainage paths are visible. For this reason, a global approach in the study of the Martian fluvial system is necessary and it is also important for the possible paleoclimatic implications as will be discussed in Section 1.3.

Another important similarity between the Martian valley networks and the terrestrial valleys is that the former often open into basins, such as depressions or impact craters, possibly forming lakes in the past (Goldspiel and Squyres, 1991; Cabrol and Grin, 1999; Fassett and Head, 2008; Goudge et al., 2012; 2015).

Very often we can see open and/or closed basin lakes associated with these valleys (Fassett and Head, 2008). Open-basin lakes are characterized by an outlet valley. The presence of this outlet means that the water filled the basin sufficiently to overtop and breach the crater rim (Fassett and Head, 2008; Goudge et al., 2015). Closed-basin lakes are, instead, basins that may have ponded water, supplied either by an inlet valley or groundwater influx, but that show no outlet valleys (Goudge et al., 2015). Examples of closed and open basin-lakes are shown in **Fig. 1.3**.

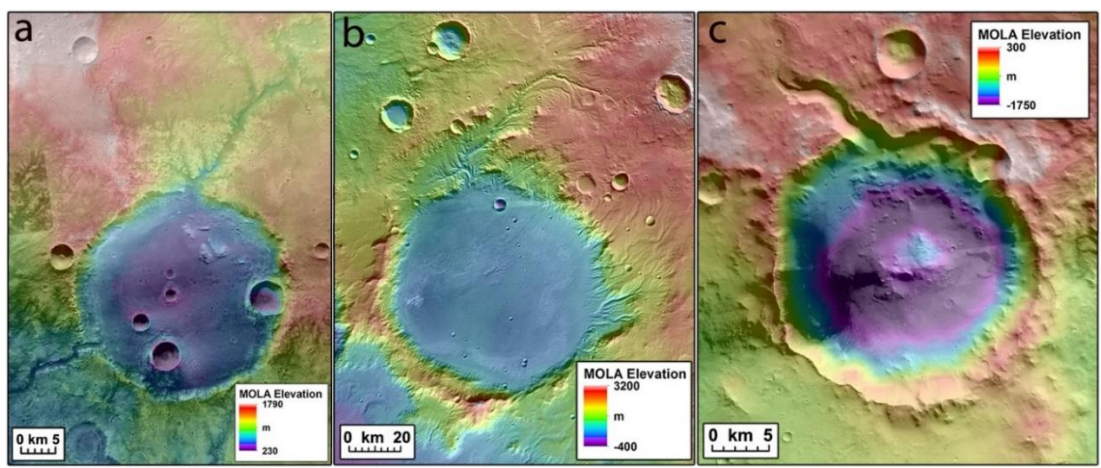

**Fig. 1.3.** Different types of paleolake basins on Mars: a) valley network-fed open-basin lake in heavily degraded crater centered at 20.5°S 87.0°E. The image shows Mars Orbiter Laser Altimeter (MOLA) gridded topography overlaid on a mosaic of Context Camera (CTX) images B05_011498_1584, B10_013621_1594, G08_021480_1594, and G14_023748_1601 and High-Resolution Stereo Camera (HRSC) nadir image h6523_0000. North is to the right. b) valley network-fed closed-basin lake in heavily degraded crater located at 18.8°S 59.2°E. Image shows MOLA gridded topography overlaid on a mosaic of CTX images B21_017868_1583 and P16_007148_1628, and HRSC nadir images h8440_0000 and h0532_0000; c) isolated inlet valley closed-basin lake in a less degraded crater at 9.6°S 144.1°E. Image shows Mars MOLA gridded topography overlaid on a mosaic of CTX images P16_007158_1703 and B02_010230_1715. The images have been obtained and reported in their paper by Goudge et al. (2016).

On the surface of the Red Planet, more than one hundred basins have been identified that could have accommodated lakes in the past (Fasset and Head, 2008; Orofino et al., 2009; Goudge et al., 2012) (see Chapter IV). This is also suggested by the presence in such basins of massive deposits interpreted as of fluvio-lacustrine sedimentary origin (Cabrol and Grin, 1999; Di Achille et al., 2006; Di Achille and Hynek 2010a, 2010b; Goudge et al., 2012).

In fact, sometimes where a valley flows into a basin, depositional structures are observed. These structures are subdivided into delta deposits and alluvial deposits and differ in the environment in which deposition occurs and in the resulting stratigraphy. Alluvial fans are formed when sediments are deposited by a stream of water at the entrance of a basin not occupied by water while the delta forms at the mouth of a river when it flows into a body full of water (Di Achille and Hynek, 2010a). Thanks to the use of topographic data obtained from the high-resolution stereoscopic camera (HRSC) onboard the European Mars Express probe, Mars has been shown to be full of fan-shaped features (Di Achille et al., 2007; Pondrelli et al., 2008; Hauber et al., 2009; Di Achille and Hynek 2010a, 2010b). Some examples of delta and fan deposits are shown in **Fig. 1.4**.

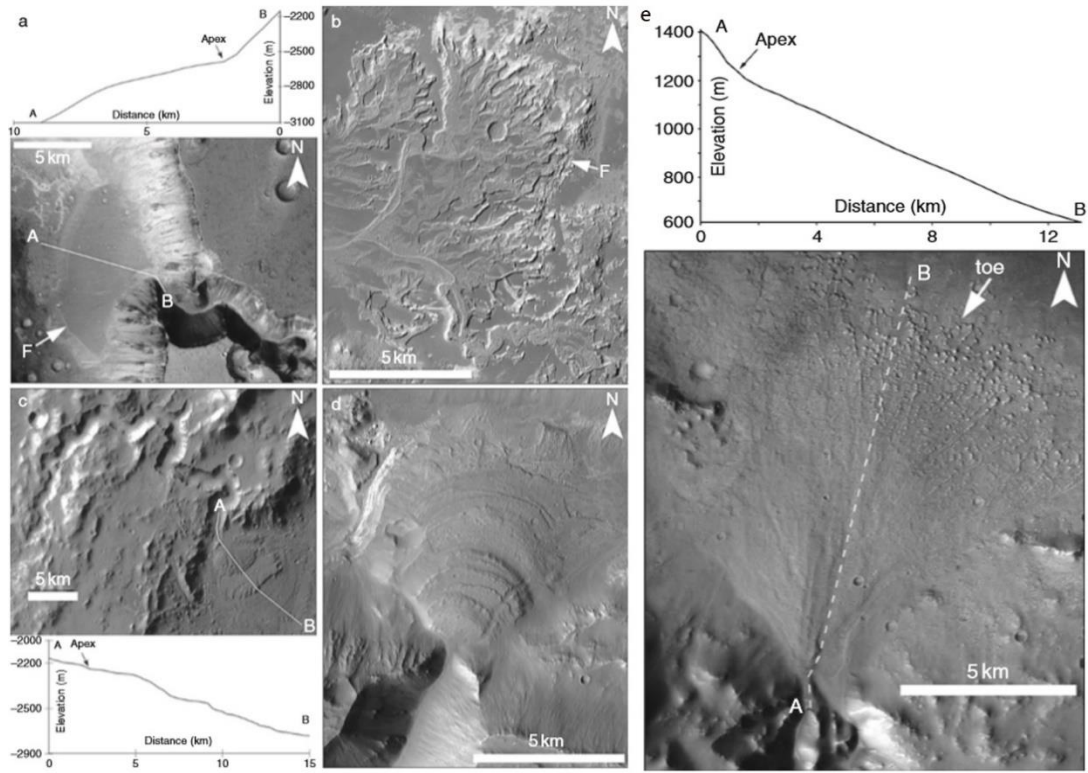

**Fig. 1.4.** On the left (panels from a to d) - morphology and morphometry of the Martian delta deposits: a) single-lobed delta deposit located in *Shalbatana Vallis*; b) multilobed deposit in the *Eberswalde crater*; c) delta fan located at the mouth of *Tyras Vallis*; d) deposit located in the *Coprates Chain* characterized by a series of concentric lobbies. On the right (panel e) - Example of alluvial Martian fan. Profiles were obtained using MOLA topographic data. Profile in a) shows a slope typical of the delta deposits. In the case of the alluvial fan (profile in e), the profile shows a steady slope (Di Achille et al., 2010).

In addition, the crater's inner edges often show terrace systems at various levels probably produced by processes of deposition or of mechanical erosion generated by wave motion. Sometimes these inner edges are incised by small channels called gullies (see **Fig. 1.5**), which tend to be arranged parallel one to another in terrains characterized by very steep gradients (Clow, 1987). The importance of the gullies is related to the fact that they may have been newly formed by the flow of liquid water although recently this hypothesis has been questioned on the basis of spectral and geological analysis (Dundas et al., 2015; Núñez et al., 2016). In fact, in a recent work, Núñez and colleagues (2016) conducted an analysis of the CRISM observations on about 100 gully sites on Mars. During their analysis they observed no hydrated minerals within the gullies (see for example **Fig. 1.5**), indicating limited interaction or no interaction of the mafic material with liquid water. This seems to suggest that a mechanism not requiring liquid water may be responsible for carving the gullies on Mars (Núñez et al., 2016).

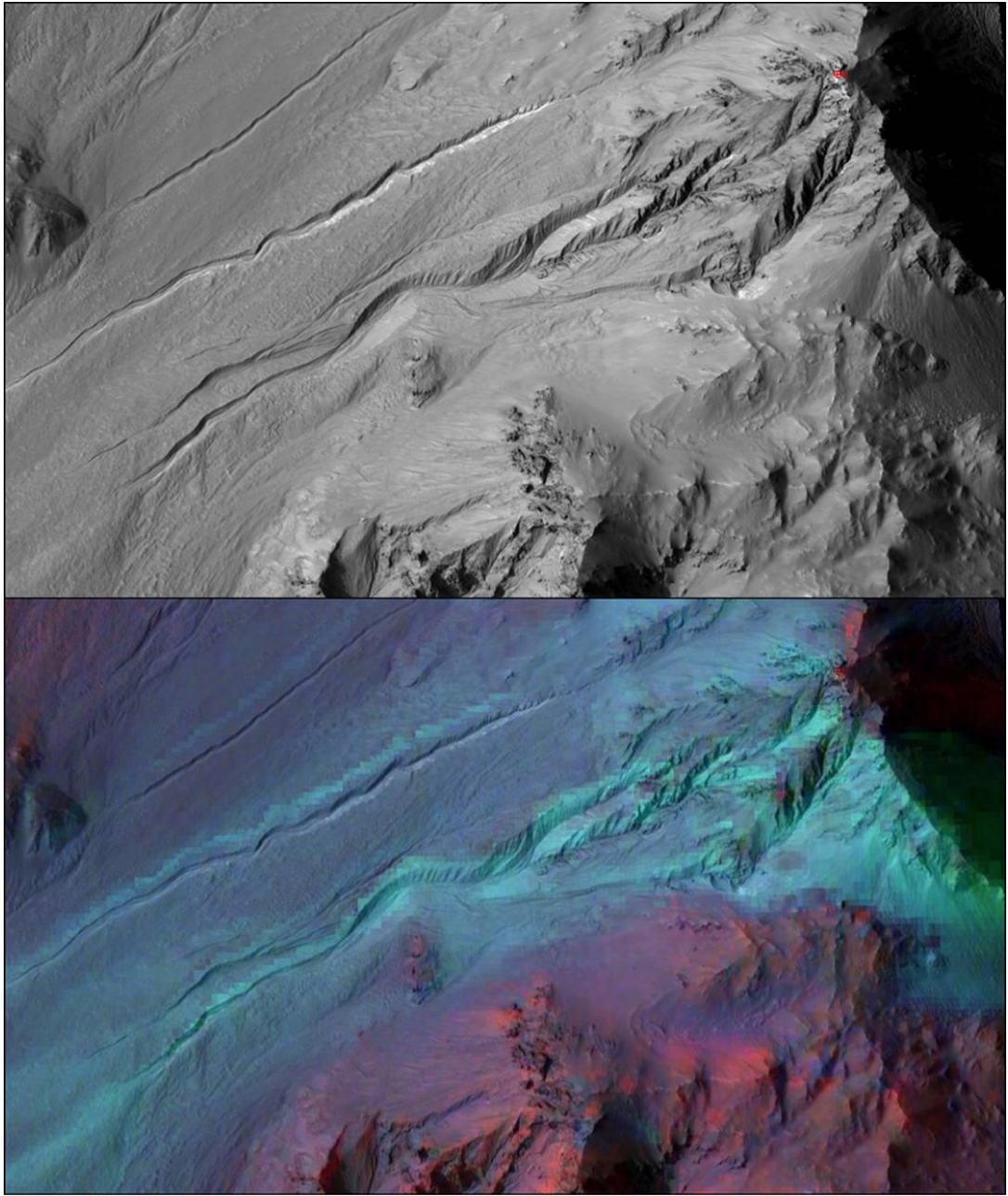

**Fig. 1.5.** Top: Highly incised Martian gullies on the eastern rim of Hale crater in a visible-light image obtained thanks to the High Resolution Imaging Science Experiment (HiRISE) camera onboard NASA's Mars Reconnaissance Orbiter. Bottom: same image as at the top with superposed color-coded compositional information, extracted from the data of the Compact Reconnaissance Imaging Spectrometer for Mars (CRISM) onboard the same orbiter. Light blue corresponds to surface composition of unaltered mafic material of volcanic origin. It is possible to observe that mafic material from the crater rim is carved and transported downslope along the gully channels. No hydrated minerals are observed within the gullies, in the data from CRISM, indicating limited interaction or no interaction of the mafic material with liquid water. The area shown in the images spans about 3 kilometers on the eastern rim of Hale Crater. Note the smaller size of the channels compared with that of the valleys shown in **Figs. 1.1** and **1.2**. Image credits: NASA/JPL-Caltech/UA/JHUAPL

## *1.2.2 Spectroscopic evidences: aqueous alteration minerals*

The fundamental question connected to the study of any Solar System object is: what is it made of? In addition, the composition of the surface of planetary objects records the history of their formation and the subsequent alteration over time. The mineralogy and chemistry of a planetary surface is also connected to other important information about the planet, ranging from its interior to the potential habitability.

For Mars, the mineralogical record provides significant clues on the geochemical environment and aqueous history of the planet.

The knowledge of Mars mineralogy mainly relies on data collected remotely, using infrared spectroscopic tools (Farmer, 1974; Clark, 1995; Ehlmann and Edwards, 2014). Molecular lattice vibrations lead to distinctive minima, diagnostic of composition, in electromagnetic radiation thermally emitted from the surface (Ehlmann and Edwards, 2014). Similarly, in the visible and near infrared wavelengths sunlight reflected from the surface of Mars has also specific features at certain wavelengths that are diagnostic of the surface composition (Ehlmann and Edwards, 2014). Therefore, thermal and reflectance spectra contain a lot of information about the superficial composition of a planet and its characteristics.

Nearly every mission to Mars was sent with an onboard instrument devoted to spectroscopy. This includes instruments like Thermal Emission Spectrometer (TES) onboard Mars Global Surveyor (Christensen et al., 2001), the imaging spectrometer Observatoire pour la Mineralogie, l'Eau, le Glace e l'Activite (OMEGA) (Bibring et al., 2006) and the Planetary Fourier Spectrometer (PFS) (Formisano et al., 2005), onboard Mars Express and the Compact Reconnaissance Imaging Spectrometer for Mars (CRISM) onboard Mars Reconnaissance Orbiter (Murchie et al., 2007). These instruments provide remote-sensing measurements of mineralogical and thermophysical properties of the Martian surface.

### *1.2.3 Mineralogy of the Martian surface*

By means of the spectroscopic data sent to Earth by these instruments in orbit around Mars it is possible to try to reveal the mineralogic composition of the Martian surface. Before that, thanks to the first telescopic observations of Mars, it was possible to observe a widespread presence of ferric oxides and it was also demonstrated that the surface has been altered from its primary mineralogic composition (Singer et al. 1979; Bell et al., 1990). Later, the data collected by Viking landers gives some evidence for salts and a component of clay minerals in the Martian soil (Clark et al. 1978; Banin et al. 1992).

With the increase in the amount and quality of the data, it has been possible to recognize an increasing variety of minerals present on Mars. So far, we know that, globally, Mars has a basaltic upper crust with widespread regionally variable deposits of plagioclase, pyroxene and olivine (Ehlmann and Edwards, 2014).

An important discovery was the detection on the Martian surface of phyllosilicates (Bibring et al., 2005; Poulet et al., 2005), a class of hydrous minerals also known as clay minerals. Actually, the term "clay" can be used either with reference to the smallest particle size fraction in sediments (< 4 µm) or to phyllosilicates mineralogy (Ehlmann et al., 2013). In this work, we use the term "clay" in the latter sense referring to hydrated or hydroxylated minerals.

### *1.2.4 The importance of clay minerals*

Processes leading to clay formation on Earth are important mechanisms in terrestrial geochemical cycles (Ehlmann and Edwards, 2014). Clay minerals, in fact, originate on Earth by several processes including near-surface weathering, precipitation in water bodies within basins, hydrothermal alteration (volcanic- or impact- induced), diagenesis, metamorphism, and magmatic precipitation. These minerals are then eroded, transported, buried, and metamorphosed in processes driven by climate and tectonics (Merriman, 2005; Meunier, 2005). In addition, ocean and river water chemistry is partly controlled by continental silicate weathering to form clays and clay-forming reactions of seawater with the basaltic seafloor (Spencer and Hardie, 1990; Ehlmann et al., 2013). Clay formation by weathering gives important feedback on atmospheric chemistry, climate and inclusion of volatiles in the crust (Kump et al., 2000; Ehlmann et al., 2013). The formation mechanism of these minerals together with the extent of their degree of alteration determine the geochemical consequences of their formation and, consequently, their chemical characteristics (Ehlmann et al., 2013). In **Table 1.1** are reported, for example, seven major mechanisms by which clay minerals are formed on Earth or transformed and the expected characteristics for each mechanism.

**Table 1.1**

Major mechanisms of formation of clay minerals on Earth (Ehlmann et al., 2013).

| Setting | Clay mineralogy | Facies and abundance (% vol.) | Setting | Accompanying minerals |
|---|---|---|---|---|
| Near-surface pedogenic | Fe/Al smectites, kaolinite | Bulk soil component (up to 95 %) | Horizons of leaching and deposition, with alteration lessening with depth | Fe/Al oxides, carbonates, silica, allophone |
| Near-surface basin | Fe/Mg/Al smectites, kaolinite, illite, chlorite | Layered deposits with sedimentary textures; clay minerals (up to 95 %) mostly detrital, *in situ* formation at very high Si, Al activities | Deposits within a river, lake, or ocean, later exposed by erosion | Other minerals eroded from rocks in basin; evaporites, e.g. chlorides, carbonates, hematite, silica, sulfates (if acidic), potassium feldspar and zeolites (if alkaline) |
| Hydrothermal (volcanic) | Fe/Mg/Al smectites, kaolinite | Bulk component (variable) | Zoned alteration surrounding fumaroles, vents, and cones | Sulfates (alunite, jarosite), ferric oxides, amorphous silica, allophane, anatase |
| Hydrothermal (impact) | Saponite, nontronite, celadonite, kaolinite | Fracture fill within breccia, alteration rinds on mineral grains (5–10 % total area) | Beneath the crater floor and rim in fractures and pore spaces of breccia; not exposed unless by erosion | Amorphous silica and altered impact glasses, carbonate, sulfates, sulfides, potassium feldspar quartz, zeolites, native metals |
| Diagenesis | Illite (from K, Al-rich precursors); Chlorite (from Mg-rich precursors), mixed-layer clays | Bulk rock; veins and pore space (variable) | Bulk rock/sediment altered; sometimes preferential alteration in pore space and veins | silica, original clay minerals |
| Metamorphism | Fe/Mg smectites, chlorite, zeolites, prehnite, pumpellyite, serpentine, epidote, actinolite | Bulk rock (variable) | 100s m to km beneath the surface; not exposed unless by deep erosion or impact excavation | Amorphous silica, zeolite, sulfite, iron oxides, garnet, original rock forming minerals |
| Magmatic | Fe/Mg smectite, celadonite | Within pores, veins of bulk rock (<15 %) | Formed during final degassing of lavas with substantial volatile content | Silica, primary minerals |

In analogy with the terrestrial case, the study of clay minerals on the Martian surface has important geological, paleoclimatic and astrobiological implications.

### *1.2.5 The presence of phyllosilicates on the Martian surface*

The presence of these minerals on the surface of a planet provide a record of water-related processes (Poulet et al., 2005). Previously in situ elemental analysis conducted by the Viking-Landers (Toulmin et al., 1977) and the identification of smectites in some SNC (Shergottite-Nakhlite-Chassigny) meteorites (Bridges et al., 2001), suggested the presence of phyllosilicates on Mars. However, the first unambiguous detection of water-bearing phyllosilicates has been reported over large areas thanks to OMEGA data (Bibring et al., 2005; Poulet et al., 2005). The latter allowed the detection of localized phyllosilicate deposits in the ancient southern highlands (Costard et al., 2006; Mangold et al., 2007). Later, with the use of a combination between multispectral mapping and targeted observations obtained thanks to the CRISM instrument, it has been possible to expand the number of detected phyllosilicates occurrences (Mustard et al., 2008). Global mapping, see **Fig. 1.6,** showed that exposures of clay minerals are widespread but apparently restricted to ancient terrains where Noachian crust is exposed (Mustard et al., 2008; Murchie et al., 2009).

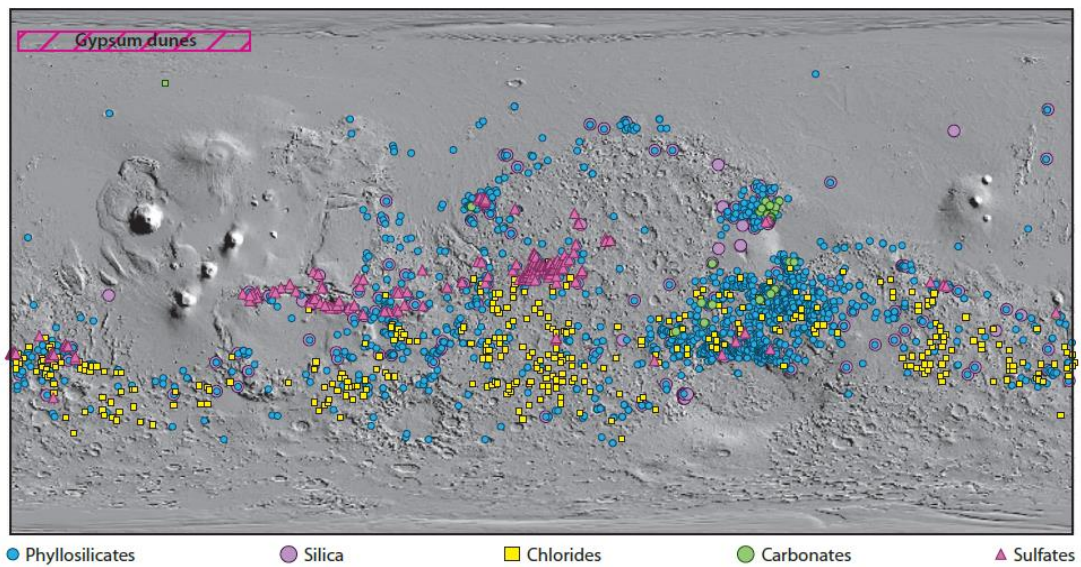

**Fig. 1.6.** Global distribution of the major classes of aqueous minerals on Mars (Ehlmann and Edwards, 2014). Blue dots: Phyllosilicate detection compiled by Ehlmann et al. (2011) and Carter et al. (2013). Purple dots: Silica detections compiled by Carter et al. (2013). Yellow dots: Chlorides compiled by Osterloo et al. (2010). Green dots: Carbonate bearing rock detections reported by Ehlmann et al. (2008b) and reviewed in Niles et al. (2013) (square indicates Phoenix lander soil carbonate). Pink triangles: sulfate detections from Murchie et al. (2009), Milliken et al. (2010), Ackiss and Wray (2012), and Carter et al. (2013). Image taken from Ehlmann and Edwards (2014). From this global representation it is possible to see that phyllosilicate minerals are widespread in almost every region where the Noachian crust is exposed while the other secondary minerals show a more regional pattern (Ehlmann and Edwards, 2014).

It was observed that the clay minerals found on the Martian surface can be divided into three types: multiple clay-bearing units likely formed by in situ alteration; mineral deposits in central peaks, walls, and ejecta of impact craters; and mineralogical units in sedimentary basins (Ehlmann et al. 2011, 2013). In the first case, stratigraphies of clay minerals have been observed. In the largest exposures of this type of clay minerals, the lowermost stratigraphic units are composed of hundreds of meters of Fe/Mg phyllosilicate–bearing basaltic materials with different morphologies ranging from brecciated to layered (Ehlmann and Edwards, 2014). In the second case, the association between clay minerals and craters is global and linked with the Noachian terrains. Clay minerals are, in fact, not widespread in the Hesperian and Amazonian units of the northern plains but are found associated with the largest impact craters, where the Noachian bedrock is exposed (Carter et al. 2010). These observations seem to suggest that phyllosilicate formation occurred during the earliest geological era of Mars, the Noachian. So it seems that, in this period of the Martian history, there were surface environments favorable to the phyllosilicate formation (Bibring et al., 2006), moderate to high pH and high water activity (Velde et al., 1995). In this view, phyllosilicate deposits on Mars are a key to investigate the aqueous activity and the possibility that habitable environments may have existed during the Noachian period.

An important discovery has been, in fact, the identification of clay minerals within sedimentary basins, specifically in fans and deltas within *Holden, Eberswalde* and *Jezero* craters (Milliken et al., 2007; Ehlmann et al., 2008a; Grant et al., 2008; Mustard et al., 2008). The latter represent the detection on Mars of hydrated silicates within sediments clearly deposited by water (Mustard et al., 2008). This seems to indicate that sedimentary phyllosilicates may have been eroded and transported although in situ formation cannot be ruled out (Mustard et al., 2008). Ehlmann and Edwards (2014) showed that groundwater played an important role in formation and diagenesis of clay minerals. In addition, carbonate-bearing rocks associated with olivine have been detected (Ehlmann and Edwards, 2014).

However, there are still several open questions about the Martian mineralogy and in particular about the genesis and formation of these phyllosilicates on Mars. For example, some still not completely answered questions are the following:

- what type of alteration and/or formation processes enabled the clay formation?
- were most clay minerals formed in the subsurface or at the surface?
- were clay minerals formed in situ?
- was the crustal or atmospheric reservoir altered, e.g. by sequestration of water or other volatile species?
- which mechanism controls the distribution of the observed deposits of minerals such as, chlorides, sulfates and carbonates, at large scale as well as at local scale in regolith?
- can clay minerals be found also in fluvial beds?
- can the presence of those minerals and carbonates be hidden by other materials that are spectrally dominant?

- what are the paleoclimatic and astrobiological implications of clay formation in the case of Mars?

As widely discussed, the global nature of mineral formation by liquid water during the first billion years has been a key discovery, but this large number of still unknown points makes the study of these mineral deposits a main goal in many remote-sensing and laboratory studies. That is the reason why, in this work, I analyzed the valleys mapped and their associated open and/or closed basin lakes to search for aqueous alteration minerals (such as phyllosilicates, hydrated silica) and/or evaporites (such as carbonates, chlorides, sulfates).

## *1.3 Paleoclimatic and astrobiological implications*

Of all the scientific arguments that have captured the popular imagination, few seem to have a timeless appeal as the idea that Mars may have hosted life. From the earliest observations with only the use of Earth-based telescopes, many hypotheses have been made on possible life forms on the Red Planet. The common point between all the work of scientists and, if we want, those of science fiction writers is to acknowledge that if there is a planet of the Solar System beyond the Earth, which hosts living organisms, this is most likely Mars.

As reported in Section 1.1, conditions on the surface of Mars today are very inhospitable for life, but all the previously discussed geological and spectroscopic evidence seem to suggest that conditions were more hospitable in the past, particularly the distant past. In fact, probably, the conditions of atmospheric pressure and superficial temperature of Mars were different from the present (Hynek et al., 2010). In particular, many authors believe that, during the Noachian Era, Mars had a warmer and wetter climate thanks to an intense volcanic activity which made the atmosphere thicker. This led to a greenhouse effect sufficient to warm the planet (see below). Later, the cooling of the core of the planet led to a progressive reduction of the volcanic activity which in turn caused a thinning of the planet's atmosphere.

A warm and wet climate during the Noachian Era is supported by many observations. While Hesperian and Amazonian terrains have well preserved impact craters, Noachian terrains are heavily eroded. Golombek and Bridges (2000) estimated that the erosion rate decreased from 3 to 6 orders of magnitude at the end of the Noachian Era. In the first billion years of Martian history not only the erosion rate was much greater but also the sedimentation and deposition rates. It is not clear where all the eroded sediments went, but there are no doubts that wind action only cannot have produced such large erosion (Hynek et al., 2010). A likely explanation is the water flow in warm climatic conditions, very different from the present ones.

This climatic change is also supported by the observation of phyllosilicates in Noachian terrains but not in younger terrains (Bibring et al., 2006). As discussed in the previous section, phyllosilicates are hydrated silicates, characterized by the presence of the hydroxyl ion OH, which are produced, on the surface of a planet, only in presence of a significant quantity of liquid water (Pollack et al., 1987; Schaefer, 1990). So, the presence of these minerals is a possible indication that in the past liquid water flowed on the Martian surface. As a result, the presence of the valley networks and of these mineral deposits in the Noachian-aged terrains have frequently been associated to a relatively warm and wet climate during the Noachian period (Craddock and Howard, 2002; Barnhart et al., 2009), requiring a vertically integrated hydrologic system (Head and Marchant, 2015).

But the hypothesis of a warm and wet Noachian Mars has also been challenged by a series of observations.

Sometimes the valleys observed on the Martian surface maintain their width downstream on the contrary to what happens in the terrestrial case where valleys tend to become less deep and wider downstream. These valleys also have shallow tributaries of small amplitude with abrupt terminations (Carr, 2006). As widely discussed in Section 2.1, these features, joined with U-shaped cross-sections, are more characteristic of structures originated from groundwater sapping (Pieri, 1980; Laity and Malin, 1985; Baker, 1990), a process which, according to Goldspiel and Squyres (2000), can virtually operate also under climatic conditions similar to the present ones.

Most valley networks show a poor integration (Carr, 1995; Stepinski and Coradetti, 2004) and even though those valleys are in some cases connected with open basin lakes, this immaturity could suggest short-term fluvial activity (Aharonson et al. 2002). In addition, Noachian degradation/erosion rates are very low by terrestrial standards (Golombek et al., 2006).

Clearly, the hydrological regimes in Noachian and post-Noachian terrains were very different. The Noachian surfaces are largely cut by valleys and show the signs of considerable erosion while the post-Noachian have not undergone significant erosion and exhibit local runoff channels. So, we know that among the Noachian era and the following period there was a noticeable change, but we do not know what was the cause. In addition, the valley networks seem to have formed by precipitation, but it is unclear under what conditions their formation took place.

In addition, recent climate modeling studies (Forget et al., 2013; Wordsworth et al., 2013; Wordsworth et al., 2015) suggest that early Mars was characterized by a cold and icy climate exhibiting adiabatic cooling and an icy highland, with mean annual surface temperatures below the melting point of water (~273 K). This new scenario contrasts with the previous one: precipitation-derived valley network formation is therefore generally inconsistent with the background cold and icy climate. Several mechanisms, invoking episodic melting of ice deposits in the highland terrains, have been proposed to account for valley network formation under these cold and icy climate conditions including: transient warming from volcanic greenhouse gas emissions (Halevy and Head, 2014), melting of ice under peak daily and seasonal temperatures, and ice melting through glacio-volcanic interactions (Cassanelli and Head, 2016). Valley network-related precipitation might be snowfall (nivial), not rainfall (pluvial) (Scanlon et al., 2013).

Moreover, as reported in the previous section, Ehlmann et al. (2011) showed that a significant part of the phyllosilicates in the Noachian crust (Fe, Mg clays) appear to be due to hydrothermal subsurface alteration. Groundwater, in fact, seems to have played an important role in the formation and diagenesis of clay minerals, and its upwelling produced large deposits of sulfates, hematite, and chlorides. In addition, carbonate-bearing rocks associated with olivine have been detected but carbonates are present in very small quantity on the Martian soil (Ehlmann and Edwards, 2014).

The main aim of this PhD work is to analyze the hydrology of the planet and to further constrain habitable environments on Mars. For this purpose, we have to: characterize the environmental context; identify places most likely to have sustained life and search for "biosignatures", namely features created only by life that can persist long after they were formed.

In this context we decided to study in detail and characterize both from a geological and a spectroscopic point of view these Martian fluvial systems and associated open and/or closed basin lakes. This will help us to have a better idea on the climatic conditions at the time of the formation of Martian valleys and on the potentiality of the Martian surface to have sustained life.

# CHAPTER II
# Global map of Martian fluvial systems and age estimations

## 2.1 State of the art

To guess the climate history of Mars as inferred from valley networks, a global approach is necessary. Up to now other global maps have been produced. Carr and Chuang (1997) produced a global manual map based on Viking photographic data. Later, Luo and Stepinski (2009) created a map of Martian valleys using automated routines based on topographic MOLA (Mars Orbiter Laser Altimeter on board of Mars Global Surveyor) data. Valleys were mapped by delineating troughs detected in the MOLA data and then manual editing this map to remove false detections.

An updating of this map was manually produced by Hynek et al. (2010) using not only MOLA data but also a photographic THEMIS (Thermal Emission Imaging System on board of Mars Odyssey) mosaic with a resolution of 200 m/pixel.

Here I present a global map of Martian valleys obtained using data at higher resolution than previously used for this purpose.

In the following section, I present the dataset used to produce our map. The methodology is described in detail, along with the different types of valleys mapped, in Section 2.3. Then, the results are presented with some comparisons with previous global maps of Martian valleys (Section 2.4). In Section 2.5 the information about the mapped valleys is coupled with age-dating determinations using the global geologic map of Tanaka et al. (2014) in order to have a rough estimation of the age distribution of these valleys. Finally, in Section 2.6, I discuss our results and the potential future improvements of the produced dataset.

## 2.2 Dataset description

I mapped all the Martian valleys longer than 20 km using photographic and topographic data obtained from different NASA missions: Mars Global Surveyor, Mars Odyssey and Mars Reconnaissance Orbiter.

Mars Global Surveyor was launched to Mars in 1996. Among the various instruments onboard the spacecraft, there was the MOLA altimeter, projected with the main goal of evaluating the height of features on the Martian surface. During its mission, MOLA collected high resolution topographic data, producing a global grid with a 30 cm/pixel vertical resolution and ~ 463 m/pixel horizontal resolution at the equator (Smith et al., 2001).

Later, in 2001, the THEMIS experiment, onboard the Mars Odyssey spacecraft, started to collect visible and infrared data of the Martian surface. THEMIS is an imaging system which combines a 5-wavelength visual camera with a 9-wavelength infrared camera. In the visible range, the instrument has a resolution of ~19 m/pixel while the thermal infrared camera has a spatial resolution of ~ 100 m/pixel and covers a wavelength range from 6.7 μm to 14.8 μm (Christensen et al., 2004). Edwards et al. (2011a, 2011b) used the data acquired during the first 7.5 years of the mission to create new daytime and nighttime global mosaics of THEMIS infrared images with a spatial resolution of 100 m/pixel or ~592 pixels per degree at the equator. To date, these mosaics represent the highest resolution and best quality global dataset of the Martian surface.

I combined the THEMIS daytime IR mosaic with MOLA topographic data to improve the identification of valley networks over the use THEMIS or MOLA alone. The mix between these datasets produces a superior data set for accurate mapping and analysis of the features displayed Martian surface. Where THEMIS data resolution was not sufficient (for small-scale systems and/or valleys with a very high level of erosion), I also used data, with a resolution up to 6 m/pixel, taken by CTX, the Context Camera of NASA's Mars Reconnaissance Orbiter, which blasted off from Cape Canaveral in 2005.

## *2.3 Methodology*

Valleys, identified in THEMIS daytime IR and/or CTX images, plus topographic MOLA data, were manually mapped as vector-based polylines within the open source QGIS (Quantum Geographic Information System) software. At low latitudes, I used an equidistant cylindrical projection, while at high latitudes, I used sinusoidal and polar stereographic projections to represent and analyze the data. I looked for topographic troughs that show a visual indication of valleys formed by fluvial processes and connected them to form networks.

In detail, each network was made by joining segments that were organized in a hierarchical system starting from the outlet and bifurcating in the upstream direction.

In our work, we preferred a manual approach rather than automated procedures for three main reasons:

a) by means of automated routines one can end up mapping every single depression on the planet. The user has then to check and manually correct the map deleting every false detection (Hynek et al., 2010). This, in our view, can result in an even much longer procedure than the manual approach. This problem of false detections is particularly evident in complex regions characterized by different faults. For example, as already acknowledged by Hynek et al. (2010), on *Alba Patera* volcano it is very difficult, even with a manual approach, to discriminate between channels carved by water flow and those that instead have a faulting origin;

b) in our case, manual mapping uses not just topographic data but also THEMIS imagery with a higher spatial resolution (~100 m/pixel) plus, if necessary, CTX data. The use of multiple datasets allows a better interpretation of the observed features;

c) mapping any feature using computer parsing of digital data is a challenging problem because, as acknowledged by Luo and Stepinski (2009), even the best algorithms lack human understanding of the problem and are, alone, unable to perform a such kind of analysis considering a broader context.

For all these reasons, we decided to adopt a manual approach even though it also has its problems and limitations. These include a high level of subjectivity in valley identification and a potential loss of valleys in regions which do not show a strong signature of incision on the surface. In fact, the identification of these features can be influenced by albedo variations and by the quality of the data (Hynek et al., 2010). As done in previous manual approaches (Carr, 1995; Carr and Chuang, 1997; Hynek et al., 2010), we tried to by-pass this problem combining imagery and topographic data. For that reason, in this work THEMIS daytime IR mosaic, plus CTX data for highly eroded systems, were combined with MOLA data.

Our approach of manual mapping is similar to those first applied by Carr (1995) and Carr and Chuang (1997) and, more recently, by Hynek et al. (2010). The valleys have been mapped using the same criteria followed by those authors: we searched for sublinear, erosional channels with a branching networks morphology characterized by the presence of small branches upslope and a size that increases downstream (Carr and Chuang, 1997; Hynek et al., 2010). In order to be included in our catalog a feature has to:

a) reflect the actual topography (as verified by longitudinal profile);
b) show V or U-shaped cross sections (as verified by cross-section profile).

Examples of longitudinal and cross-section profiles are shown in **Fig. 2.1**.

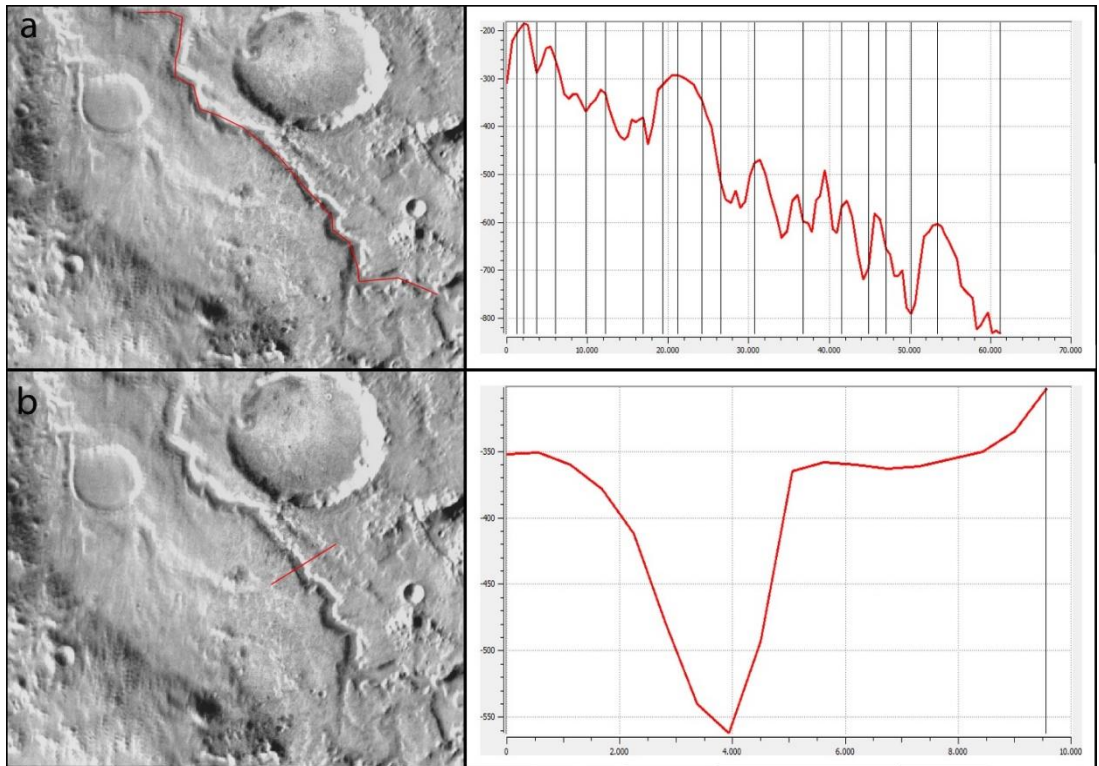

**Fig. 2.1.** a) example of longitudinal profile made tracing a line (red) along the entire route of the valley shown to the left; b) example of cross-section obtained from the red line in the left image. The images are both based on THEMIS mosaic (~100 m/pixel of resolution) and are centered around 1.0°N 90.1°E.

For this reason, during the mapping procedure I analyzed longitudinal and cross section profiles of the mapped valleys using the profile tool of QGIS.

Using QGIS, I stored the data as vectors in a shapefile containing the whole map. An attribute table was also created containing for each valley the coordinates of the system, the type of valley, and other data useful also for further analysis, like the total length and the estimated age (see below). The vector format permits storage at a resolution equivalent to that of the original mapping, yet allows display at any other resolution.

In the literature, the fluvial systems observed on the Martian surface have been described with several terms (as for example valley networks, outflow channels, runoff channels). To avoid confusion, in our work the mapped valleys were divided, on the basis of their morphology, into six different groups defined as follows: valley networks (well-developed systems with many tributaries); single valleys (systems with one or at most two tributaries); longitudinal valleys (structures characterized by a long main branch and few tributaries); valleys on volcanoes (small valley networks and single valleys which are located on volcanoes) and valleys on canyons (small tributaries of huge channels such as *Kasei Vallis* and of large crustal fractures such as *Valles Marineris*); small outflow channels (systems with a morphology very similar to that of the outflow channel but at smaller scale). The outflow channels are huge channels carved by catastrophic events characterized by a sudden release of abundant liquid water. They can have maximum width around 100 km and lengths between 1000 and 2000 km while the depth is usually greater than 1 km (Baker et al., 1992). Outflow channels generally start in chaotic terrains (fragmented and amassed regions) as a result of the sudden release of water from the subsurface probably due to the rapid melting of the Martian permafrost subsequent to the heat due to the volcanic activity (Masursky et al., 1977). A typical large outflow channel, *Mangala Vallis*, is shown in **Fig. 2.2**.

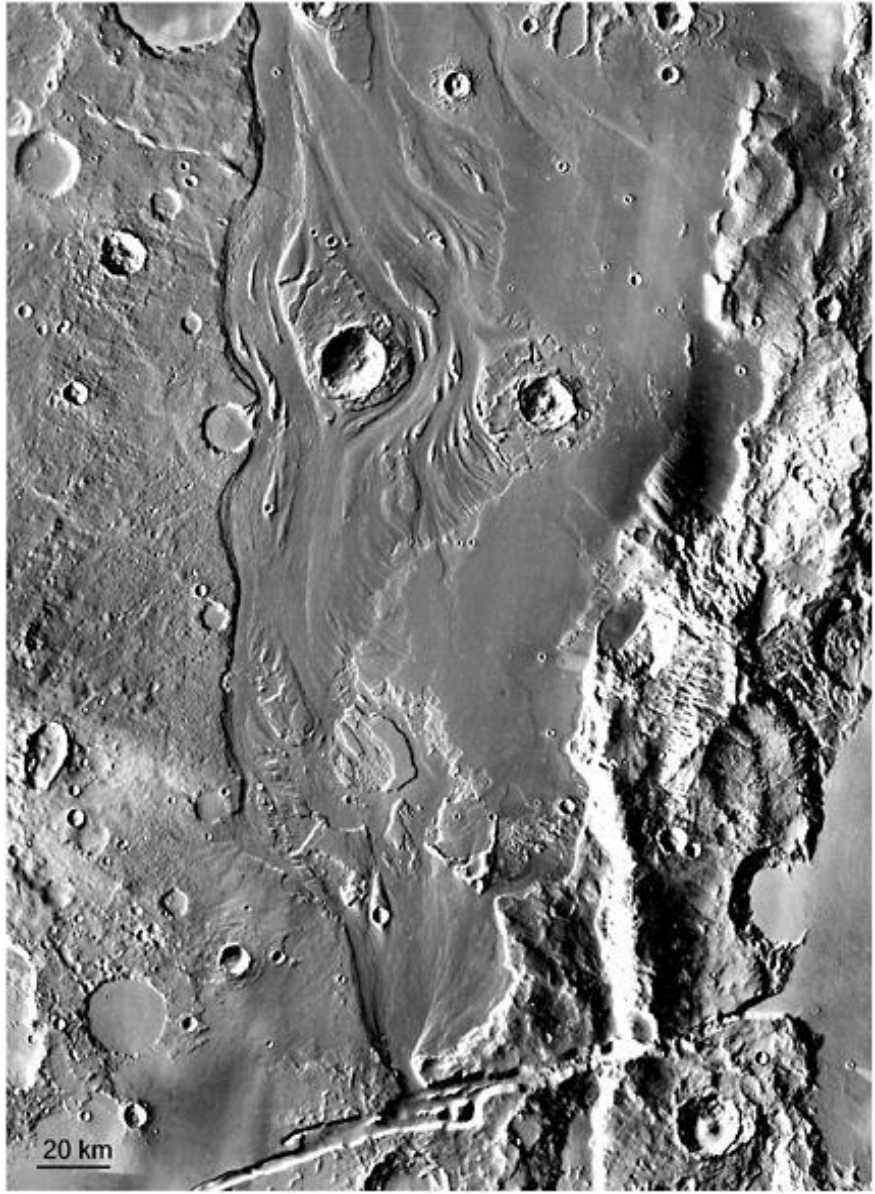

**Fig. 2.2.** *Mangala Vallis* (at 18.1°S 210.0°E), large outflow channel that starts at a 7 km gap in a graben wall (bottom center) and then extends hundreds of kilometers northward. Image credits: NASA/JPL/Arizona State University.

We decided to not include large outflow channels in our map because the duration of the water flow in these structures was ephemeral, so from a paleoclimatic point of view the largest of these structures are not diagnostic of ancient climate.

The distinction between the above reported groups of valleys was based on the analysis of the drainage pattern structures. These patterns, in fact, can be diagnostic since they are determined by the climate, the topography and the composition of the rocks. Consequently, we can see that a drainage pattern is a visual summary of the characteristics of a particular region both geologically and climatically. When we look at a drainage pattern we can draw conclusions about the structure formation and release of water as well as the climate.

We note, however, that a drainage feature cannot always be unambiguously assigned to one of the six above mentioned categories, and, in fact, there are some systems that show ambiguous features. *Ma'adim Vallis* (22.0°S, 177.3°W), for example, has network characteristics in its upper reaches but resembles to an outflow channel for part of its length (Irwin et al., 2002). This suggest that the valley probably formed by a combination of large flood and slow erosion. Also, *Mawrth Vallis* (22.6°N, 16.5°W) has characteristics of both outflow channels and networks.

Following Baker et al. (1992), we decided to include the *Ma'adim Vallis* in the group of longitudinal valleys while we consider *Mawrth Vallis* a small outflow. In fact, *Ma'adim Vallis* show a more branched structure and a long main branch. *Mawrth Vallis* consists mainly in a wide and long main branch and it is also located close to the outflow channels around *Chryse Planitia* (**Fig. 2.3**). At this point, to better understand our choice, it is important to see in more detail the different types of valleys mapped. For such reason, in the following subsections I will summarize the main features of these structures.

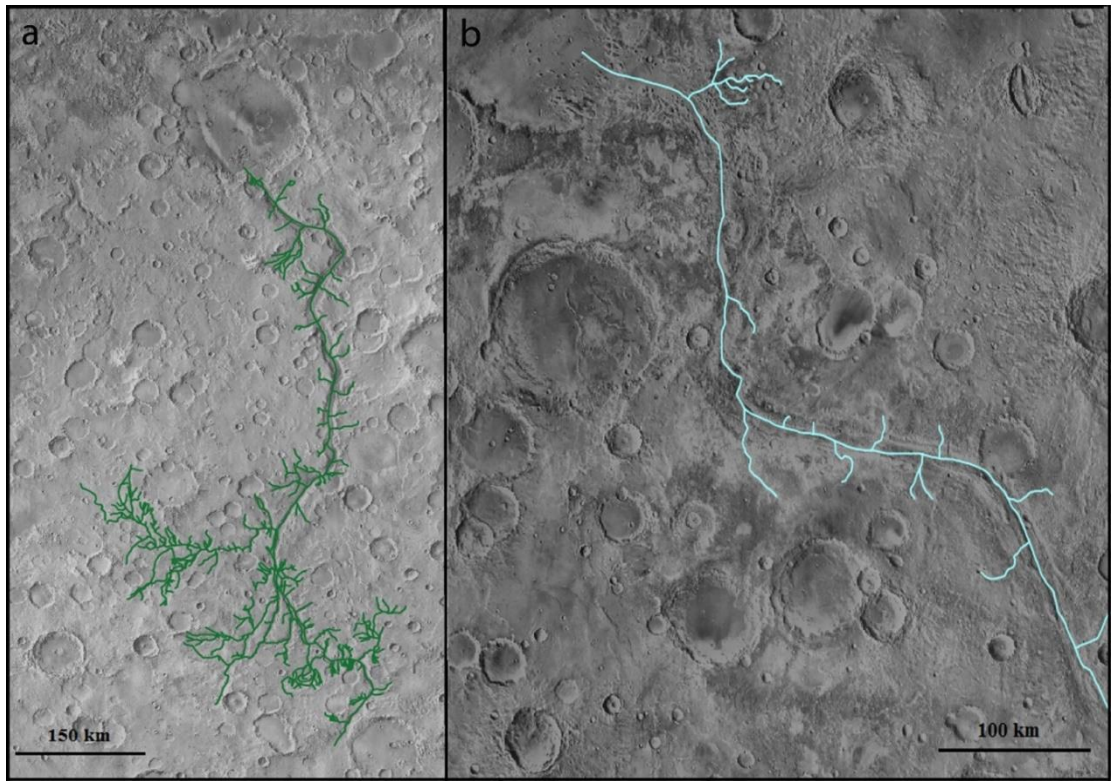

**Fig. 2.3.** a) *Ma'adim Vallis* located around 22.0°S 177.3°E, typical example of longitudinal valley. b) *Mawrth Valllis* at 22.6°N 16.5°W, example of small outflow channel. In both images the base map is the THEMIS daytime IR mosaic with a resolution of ~100 m/pixel.

### *2.3.1 Valley networks*

Most of the cratered uplands of Mars show a dissection by prevalent networks of branching valleys with many tributaries (Hynek et al., 2010), no more than a few kilometers wide but up to hundreds and even thousands of kilometers long (Carr, 2006). These structures were first observed in 1972 during the Mariner 9 mission (Masursky, 1973) and are known as valley networks.

Valley networks have a dendritic drainage pattern, which is very common on Earth and is characterized by the presence of many branches that usually are very short. This characteristic makes the dendritic drainage pattern very efficient because the branch structure allows a quick water flow from the source or sources to downslope (Carr, 1996).

Martian valley networks often begin near drainage divides (Craddock and Howard, 2002), and were interconnected across great distances, at least during their period of peak activity (see, e.g., Irwin et al., 2005b; Fassett and Head, 2008). Owing to their prevalent morphologic features, such as meandering branch and densely dendritic forms, these valleys seem to be close analogs to terrestrial fluvial valleys (Carr, 1995). In addition, very often these fluvial structures are related to possible paleolakes (Fassett and Head, 2008). From this point of view, their formation may imply a runoff origin.

However, as reported in Section 1 their morphology is sometimes ambiguous, and the interpretation of their features can be difficult. For example, most of them show steep walls and cross-sectional shapes that range from V-shaped in the upper reaches to U-shaped or rectangular in the lower reaches. Usually the valleys start with stubby, alcove-like terminations (Carr, 1995). These features led many researchers to think that valley networks on Mars were originated mainly by groundwater flow (Goldspiel and Squyres, 1991, 2000; Carr, 1995, 1996).

In addition, groundwater-driven valley erosion does not explain alone many other features of these valleys, such as the dendritic and high-order tributaries (Hynek et al., 2010). So, it is possible that some valleys formed as the result of groundwater discharge but, in any case, precipitation-based recharge seems to have been necessary to close the hydrological cycle. In fact, some authors (Goldspiel and Squyres, 1991; Gulick, 2001) suggested that subsurface water reservoirs would need to be refilled several times to produce the observed eroded volume.

However, the morphology of these valleys may not be pristine. It is, in fact, likely that after the incision of the valley its structure changed owing to resurfacing processes such as aeolian deposition, dust and volcanic mantling and impact crater ejecta. These phenomena may have caused a removal of the smaller tributaries of several Martian valleys (Craddock and Howard, 2002; Irwin and Howard, 2002; Hynek and Phillips, 2003; Irwin et al., 2008). In addition, it is important to observe that, usually, in nature, it is rare that a single process operates to the exclusion of the others. So, it is possible that a single valley is the result of different geomorphological processes. Each one of these processes can leave its own imprint or modification to the system's original morphology. The degree of modification is usually related to the entity of the environmental change. Sometimes these changes can have an impact so dramatic to totally erase or obliterate the original surface morphology (Gulick and Baker, 1990). In this point of view sometimes it can be complicated to asses a valley origin. In any case the formation of valley networks seems to require, at least, some periods when precipitation on the surface was possible and water was stable or metastable at the Martian surface (Craddock and Howard, 2002; Hynek et al., 2010).

### *2.3.2 Longitudinal valleys*

Longitudinal valleys (or sinuous valleys) are much larger than valley networks. They are narrow and sinuous and have lengths of hundreds of kilometers and widths of a dozen kilometers (Baker et al., 1992). Unlike the outflow channels, the valleys of this type never start from areas of chaotic terrain. Different hypotheses have been advanced about their origin. Some researchers believe that these channels formed by runoff (e.g. Masursky, 1973); many other authors, instead, claim that these valleys have been generated by basal sapping processes (e.g. Baker et al., 1992).

Basal sapping is subdivided into ground-ice sapping or ground-water sapping. The latter is observed when a collapse of the terrain occurs due to the flow of water that gradually erodes the above ground until it collapses (Craddock and Maxwell, 1993). In the ground-ice process sapping is, instead, the sublimation of the ice that causes the collapse of the ground.

In **Figs. 2.3** and **2.4** it is shown an example of longitudinal valley, *Ma'adim Vallis*, with its long and wide main branch.

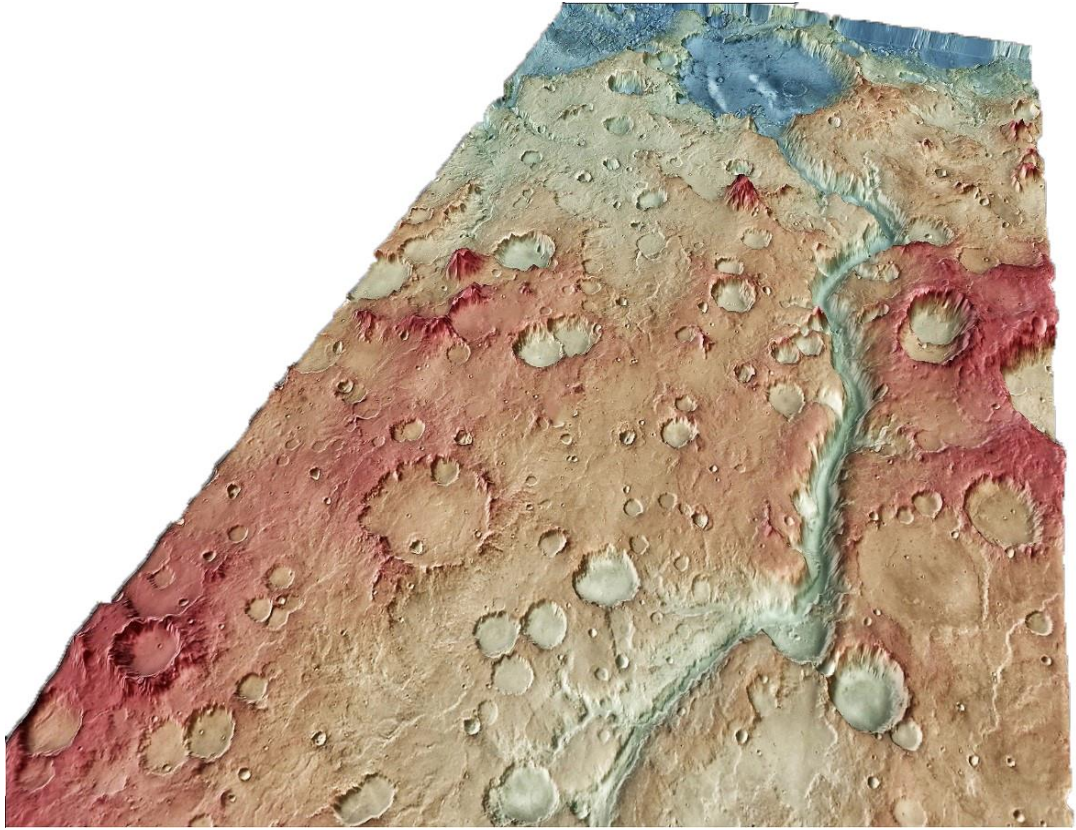

**Fig. 2.4.** 3D Rendering of the *Ma'adim Vallis* (at 22.0°S 177.3°W) obtained extracting a DEM from the MOLA mosaic ( ~ 460 m/pixel) and combining this DEM with THEMIS images ( ~ 100 m/pixel). The crater at the top of the image (*Gusev crater*) has a diameter of ~ 166 km. The whole image has been obtained using the tools of QGIS.

### *2.3.3 Valleys on volcanoes*

Many major Martian volcanoes show the presence of valleys (a typical example is shown in **Fig. 2.5**). These valleys have a radial pattern where the streams flow from a central point close to the center of the caldera. Unlike the dendritic systems where the pattern is determined by the direction of the slope of the land and the stream flow in more or less the same direction, in the radial pattern the flow has multiple directions starting from the highest point.

According to some authors (i.e. Gulick and Baker, 1990), many Martian valleys located on volcanoes had a two-stage evolutionary sequence based at first on surface runoff processes and then on groundwater sapping which expanded these structures.

In fact, comparisons of valley features on the Martian volcanoes with those of terrestrial valleys for which the origin is known, showed that a fluvial origin is compatible with most Martian valley features (Gulik and Baker, 1990). However, several lava tubes and channels of volcanic origin have been observed for example on *Alba Patera* or around the caldera of *Apollinaris Patera*. This is a sign of the fact that lava processes were also important in the formation of the channels (Gulik and Baker, 1990). However, the same authors (Gulik and Baker, 1990), analyzing the number of fluvial morphologic features present on six Martian volcanoes, concluded that, even if lava flows and volcanic activity could have initiated the formation of many valleys, then fluvial processes, like those on Earth, have long modified the original morphology with fluvial imprints which have obscured the primitive valley features.

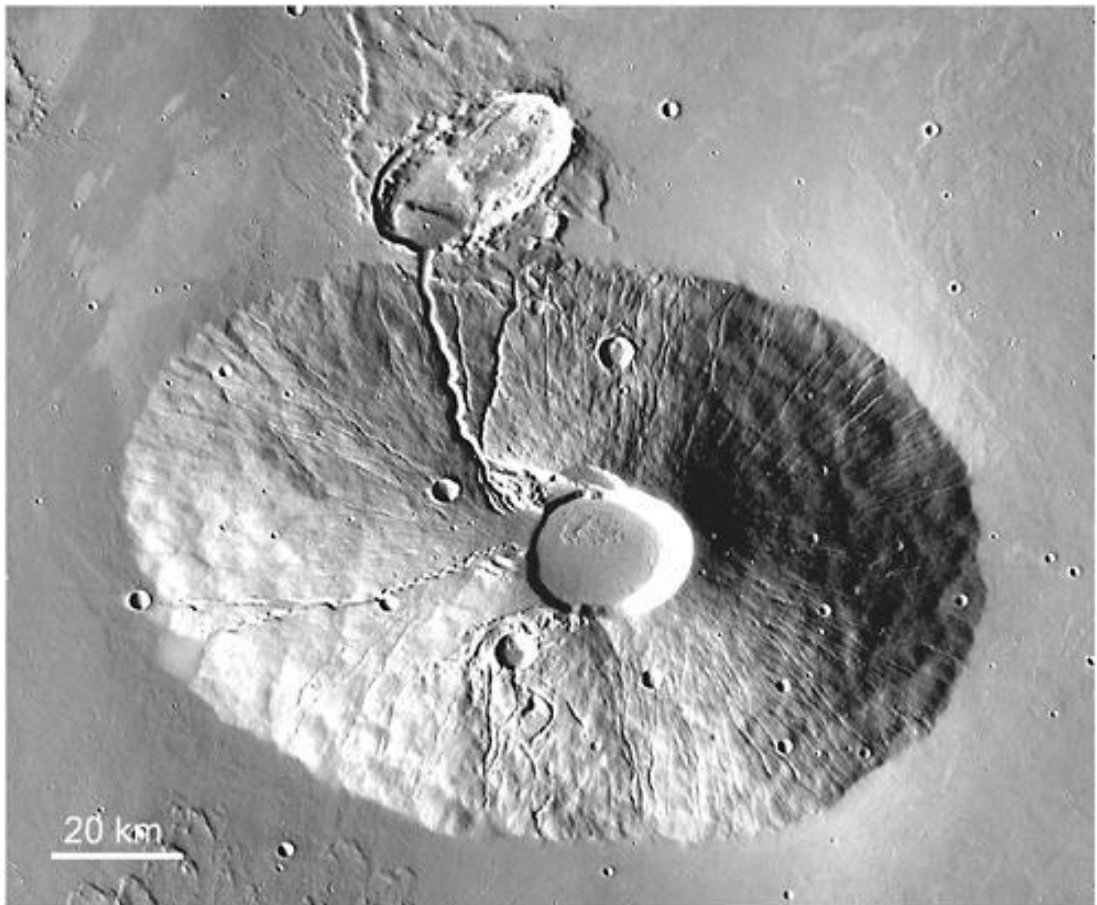

**Fig. 2.5.** *Ceraunius Tholus* volcano (centered at 24.3°N 262.7°E) which shows the presence of valleys originated as the result of melting of surface or subsurface ice by volcanic heat. Formation of the valleys may have been associated also with hydrothermal activity. Image credits: NASA/JPL/Arizona State University.

## *2.3.4 Valleys on canyons*

Many small and short valleys have been observed on outflow channels and on crustal fractures of the Martian surface. We included in this category the small tributaries of the huge outflow channels, like *Kasei Vallis*, and small valleys located on crustal fractures such as those of the fault zone of *Valles Marineris*. In general, the valleys located on both of these structures display theater-headed tributary terminations and primitive, angular, dendritic-drainage patterns. These features provide a strong argument for a sapping origin (Kochel et al., 1985; Baker et al., 1992). In fact, sapping valleys will often form along fractures and joints in the bedrock, resulting in a rectilinear and controlled pattern (Baker et al., 1992). A sapping pattern is often simple and few integrated like those observed for these valleys. However, in some cases, it is possible to see small dendritic valleys located on the floor of Martian canyons. For examples, some valleys located on *Valles Marineris* (on the plateau west of *Echus Chasma* and on the inner plateau west of *Melas Chasma*) show a network system with heads randomly located on the plateau (**Fig. 2.6**). According to Mangold et al. (2004), these features are characteristic of a surface runoff derived by rainfall.

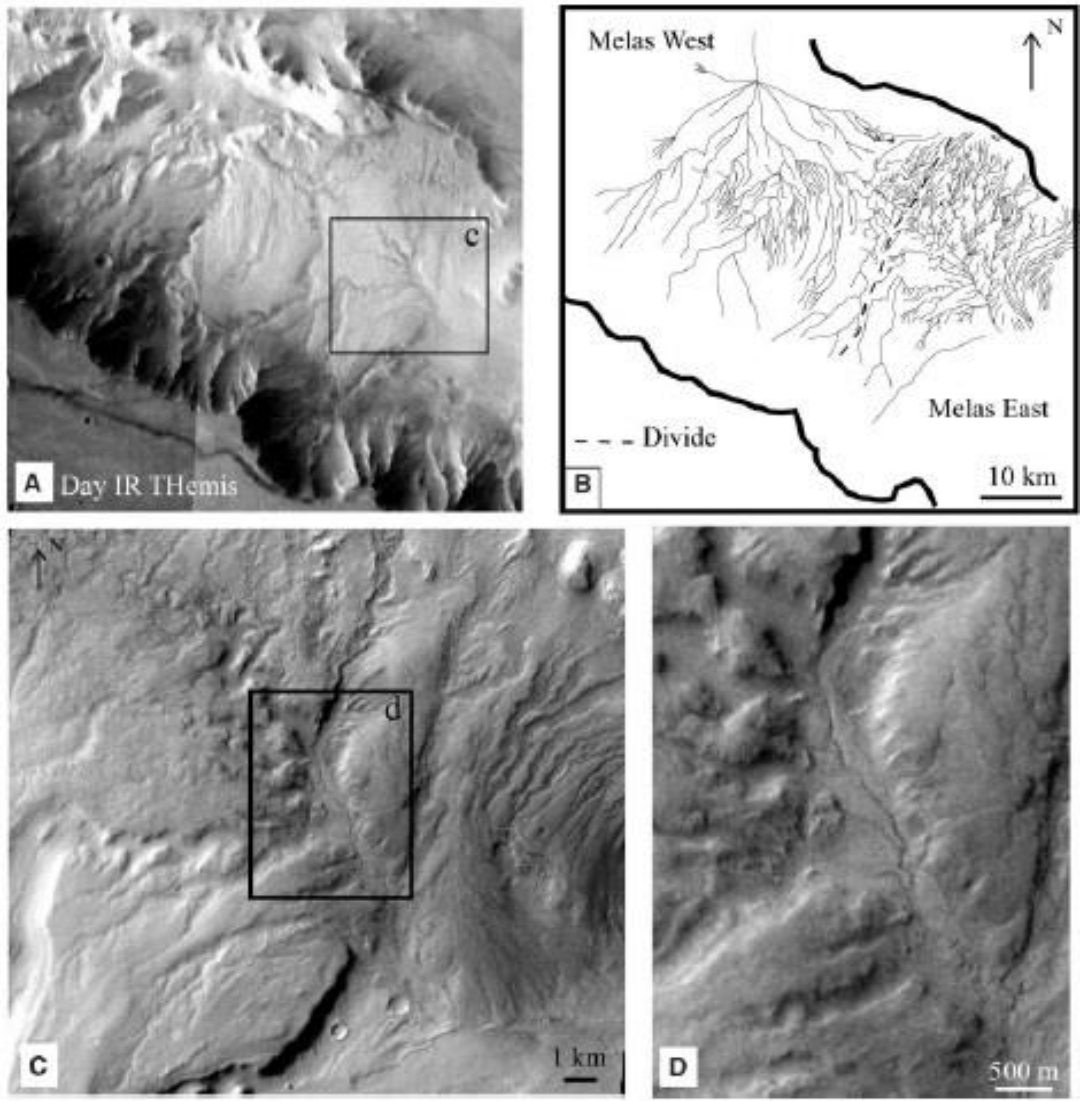

**Fig. 2.6.** Example of dendritic valleys located on the West inner plateau of *Melas Chasma* (10.0°S 77.5°W). (A) THEMIS IR daytime images (I06631017 and I06227001). (B) Valley network map of the area shown in (A) obtained combining IR and visible-light THEMIS images. Valleys, located around 2 km above the floor of *Melas Chasma*, are separated in West and East drainage basins (Mangold et al., 2004). (C) Dendritic valleys present on the East *Melas Chasma* and visibile in THEMIS visible-light image (V3249001). (D) Close-up over the central valley with inner channels. Image credits: Mangold et al. (2004).

## *2.3.5 Small outflow channels*

To the category of the outflow channels belong channels with a very large range of sizes: as already mentioned, the largest is *Kasei Vallis* and it is over 400 km across and 2.5 km deep, but other outflow channels may be less than 1 km across (Carr, 1996).

We remind that we have not mapped large structures such as *Kasei Vallis* or *Ares Vallis* but we introduced, instead, in our map the category of small outflow channels, i.e. channels that have a similar morphology of that of the real large outflow but on a smaller scale. The total length of the mapped small outflow channels varies in the range between 200 and 4000 km with a median value of 900 km. For example, we classified *Mawrth Vallis* as small outflow (see **Fig. 2.3**). Most seem to have originated by eruption of groundwater due to impacts, volcanic or tectonic events or by catastrophic drainage of lakes (Carr, 1996). In any case their presence across the whole planet surface indicates the abundant presence of water and water ice in the Martian crust (Carr, 1996).

### *2.3.6 Single valleys and systems of them*

Finally, many single small segments apparently disconnected one to another, are present on the Martian surface. In our map these segments were classified as single valleys and systems of them. With the latter we intend a group of single segments located very close one to each other. These segments do not seem to belong to any of the categories above mentioned and probably were once part of more integrated systems that were then eroded. In these cases, it is difficult to recognize their original morphology and to include them in one of the previous groups.

## *2.4 Mapping results*

Overall, the previous global maps of Luo and Stepinsky (2009) and Hynek et al. (2010) are similar to our map. However, at a fine scale, the maps differ significantly. With respect to the previous global maps (Carr, 1995; Scott et al., 1995; Carr and Chuang, 1997; Luo and Stepinsky, 2009; Hynek et al., 2010) data of higher image quality (new THEMIS mosaic and CTX data) and topographic information (MOLA data) allowed us to identify more tributaries for a large number of systems, to remove false positive and to find some new systems.

In **Fig. 2.7** you can see examples of mapped systems in comparison with the map of Hynek et al. (2010). In **Fig. 2.7a** and **b** is shown a valley located around 32.0°S 162.1°E in *Terra Cimmeria*. Using the global THEMIS mosaic with a resolution of ~ 200 m/pixel (**Fig. 2.7a**) only few parts of the system are visible, while, thanks to the new global mosaic (~ 100 m/pixel of resolution) it has been possible to see in more detail the valley morphology and to recognize the presence of other small tributaries (**Fig. 2.7b**). In **Fig. 2.7b, d, e** and **f** two other examples are shown: a small system located around 41.0°S 43.2°E (**Fig. 2.7b** and **d**) and another at 24.9°S 7.1°E (**Fig. 2.7e** and **f**). In both cases we can observe that in our map each system appears more developed with a higher number of tributaries.

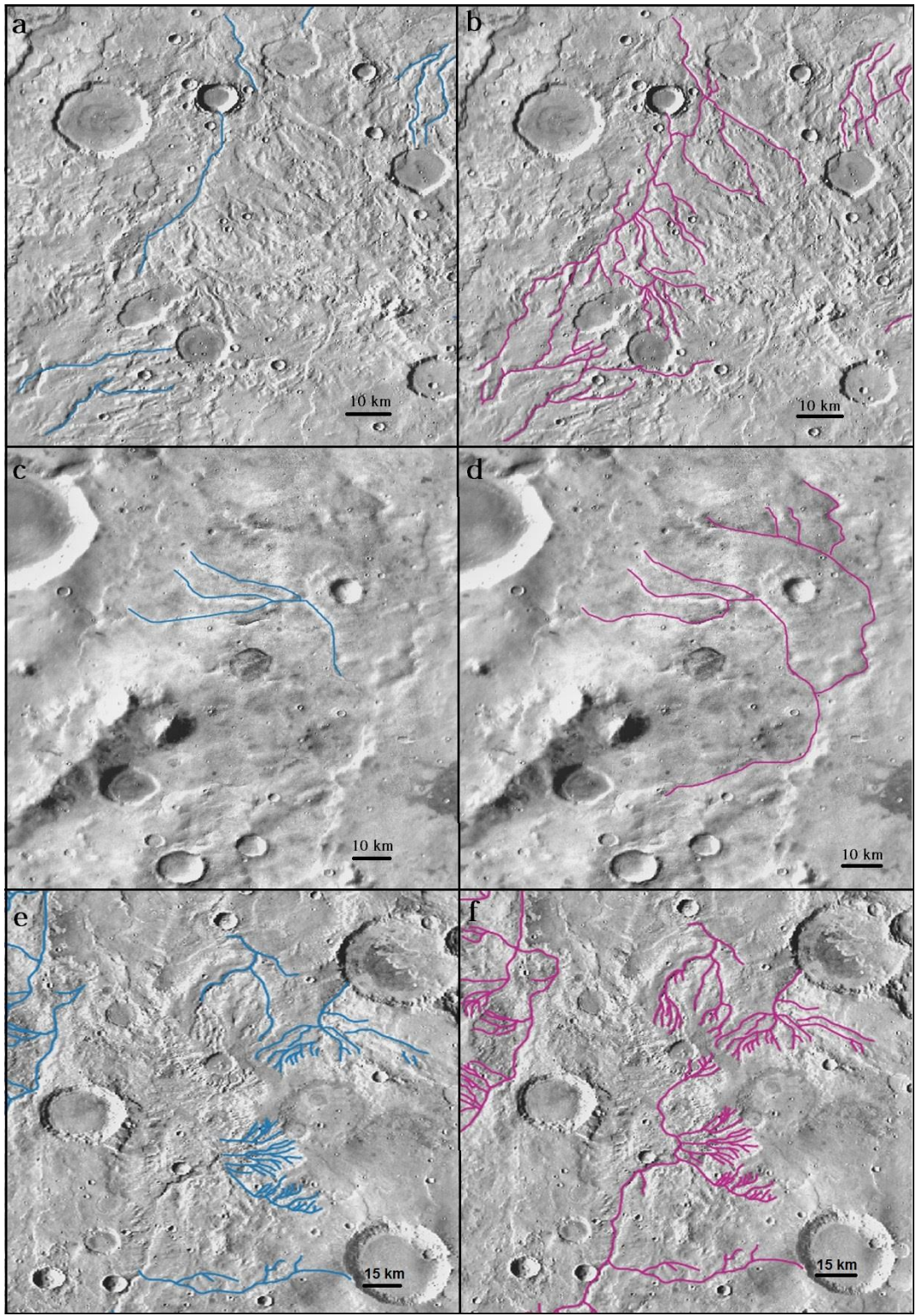

**Fig. 2.7.** Comparison between the valleys mapped by Hynek et al. (2010) using THEMIS data with a resolution of ~200 m/pixel (a, c, e) and those identifiable from THEMIS data at ~ 100 m/pixel (b, d, f). In a) and b) valleys centered around 32.0°S 162.1°E are shown; in (c) and (d) one can see some small system located around 41.0°S, 43.2°E; finally, in (e) and (f) some valleys at 24.9°S, 7.1°E are shown.

Globally, with respect to the previous global map of Hynek and colleagues I found new tributaries for 919 systems and 204 new small systems. At the same time, different segments (about 360) of the Hynek and colleagues map have been removed because they revealed to be false detection.

The complete map of Martian valleys is shown in **Fig. 2.8** where the different types of valleys are reported with distinct colors.

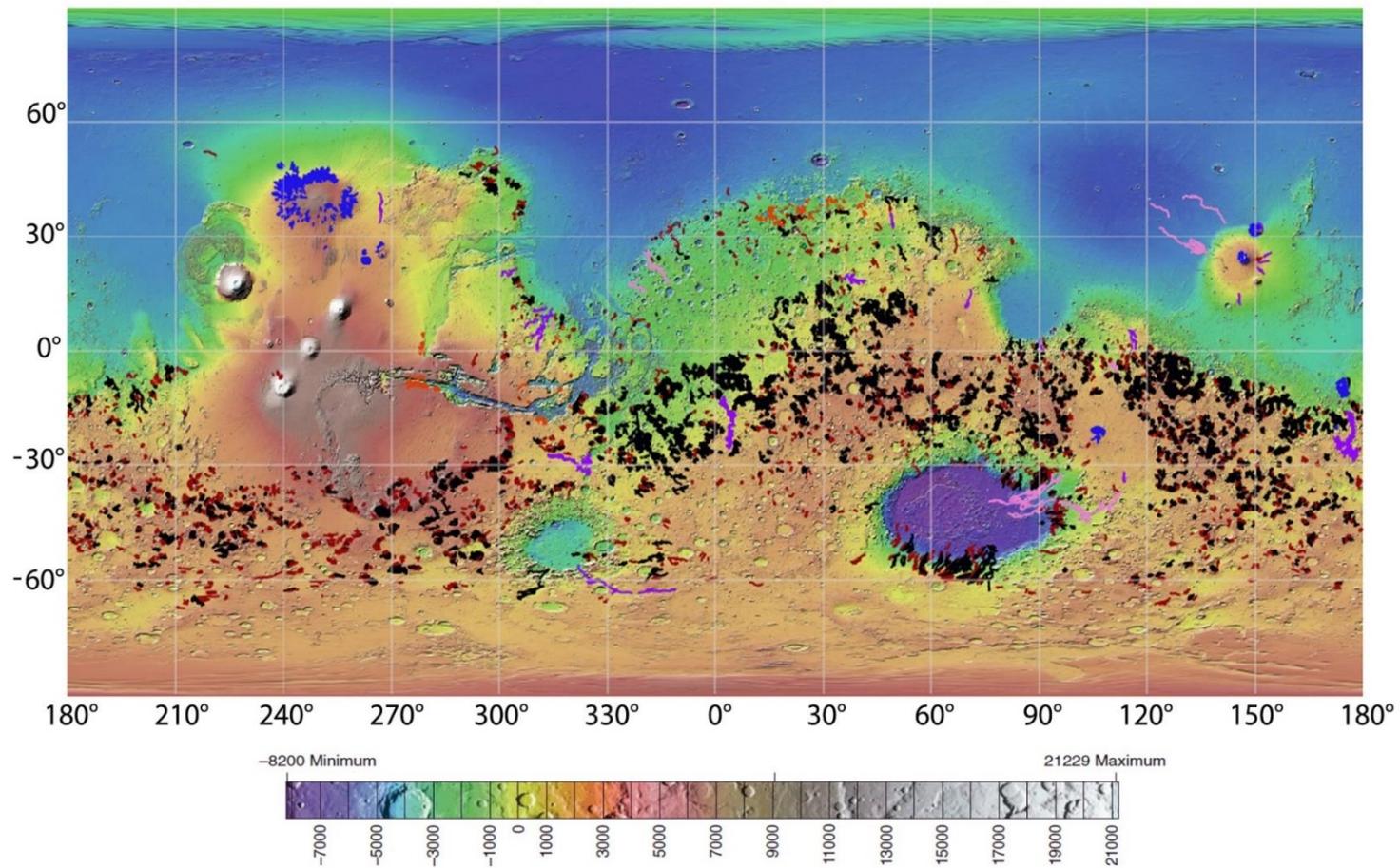

**Fig. 2.8.** Mapped Martian fluvial systems in different colors: valley networks (black); single valleys (red); longitudinal valleys (purple); valleys on volcanoes (blue); valleys on canyons (orange); small outflows (pink). The background map is based on MOLA Global Colorized Hillshade (credit: USGS Astrogeology Science Center, Goddard Space Flight Center, NASA). The whole image was obtained through the software QGIS.

As shown in **Figs. 2.8** and **2.9**, and reported in **Table 2.1**, the majority of valleys present on the planet surface are single structures (single valleys or disconnected systems of them). This is probably due to the fact that, as reported in Section 2.3.6, these single segments were part of more integrated systems that were then eroded. So now we can see only diffuse segments apparently not connected each other.

It is important to note that all the single segments associated to a specific epoch of the Martian history and located very close to each other were grouped together in a single geometry thus forming a system of single valleys. In this view, the number in **Table 2.1**, is not the number of each single segment but represents instead the number of systems of single valleys according to their position and age. In fact, as we will see in detail in the next section, to improve the information associated with our global map, we tried to assign an age to each valley assuming that a valley is as old as the terrain on which it has been carved (Carr, 1996; Hynek et al, 2010). To do that we used the recent and detailed global geologic map produced by Tanaka et al. (2014).

Using a similar approach, all the valleys present on a single volcano were mapped together in one geometry. So also in that case, the number of valleys of this kind, reported in **Table 2.1**, actually is not the real one but, rather, is the number of volcanoes incised by fluvial valleys (except for the valleys on *Apollinaris Patera* which are divided in two different groups associated with different periods of Martian history). In the same way, mapping small valleys associated with outflow channels, we grouped all the valleys connected to a crustal depression or an outflow channel in relation with their age. So, for one outflow we can have different group of valleys.

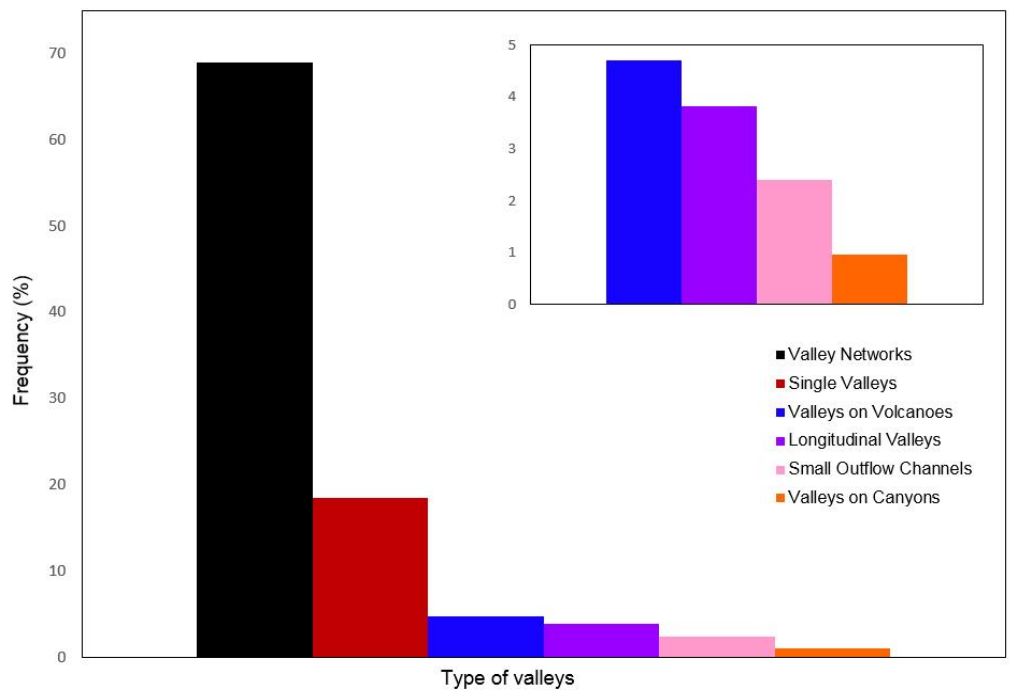

**Fig. 2.9** Distribution of the mapped Martian valleys (in relation with their total length) grouped by typology. Inset: detail of the distribution of the last four classes.

**Table 2.1**
Number, total length and percentage (in total length) of the mapped valleys

| Type | Number | Total length (km) | Percentage (%) |
|---|---|---|---|
| *Valley Networks* | 1638 | 537491 | 69 |
| *Single valleys and systems of them* | 2037 | 143270 | 19 |
| *Valleys on Volcanoes* | 10 | 36824 | 5 |
| *Longitudinal Valleys* | 22 | 29539 | 4 |
| *Small Outflow Channels* | 14 | 18493 | 2 |
| *Valleys on Canyons* | 51 | 7942 | 1 |
| TOTAL | **3772** | **773559** | **100** |

## *2.5 Age distribution and geographic location*

Thanks to global maps we have unique and useful tools to study and estimate the spatial and temporal sequences of geologic events that dominated the surface of a rocky planetary body. In addition, the combination of different kinds of maps allows a more complete and thorough analysis and understanding of these phenomena. For this reason, we tried to improve our dataset combining the information about the age of the Martian terrains obtained from the geologic map of Tanaka and colleagues (2014) with our global map of Martian valleys in order to have an estimation of the maximum age of these structures. This is a rough estimation, but according to Carr (1995) allow us to assign an age also to those valleys that are too small for age determination from superposition of impact craters and to have an idea of the global distribution of these valleys in relation with their age.

Previously, other geologic maps have been produced and used for such kind of analysis. The first geologic map was realized using Mariner 9 images with a resolution of 1-2 km/pixel and creating a photomosaic of them at 1:25,000,000 scale (Scott and Carr, 1978). Then, thanks to the Viking Orbiter data a series of three local 1:15,000,000-scale maps were generated with a resolution ranging from 100 to 300 m/pixel (Scott and Tanaka, 1986; Greeley and Guest, 1987; Tanaka and Scott, 1987).

Carr (1995) and Hynek et al. (2011) used the Viking data-based geologic map (Greeley and Guest, 1987; Tanaka and Scott, 1987) for the age evaluation of their mapped valleys. The improvement in the quality and resolution of orbital topographic and imaging data allowed a more detailed analysis of the Martian surface. Using MOLA (~ 463 m/pixel resolution), THEMIS (~100 m/pixel resolution mosaic) and where necessary CTX data (with a resolution up to 5-6 m/pixel), Tanaka et al. (2014) produced a new global geologic map at a 1:20,000,000 scale. The Martian surface has been divided into different units. An age range has been assigned to each map unit according to the Martian chronostratigraphic periods (Noachian, Hesperian and Amazonian - Scott and Carr, 1987). These periods has been subdivided into eight epochs (Early, Middle and Late Noachian; Early and Late Heperian, Early, Middle and Late Amazonian - Tanaka, 1986), based on stratigraphic relations and crater-density determinations. Each epoch has crater-density defined boundaries that are fit to widely used crater production and chronology functions (Hartmann and Neukum, 2001; Ivanov, 2001; Neukum and Ivanov, 2001; Hartmann, 2005; Werner and Tanaka, 2011; Tanaka et al., 2014).

The results obtained, combining the age information extrapolated from the geologic map of Tanaka et al. (2014) with our map, are shown in **Figs. 2.10** and **2.11**. We have grouped all the mapped valleys in three large categories corresponding to the three epochs of Martian history (Noachian, Hesperian and Amazonian). In addition, we included two other groups: Noachian – Hesperian and Hesperian – Amazonian. In the first case (Noachian – Hesperian), the majority of the valleys are located in the units defined by Tanaka et al. (2014) as "Hesperian and Noachian basin unit", "Hesperian and Noachian highland undivided unit" and "Hesperian and Noachian transition unit"; for these valleys we do not have a precise indication of the epoch because, according to the authors, they span the entire Noachian-Hesperian epochs. The valleys of the category Hesperian – Amazonian, instead, belong to the so called "Amazonian and Hesperian impact unit"; also in this case we do not have a precise indication of the age of the terrains in which the valleys are incised because, according to the map of Tanaka et al. (2014), they span the entire Hesperian-Amazonian periods.

It is important to note that, with this method of age estimation from the youngest unit that a valley cuts, we obtain for each valley an indication about its age that represent a maximum value. The determination of the true relative age of a valley is instead based on crater counting (Tanaka, 1986; Fassett and Head 2008; Hoke and Hynek, 2009). So far, some of the larger valleys systems have been dated using crater counting statistics assigning in this case an age which represents a minimum value of the actual age of the valley.

In addition, the maximum age assigned with this approach is affected by the uncertainties in the geologic mapping. As already mentioned, in some cases, extensive geologic deposits can be difficult to assign to a specific epoch (Carr, 1996; Tanaka, 1986) so for these units we do not have a clear indication of their age.

However, even with these intrinsic uncertainties, this method can give us a good indication of the valley maximum age distribution and it is also useful for the age estimation of many valleys that are too small for ages to be determined from superimposition of impact craters (Carr, 1995; Hynek et al., 2010). Moreover, so far, the global geologic map of Tanaka et al. (2014) is the most accurate planetary dating of Martian surface.

As previously mentioned, our approach is similar to that applied by Carr (1995) and Hynek et al. (2010) but in contrast to these previous works we used the new and, so far, more detailed geologic map of Tanaka et al. (2014). Considering the same ages breakdown, we can compare the results here obtained with those reported by Carr (1995) and Hynek et al. (2010) (**Table 2.2**). In comparison to the map of Hynek et al. (2010) we obtained that 94% of the mapped valleys lie within Noachian terrains, while 4% are carved in Hesperian terrain and only 2 % in the Amazonian one.

**Table 2.2**
Comparison of valleys age obtained in our work and for the map of Carr (1995) and Hynek et al. (2010)

| Valley's Age | Carr (1995) | Hynek et al. (2010) | This work |
|---|---|---|---|
| *Noachian* | 90% | 91% | 94% |
| *Hesperian* | 5% | 6% | 4% |
| *Amazonian* | 5% | 3% | 2% |

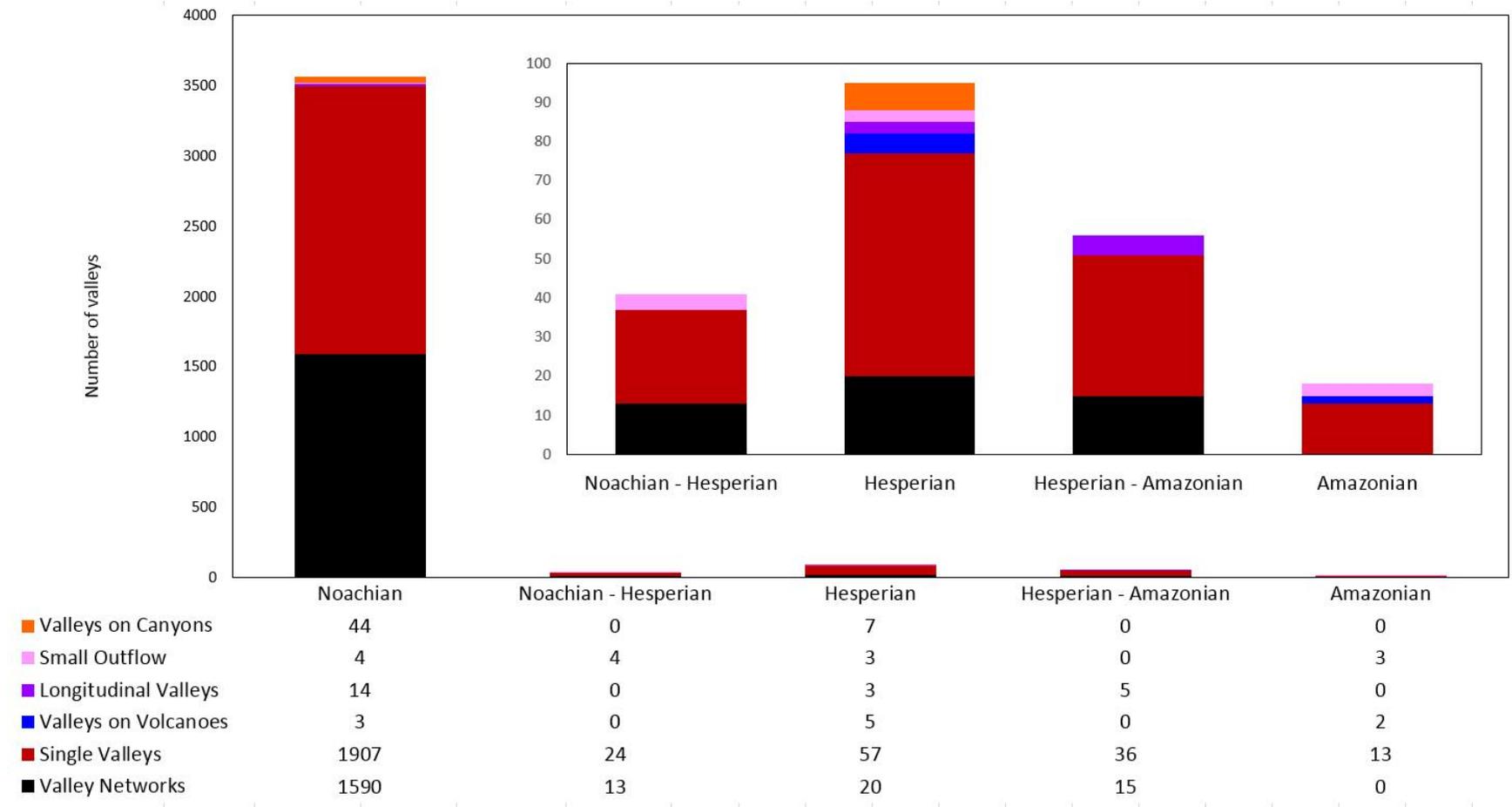

**Fig. 2.10.** Distribution of the mapped Martian valleys grouped by typology and by epoch. Same color code as **Figs. 2.8** and **2.9**. Inset: detail of the last four classes.

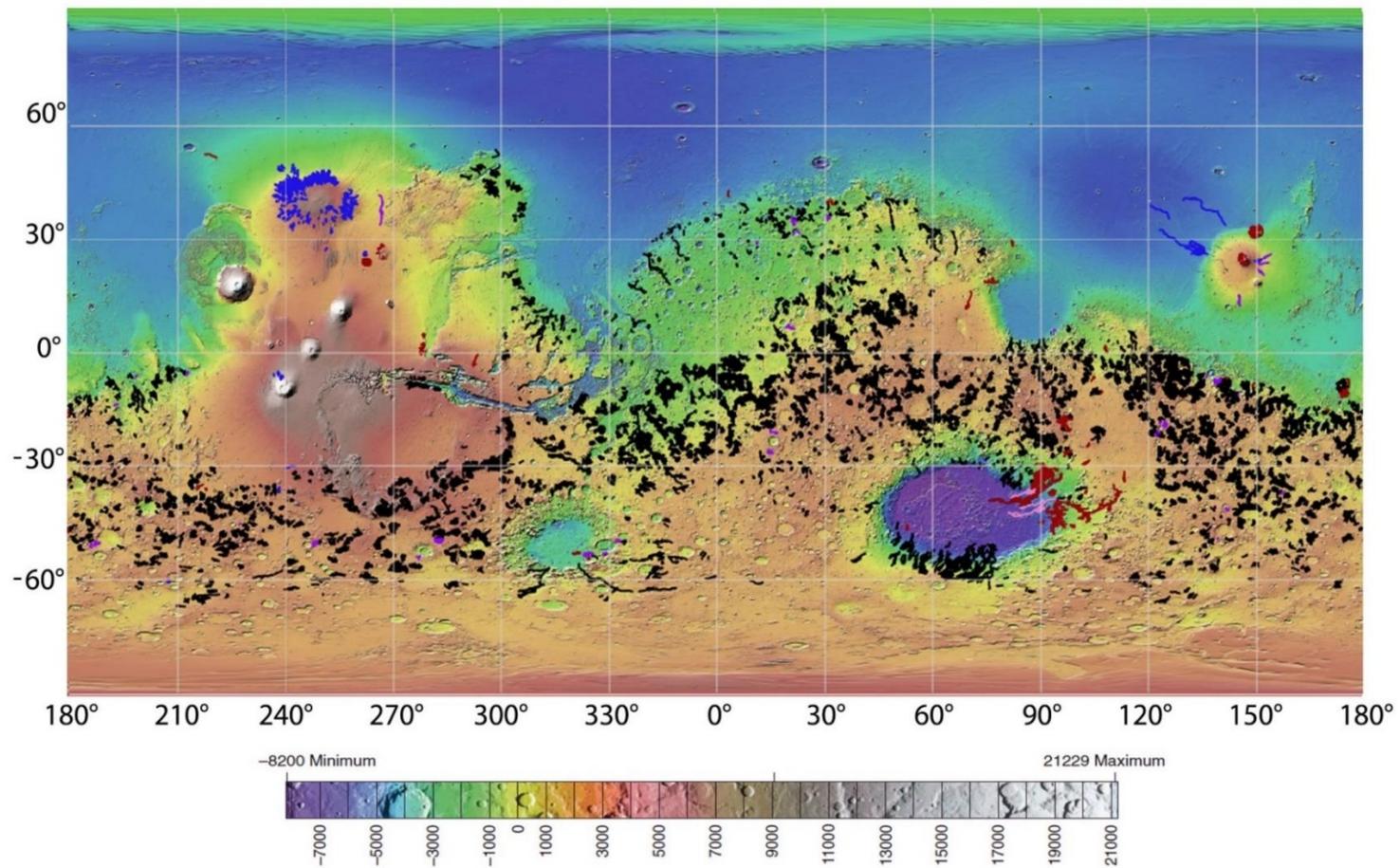

**Fig. 2.11.** Mapped Martian fluvial systems in different colors according to various age classes: Noachian (black); Noachian-Hesperian (pink); Hesperian (red); Hesperian-Amazonian (purple); Amazonian (blue). The background map is based on MOLA Global Colorized Hillshade (credit: USGS Astrogeology Science Center, Goddard Space Flight Center, NASA). The whole image was obtained through the software QGIS.

**Table 2.2** shows that the progressive improvement in the resolution of the photographic data leads to an increase of the Noachian valleys with respect to those identified by Carr (1995) and Hynek et al. (2010). In any case, according to these previous works, the vast majority of ancient valleys are incised in the Noachian units while the youngest networks are mostly where there are steep slopes, as on crater and canyon walls, and/or where high heat flows are expected as on volcanoes. The results here obtained confirm what was previously suggested: it appears that the rate of valley formation declined with time. Formation was widespread at the end of the Noachian but then become progressively more restricted with time to areas of steep slopes and/or high heat flows. In fact, in Hesperian and Amazonian terrains it is more likely to find outflow channels rather than valley networks (Carr, 1995; Hynek et al., 2010).

Fluvial valleys cut Martian volcanoes of all ages: as already acknowledged by Gulik and Baker (1990), this suggests that likely in some isolated regions of the planet surface there were active water flows until mid to late Amazonian. This means that the formation of the valleys on volcanoes can be related to local climatic changes due to volcanic activity and hydrothermal circulation (Gulik and Baker, 1990). As it happens on Earth, it is possible that locally warmer environments develop along regions of hydrothermal activity and this induces enhanced precipitation in these areas (Gulik and Baker, 1990). Another alternative scenario, proposed by Hynek et al. (2010), attributes the formation of valleys on volcanoes to periods characterized by a temporarily warm and wet climate characterized by rainfall.

In any case, it is important to note that, due to geological resurfacing events subsequent to the formation of valley networks (Hynek et al., 2010), the observed geographic distribution is likely an underestimation of the original distribution and partly represents the overprinting effects of later geological history. For example, recent mantling and terrain softening (Squyres and Carr, 1986; Kreslavsky and Head, 2000; Mustard et al., 2001) makes it increasingly difficult to recognize and map valley network drainage patterns at the higher latitudes of the southern hemisphere (south of 30°S).

If we consider in more detail all the valley networks and analyze their spatial distribution in relation with their total length and their age, we obtain the results shown in **Fig. 2.12**. In this case we considered all the eight epochs: Early, Middle and Late Noachian; Early and Late Hesperian, Early, Middle and Late Amazonian.

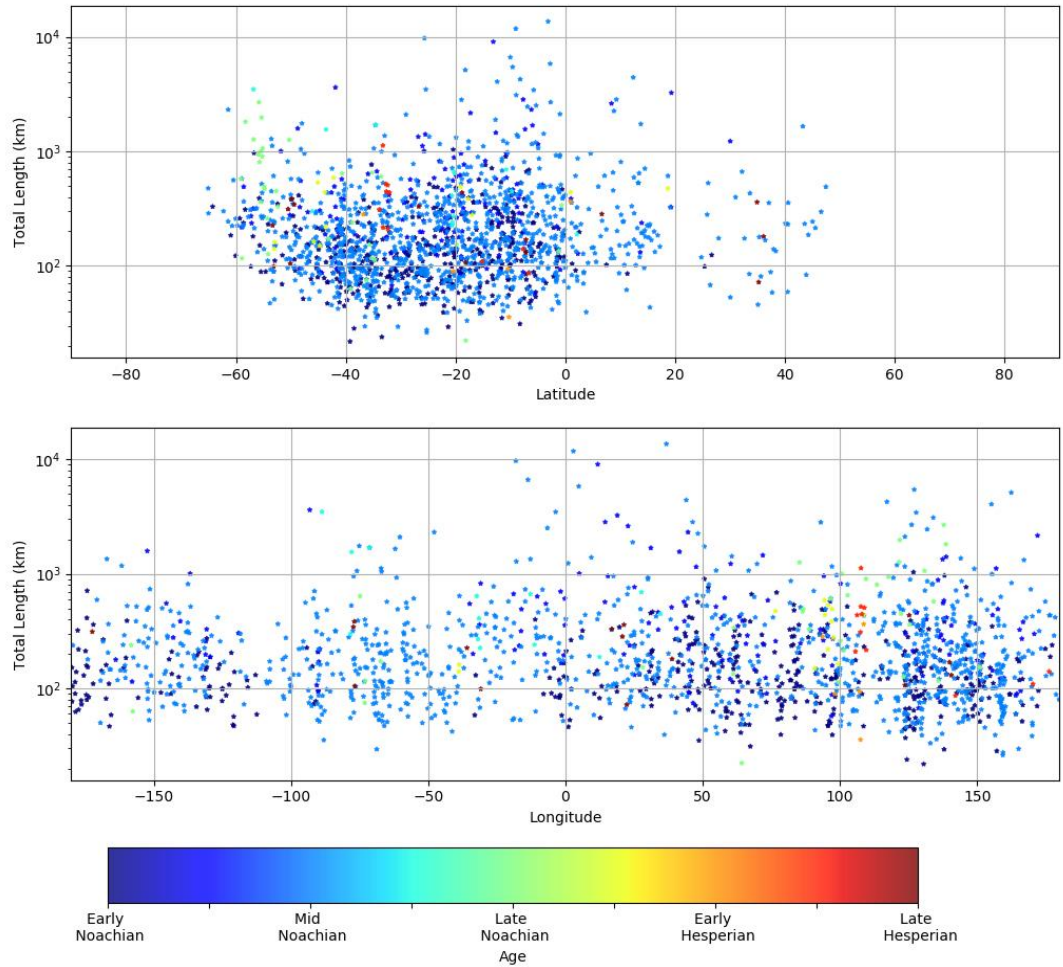

**Fig. 2.12.** Geographic distribution (with latitude – upper panel; with longitude – lower panel) of the mapped Martian valley networks with age and total length. The coordinates correspond to the central point of each mapped system.

As we can see in **Fig. 2.12**, all the valley networks are more or less evenly distributed with longitude (even though a greater concentration to the east longitudes is visible) while regarding the latitude, as already noted by other authors (e.g. Hynek et al. (2010) and references therein), these valleys are mainly located between 60.0°S and 20.0°N with few of them between 20.0°N and 50.0°N. The oldest valleys are mainly located in the south-east of the planet surface and are entirely Noachian.

As previously discussed, the Martian valley networks density is greater in the Noachian terrain. This supports the hypothesis widely discussed (Carr, 1995; Hynek et al., 2010) that most valleys originated during this phase of the Martian history. Then the activity decreased drastically as few valleys formed in the following period.

In some cases, we also observed well developed systems in two of these distinctive and consecutive epochs. For example, *Warrego Valles* is a well-developed system in the two distinct units: Early and Middle Noachian (**Fig. 2.13**). Following the criteria, above reported, of assigning an age on the basis of the youngest unit cut by a valley, we attribute a Mid-Noachian age to this system.

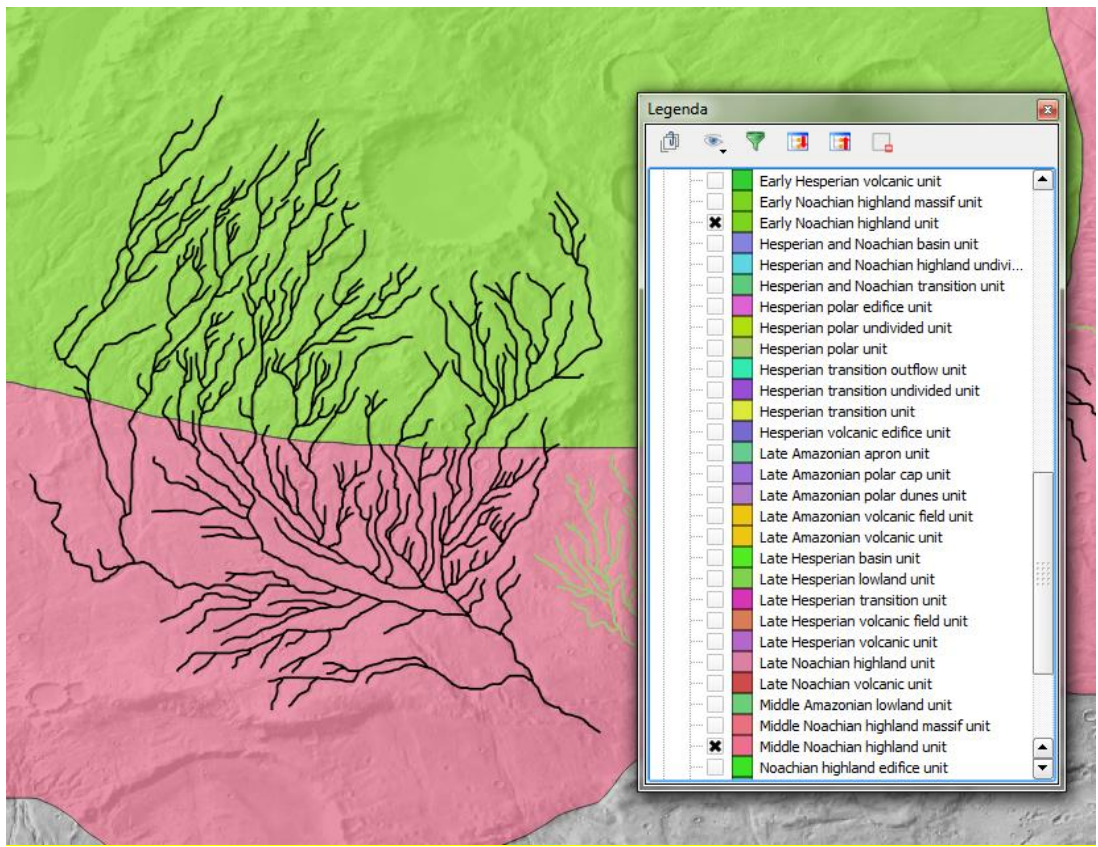

**Fig. 2.13.** Example of a valley well developed in two distinct units (Early Noachian highland unit in green; Middle Noachian highland unit in pink) of the map of Tanaka et al. (2014). The valley showed is *Warrego Valles* located around 42.2°S 93.0°W.

## *2.6 Summary and discussion*

In this work we updated previous global maps of Martian fluvial systems using the, so far, best resolution THEMIS mosaic (~100 m/pixel) plus CTX data (~6 m/pixel) and MOLA data (~463 m/pixel).

With respect to previous works, these data along with a manual approach and a thorough analysis allowed us to better map these structures at a finer scale and consequently to identify new tributaries for a considerable number of systems, some small valleys not previously mapped, and to remove false positives. To improve our map, I associated with the mapped valleys an attribute table containing valuable information such as coordinates, total length of the systems, and an estimated maximum age. For the latter, I combined the data obtained by our map we those of the map of Tanaka et al. (2014) which represents, to date, the most accurate dating of the planet surface. As already discussed, in this way we obtained an indication of the maximum age of each valley since we assumed that a valley is as old as the terrain on which it has been carved (Carr, 1995; Hynek et al., 2010). Even though this assumption may seem too strong, it is, actually, a good choice to have a global idea of the age distribution of these valleys and it is also a way to assign an approximative maximum age to valleys that are too small for age determination by means of crater counting techniques (Carr, 1996).

Several additional analyses may be done to improve our global dataset of Martian valleys. First of all, we are planning to evaluate the drainage densities of the mapped systems and to compare the values obtained with those of the previous manual maps (Luo and Stepinsky, 2009; Hynek et al., 2010). In addition, it will be useful to compare our dataset with other datasets of the Martian surface in order to better understand the origin of these features and their connection with the potential habitability of the planet. To allow further analysis, our updated global map is released to the scientific community on Zenodo (Alemanno and Orofino, 2017) and is also included in the Open Planetary Map (OPM) platform (Manaud et al., 2017). The Open Planetary is a team of planetary scientists, scientific programmers and planetary science enthusiasts who collaborate to create an easy way of finding solutions to problems related to planetary science data archiving, discovery and analysis. The main aim of this project is to create planetary interactive maps to allow collaborative learning, social interactions, and reusability of data. Our map was the first pioneer of OPM for data sharing and it is displayable at the following link:

https://openplanetary.carto.com/u/nmanaud/builder/7d0324f0-539e-45aa-87dd-d191e0858495/embed.

# CHAPTER III
# Estimate of the water flow duration in large Martian fluvial systems

## 3.1 State of the art

As discussed in detail in Chapter I, today Mars has a thin $CO_2$ atmosphere and is very cold and dry with respect to the Earth; nevertheless, it is likely that the planet may have experienced a warmer and wetter climate in the past due to the presence of a thicker atmosphere able to generate an important greenhouse effect and stabilize liquid water on the Martian surface (Sagan et al., 1973; Pollack et al., 1987; Kargel, 2004; Carr, 2006). This possibility, however, remains rather speculative since paleoclimatic models (Kasting, 1991; Haberle, 1998; Wordsworth et al., 2013; Urata and Toon, 2013; von Paris et al., 2013; Ramirez et al., 2014; Ramirez and Kasting, 2017) fail to give an univocal explanation for how and how long a warm and wet climate could have been sustained on early Mars. Moreover, physicochemical models (Kass and Yung, 1995; Johnson et al., 2000; Leblanc and Johnson, 2001; Manning et al., 2006; Lammer et al., 2013; Hu et al., 2015) do not explain the fate of the thick ancient atmosphere; in fact, they do not agree about the $CO_2$ mass still present on Mars in various reservoirs (polar deposits, regolith, and carbonate rocks) and/or lost by the planet in the last four billion years, even if very recent results of NASA's MAVEN mission strongly indicate that this loss could have been massive, mainly due to the atmospheric sputtering by the solar wind (Jakosky et al., 2017).

For the above-mentioned reasons, the study of the water flow within the valley networks is fundamental to understanding the formation mechanisms of these fluvial systems and to determining if and how long the planet has been characterized by a clement climate.

To achieve the latter goal several authors (Komar, 1979; Jaumann et al., 2005; Kleinhans, 2005; Kraal et al., 2008; Hoke et al., 2011; Morgan et al., 2014; Palucis et al., 2014) implemented sediment transport models, scaling to the Martian case semi-empirical relations previously found for terrestrial rivers.

Sediment transport equations have been derived from first order physics laws, and calibrated on terrestrial data in order to include the influence of several factors such as the size of the particles, the density, the superficial roughness, the fluid turbulence and the gravity (Meyer-Peter and Muller, 1948; Ribberink, 1998; Van Rijin, 1984). Some of these relations have been then applied to the Martian case (Komar, 1979; Wilson et al., 2004; Kleinhans, 2005).

In particular, models of sediment transport have been analyzed by several authors with the main goal of evaluating water and sediment discharge for Martian fluvial systems.

In the next subsections (3.1.1 and 3.1.2), I discuss the models proposed by Jaumann et al. (2005) and Hoke et al. (2011), since we will apply a similar approach to our valleys with visible interior channels. These methods, as we will see, investigate the formation timescales of Martian valley networks through the estimation of the amount of water that carved the valleys. To do that, they use empirical relationships between flood discharge and geometric parameters of the interior channel.

### *3.1.1 Jaumann et al. (2005) method*

Jaumann et al. (2005) applied a method based on the evaluation of water and sediment discharge through the analysis of the interior channel of a valley. For that reason, they choose a valley located in the western part of *Libya Montes* (between 1.4°N to 3.5°N and 81.6°E to 82.5°E) with a visible and preserved interior channel. At first, they derived some morphometric parameters, such as the width, the depth and the slope of the interior channel, by means of cross section and longitudinal profiles taken using MOLA and HRSC topographic data. These morphometric parameters were then used to evaluate the hydraulic parameters of the channel. From these parameters, they estimated the maximum water discharge (Q) by using a modified Manning's equation for steady, uniform flow taking into account the lower gravitational acceleration on Mars (Komar, 1979; Manning, 1981; Wilson et al., 2004).

To determine the formation timescales of the valley, they estimated the sediment discharges from Q ($Q_s = q \cdot Q$, with q sediment load) making simple assumptions based on analogies with terrestrial rivers. Jaumann et al. (2005) consider terrestrial transport loads on the order of 0.03 kg/m$^3$ to a few kg/m$^3$ but that may reach in semiarid rivers extremely high concentrations of about 40% sediment by weight (Leopold et al., 1964; Beverage and Culbertson, 1964; Nordin and Beverage, 1965). They assumed that Martian rivers would transport sediments more efficiently than terrestrial ones per unit discharge. This is probably owed to the low settling velocity and low critical velocity for suspension. They reported that according to previous analysis Martian rivers could reach sediment concentrations of 60-70% by weight (Pieri, 1980; Komar, 1980), equivalent to 600-700 kg/m$^3$. Based on these observations, Jaumann et al. (2005) assumed that for Mars a sediment load (q) of about 5 kg/m$^3$ seems to be reasonable. Although this sediment concentration is a rough estimation it can be used to constrain the duration of the valley formation based on an order-of-magnitude calculation (Jaumann et al., 2005). Once $Q_s$ was obtained, they evaluated the formation time T of the valley by the ratio between the eroded volume ($V_s$) and $Q_s$, finding T $\approx$ 2000 yr for continuous flow.

### *3.1.2 Hoke et al. (2011) method*

Hoke et al. (2011) estimated the formation timescales of seven Martian valley networks using three different models of sediment transport (Meyer-Peter and Muller, 1948; Van Rijn, 1984; Ribberink, 1998), the Darcy-Weisbach equation (D-W) and several parameters. Hoke and colleagues' (2011) approach consists of evaluating the parameters necessary for the application of the three above mentioned models. The eroded volume has been estimated by these authors using MOLA and HRSC data and the software ArcGIS, a Geographic Information System usually used for geospatial analysis.

As far as the grain size is concerned, the authors used a value obtained by Kleinhans ($D_{50}$ = 0.1 m). The channel widths were obtained using the CTX data with a resolution of 6 m/pixel.

For each channel, the width has been evaluated by the average of three values obtained at the end of its course. Where the interior channel of a valley is not visible, Hoke and colleagues used a value obtained for other valley networks since they assumed that globally all the valleys' widths are more or less similar (Hoke et al., 2011). Another important parameter is the channel depth which cannot be measured directly but can be derived indirectly analyzing the channel morphology. Therefore, there is a large uncertainty connected with this parameter determination. In the attempt to have a good estimation of the flow depth, Hoke and colleagues followed two different procedures (Hoke et al., 2011):

1) using the fixed width-depth ratio (w/h=58) based on the dimensions of terrestrial channels (Finnegan et al., 2005);
2) using the relation which connects the flow depth to the shear stress ($\tau$) (Hoke et al., 2011).

Once the flow depth is obtained it is possible to evaluate the flow velocity. The latter has been evaluated by Hoke et al. (2011) by means of the D-W equations and used to determine the sediment discharge with the following relation:

$$q_{b/s} = \varphi_{b/s}(Rg)^{1/2}(D_{50})^{3/2},$$

where $q_b$ is the *bed load* rate and $\varphi_b = 8(\vartheta - \vartheta_b^*)^{1.5}$ (Meyer-Peter e Muller, 1948), and $q_s$ is the *suspended load* rate and $\varphi_s = \frac{0.1}{f}\vartheta^{2.5}$ (Engelund e Hansen, 1967).

It is important to note that Hoke and colleagues assumed that the sediment transport in the Martian valley networks is dominated by bedload processes (Hoke et al., 2011). So, in this model the authors neglected the suspended load.

As already mentioned, Hoke and colleagues followed three different models to evaluate $\varphi_b$, i.e. Ribberink (1998), Meyer-Peter and Muller (1948) and Van Rijn (1984).

Once the $\varphi_b$ function is obtained it is possible to evaluate the sediment discharge Qs = $q_b \cdot$ w (where w is the width of the channel inside the valley) and, from this parameter, the valley networks formation timescales T using the relation:

$$T = \frac{V_s}{(1-\lambda)Q_s}$$

where $\lambda = 0.3$ is the mean porosity of the transported material.

With this method Hoke et al. (2011) evaluated the formation timescales for a sample of 7 Martian valley networks obtaining values ranging typically from $10^4$ to $10^7$ yr for an intermittency of the erosion processes equal to 5%.

In the present work we decided to adopt a modified version of the model by Jaumann et al. (2005) with the main goal of evaluating the duration of the fluvial activity for a sample of Martian valleys with and without visible interior channels.

In the following section, I present our sample and discuss how the geometric parameters required for the application of the model were evaluated. The model is described in detail, along with its results, in Section 3.3. Then, some comparisons are presented with the terrestrial cases and with previous works on Martian valleys (Section 3.4). In Section 3.5 the formation timescales obtained for some valleys in our sample are coupled with age-dating determinations from crater density analysis (Fassett and Head, 2008; Hoke and Hynek, 2009) in order to assess the time of onset of these structures and the number of valleys that were active at the same time. Finally, in Section 3.6, I discuss our results and their implications from a paleoclimatic and astrobiological perspective.

## 3.2   *Choice of the sample and determination of the geometric parameters*

Our sample of 63 Martian valleys includes Martian fluvial systems with a main branch longer than 150 km and with a total length (including all the tributaries) greater than 600 km, geographically widespread on the planet surface. We excluded from our selection highly eroded valleys since the estimate of some parameters necessary for the determination of the formation timescales (see below) would have been rather uncertain. Most of the valleys in our sample show features consistent with formation by precipitation, including densely spaced dendritic forms with interior channels that increase in width and depth downstream; sinuous main trunks and major tributaries that occasionally also exhibit multiple interior channels, braiding, and terracing.

Therefore, these valleys could give valuable information on the climate of the Red Planet during their formation/activity.

Our sample (see **Fig. 3.1** and **Table 3.1**) was divided in two groups: a) valleys with an interior channel (13 valleys); and b) valleys with no visible interior channels (50 valleys). A typical valley system of the first group is shown in **Fig. 3.2**.

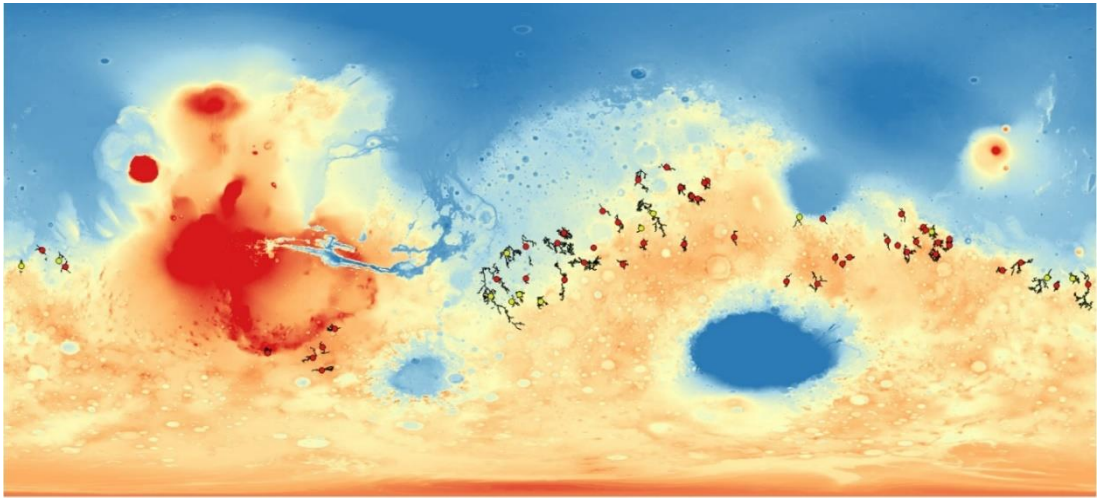

**Fig. 3.1.** Map of Mars based on MOLA data showing the valleys studied in this work. Valleys with an interior channel: yellow points; valleys without visible interior channel: red points. The base map is in false colors ranging from red for the higher areas to blue for the most depressed.

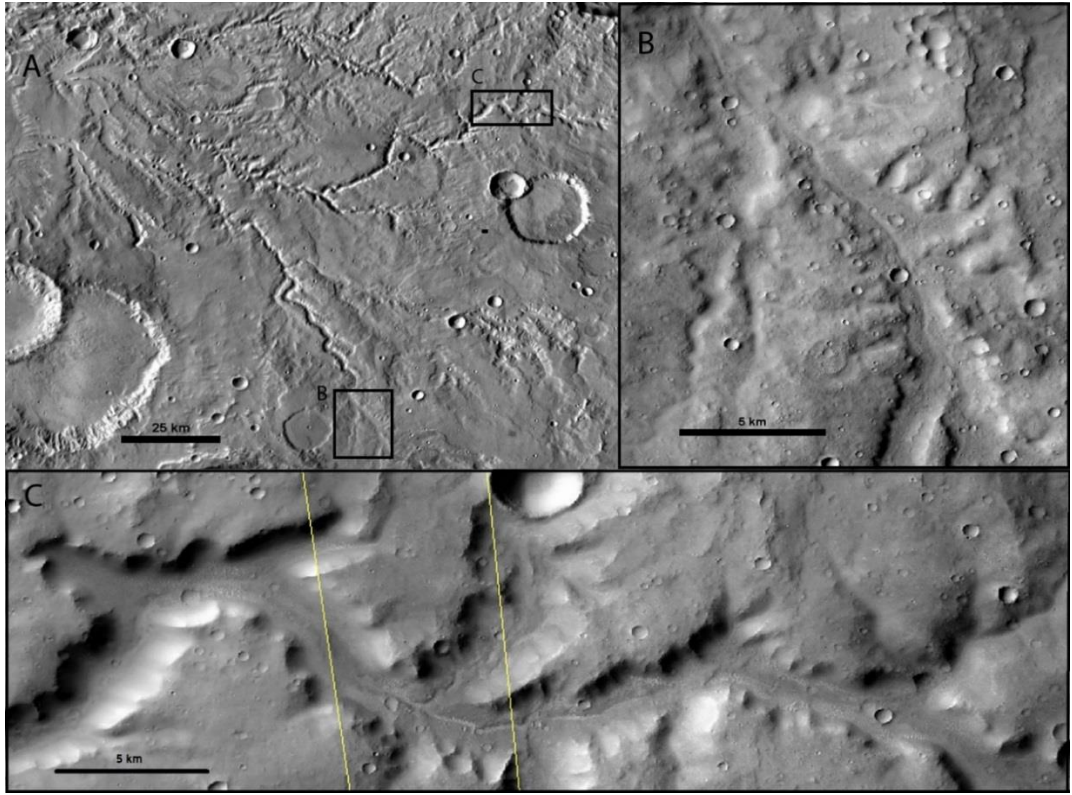

**Fig. 3.2.** (A): Photomosaic of *Parana Valles* (Valley ID 1) obtained using THEMIS imagery with a resolution of ~100 m/pixel. Two different portions of the valley system, where the interior channel is visible, are shown in the black boxes B and C. (B): Gray scale CTX image P03_002285_1562_XI. (C): Mosaic of the CTX images: B20_017409_1562_XN and J03_045971_1556_XN, whose boundaries are indicated by the yellow lines.

To estimate the formation timescales of the analyzed fluvial systems, some geomorphometric parameters have to be evaluated, such as area and volume for all the valleys in the sample, along with slope and width of the interior channel for the first group of valleys. To determine such parameters, we used QuantumGIS (QGIS), an open source Geographic Information Software, as a tool to create a map of Martian valleys (see Chapter II), mainly obtained by combining MOLA topographic data with THEMIS daytime infrared imagery. Also for these analyses, the global topographic MOLA data used are those with vertical resolution of ~ 30 cm/pixel and a horizontal resolution of ~ 463 m/pixel (Smith et al., 2001) and the THEMIS mosaic used has a resolution of ~ 100 m/pixel (Christensen et al., 2004). Where this spatial resolution was not sufficient (for very highly degraded systems), we also used CTX photographic data with a resolution up to 6 m/pixel.

**Table 3.1**
List of the valleys in our sample. The first thirteen are the valleys with visible interior channels. Along with the ID number, the coordinates of the middle point, the length of the main branch ($L_b$) and the total length of the system ($L_T$) are reported for each valley.

| Valley ID | Valley Name | Lat. (°N) | Long. (°E) | $L_b$ (km) | $L_T$ (km) |
|---|---|---|---|---|---|
| 1 | *Parana Valles* | -22.9 | 349.0 | 270 | 2961 |
| 2 | *Naktong Vallis* | 5.0 | 33.0 | 1100 | 10582 |
| 3 | *Zarqa Vallis* | 1.6 | 80.8 | 360 | 867 |
| 4 | | -0.9 | 30.0 | 220 | 929 |
| 5 | *Licus Vallis* | -2.4 | 126.1 | 350 | 2686 |
| 6 | | -9.8 | -14.0 | 390 | 6631 |
| 7 | | -12.0 | -161.8 | 500 | 1469 |
| 8 | | -13.6 | -174.3 | 292 | 712 |
| 9 | *Durius Vallis* | -17.7 | 172.1 | 280 | 2161 |
| 10 | *Al-Qahira Vallis* | -18.0 | 165.5 | 650 | 5127 |
| 11 | *Samara Vallis* | -23.8 | 340.9 | 1680 | 10515 |
| 12 | | -25.2 | -12.4 | 320 | 1345 |
| 13 | | -25.2 | -3.1 | 270 | 3474 |
| 14 | *Indus Vallis* | 19.2 | 38.7 | 420 | 1187 |
| 15 | *Scamander Vallis* | 15.8 | 28.5 | 720 | 3245 |
| 16 | | 14.7 | 51.8 | 270 | 706 |
| 17 | *Cuscus Valles* | 13.6 | 50.5 | 260 | 2667 |
| 18 | | 12.0 | 43.0 | 250 | 2420 |
| 19 | | 9.4 | 46.6 | 220 | 2834 |
| 20 | *Locras Valles* | 8.6 | 47.7 | 300 | 2904 |
| 21 | | 4.5 | 17.4 | 410 | 957 |

**Table 3.1** – continued

|     |              |       |        |     |       |
|-----|--------------|-------|--------|-----|-------|
| 22  |              | 3.3   | 115.7  | 400 | 805   |
| 23  |              | 1.8   | 90.0   | 240 | 326   |
| 24  |              | 0.0   | 23.0   | 440 | 2276  |
| 25  |              | -0.5  | 123.2  | 230 | 801   |
| 26  | *Verde Vallis* | -0.7  | 30.1   | 230 | 929   |
| 27  |              | -1.6  | 124.0  | 250 | 1163  |
| 28  |              | -3.0  | 5.0    | 390 | 5820  |
| 29  | *Naro Vallis*  | -4.0  | 60.6   | 345 | 1425  |
| 30  |              | -4.9  | 131.8  | 240 | 863   |
| 31  | *Tinto Vallis* | -5.1  | 111.4  | 370 | 1003  |
| 32  |              | -5.6  | 120.8  | 290 | 1193  |
| 33  |              | -5.8  | 128.1  | 240 | 3061  |
| 34  |              | -6.0  | 45.0   | 370 | 2315  |
| 35  |              | -6.2  | 31.0   | 310 | 1686  |
| 36  | *Tagus Valles* | -6.5  | 114.4  | 160 | 643   |
| 37  |              | -7.0  | 3.0    | 900 | 11477 |
| 38  |              | -7.3  | 131.6  | 380 | 2454  |
| 39  |              | -7.4  | -7.6   | 590 | 2149  |
| 40  | *Brazos Valles* | -7.6  | 18.9   | 150 | 2835  |
| 41  |              | -8.3  | -167.1 | 550 | 1349  |
| 42  |              | -9.2  | 117.1  | 450 | 4261  |
| 43  |              | -9.7  | 127.0  | 250 | 5446  |
| 44  |              | -10.5 | 98.8   | 170 | 977   |
| 45  |              | -10.8 | 94.1   | 160 | 961   |
| 46  | *Evros Vallis* | -12.0 | 13.9   | 780 | 9053  |
| 47  |              | -12.4 | 155.4  | 550 | 4057  |

**Table 3.1** - continued

| | | | | | |
|---|---|---|---|---|---|
| **48** | | -12.6 | 23.9 | 270 | 1888 |
| **49** | | -12.7 | 96.4 | 160 | 977 |
| **50** | | -13.5 | -159.6 | 370 | 525 |
| **51** | | -15.0 | 149.0 | 260 | 1107 |
| **52** | | -17.5 | -7.9 | 270 | 729 |
| **53** | | -18.0 | 78.4 | 400 | 1232 |
| **54** | *Loire Vallis* | -18.1 | 343.3 | 950 | 6237 |
| **55** | *Vichada Valles* | -18.6 | 88.1 | 500 | 2821 |
| **56** | *Marikh Vallis* | -19.2 | 3.9 | 1000 | 5362 |
| **57** | | -20.0 | 167.0 | 280 | 1322 |
| **58** | *Ma'adim Vallis* | -22.0 | 177.3 | 900 | 7158 |
| **59** | | -34.6 | -71.3 | 290 | 1725 |
| **60** | | -40.6 | -74.6 | 340 | 641 |
| **61** | *Warrego Valles* | -42.2 | -93.0 | 160 | 3592 |
| **62** | | -43.8 | -78.4 | 440 | 1553 |
| **63** | | -48.0 | -74.9 | 460 | 1754 |

## *3.2.1 Valleys volume and area*

To determine the eroded volume and area, each fluvial valley was mapped through polygonal shapes using instruments of the QGIS toolset. For each valley, a polygonal shapefile was produced using the "pencil" instrument 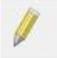 followed by the "add element" instrument 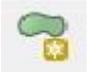 which allowed me to approximate the valley with polygonal shapes by tracing the vertices along the entire route of the valley system. The various polygons were joined in one big polygon for each valley using the "join

elements" instrument 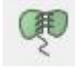. Each shape was manually drawn along the outer walls of visible valleys from topographic MOLA data and THEMIS daytime infrared images (100 m/pixel).

The result of this operation for a valley (*Evros Vallis* – Valley ID #46) is shown in **Fig. 3.3**.

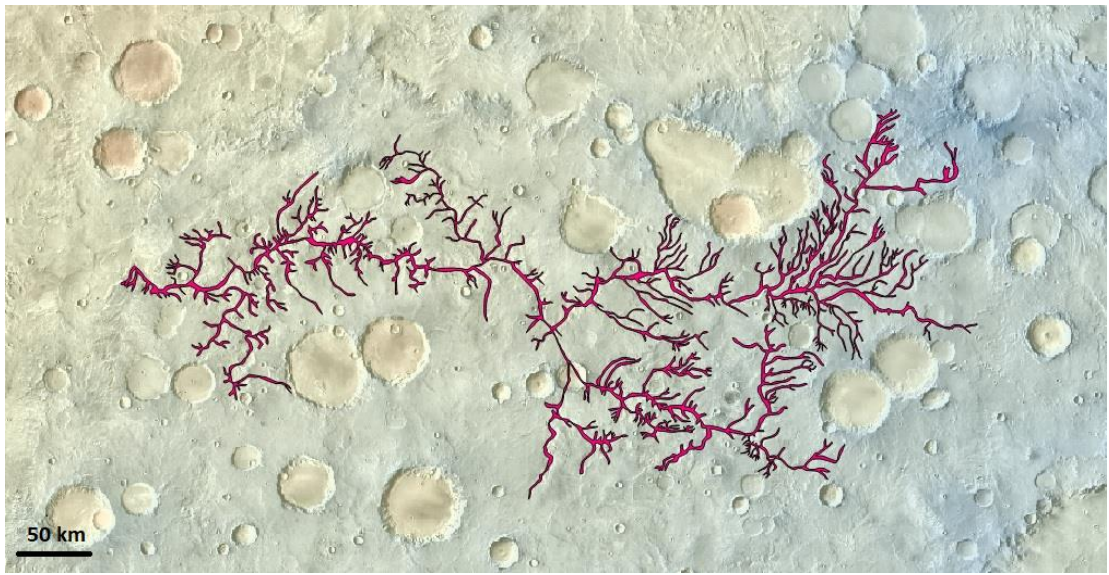

**Fig. 3.3.** The image centered around 12.0°S 13.9°E shows the polygon (in red) obtained by means of QGIS software for the *Evros Vallis* (Valley ID #46). The image is in false color ranging from brown for the most depressed areas to blue for the highest.

Once the above figures were made, I obtained the area of the shape through the Field Calculator of QGIS. The volume calculation was, instead, performed by the Grid Volume tool which allowed me to determine volume compared to a reference plane properly chosen. The input surface must be a raster, a TIN (Triangular Irregular Network), a digital data structure which is used in GIS to represent a surface, or a DEM (Digital Elevation Model) of the terrain.

I created for each valley a DEM. To do that I used the polygonal shapefiles realized and the MOLA data. We extracted from the MOLA mosaic the data for a valley using as mask the shapefile previously obtained.

The Grid Volume algorithm calculates the volume between a reference plane (Plane Height) and the surface. In principle the algorithm can work in two configurations: ABOVE or BELOW according to the position of the surface with respect to the Plane Height.

In the ABOVE configuration, the Plane Height corresponds to the minimum height of the surface (see **Fig. 3.4**). Volume and area calculations will represent the region of space between the specified Plane Height and the portions of the surface that are above the plane.

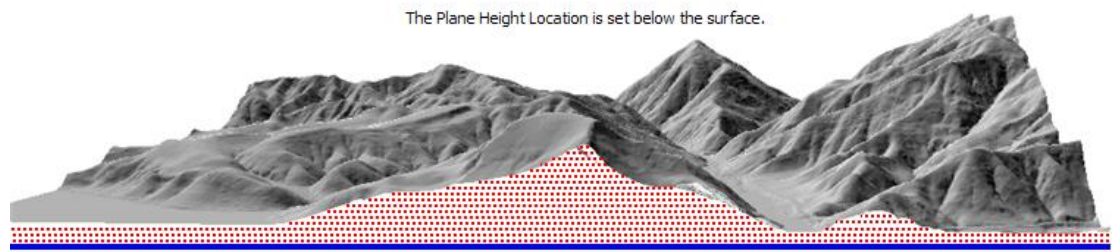

**Fig. 3.4** Example of the ABOVE configuration. The Plane Height (in blue) is located at the minimum height of the surface. As one can see the volume calculated (in red) it is that of the portion between the Plane Height and the surface. This and the following image have been obtained from http://resources.arcgis.com site.

In the BELOW configuration, the Plane Height is set to the maximum height from the surface (see **Fig. 3.5**). In this case the volume will represent the region of space between the specified Plane Height and portions of the surface that are below the plane.

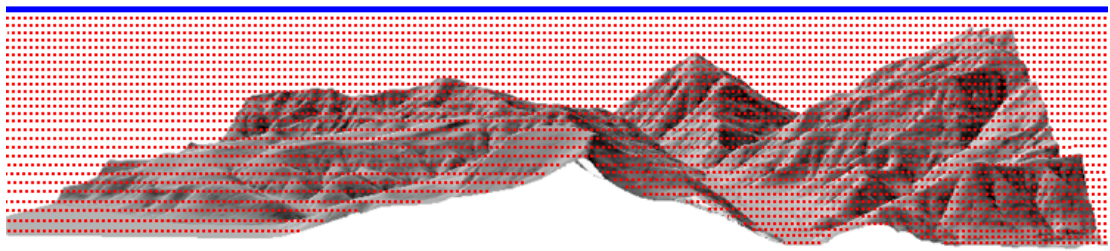

**Fig. 3.5** Example of the BELOW configuration. In this case, the Plane Height (in blue) corresponds to the maximum height of the system.

For the measurements, used in this work, I applied the configuration BELOW in which the Plane Height is set at the maximum height of the surface. In fact, having to determine the volume of water at the time present in the Martian valley systems, this is the most appropriate configuration to evaluate the volume of the "missing piece" (that is the eroded volume by water) and then to obtain, as we will see, the formation timescales of the valleys.

Each volume thus obtained has then been improved applying a correction which takes into account the slope of a valley. To each volume, I subtracted a quantity given by the product between the area of the valley and *h/2* where *h* is the difference between the maximum and the minimum elevation of the system. According to Hoke et al. (2011) it is assumed that infill of material since the end of valley network formation is negligible compared to the total volume of the valley.

For some valleys the volumes thus obtained were compared with volumes obtained from other authors and reported in literature (Goldspiel and Squyres, 1991; Cabrol et al., 1996; Hoke et al., 2011; Jung et al., 2012; Matsubara et al., 2013; Luo et al., 2015, 2017).

First of all, I compared the volumes obtained for seven of our valleys also included in the sample analyzed by Hoke et al. (2011) and for comparison by Luo et al. (2017). The results are shown in **Table 3.2**.

**Table 3.2**
Comparison between the volumes obtained in this work ($V_1$) and those obtained by Hoke and colleagues (2011) ($V_2$) and Luo et al. (2017) ($V_3$) for some valleys of our sample

| Valley ID | Valley Name | Lat.(°N) | Long.(°E) | $V_1$ (m³) | $V_2$ (m³) | $V_3$ (m³) |
|---|---|---|---|---|---|---|
| 2 | *Naktong Vallis* | 5.0 | 33.0 | 7.80E+12 | 8.48E+12 | 3.01E+12 |
| 18 | | 12.0 | 43.0 | 1.36E+12 | 1.70E+12 | 7.88E+11 |
| 24 | | 0.0 | 23.0 | 1.46E+12 | 1.50E+12 | 7.19E+11 |
| 28 | | -3.0 | 5.0 | 3.53E+11 | 2.80E+11 | 4.16E+11 |
| 34 | | -6.0 | 45.0 | 1.81E+12 | 2.10E+12 | 7.49E+11 |
| 37 | | -7.0 | 3.0 | 6.22E+11 | 7.20E+11 | 1.38E+12 |
| 46 | *Evros Vallis* | -12.0 | 13.9 | 5.37E+12 | 9.60E+12 | 1.48E+12 |

Hoke and colleagues evaluated the volumes using ArcGIS software. As in the present work, they extracted the valley topography using a mask obtained on THEMIS data, but they used the old mosaic with a resolution of 256 m/pixel. In the volume calculation, they also took into account the local slope of the surroundings surface. The agreement with our results is satisfactory, with maximum discrepancies around 20%; probably most of the differences are due to the different resolution of THEMIS data.

Later, some authors (Rodriguez et al., 2002; Luo et al., 2011), applied to DEM data a morphological image processing technique, known as Black Top Hat transformation (BTH) to evaluate the eroded volume of the valleys. This technique allows one to extract valley depth at pixel level. In principle, it works by constructing a pre-incision surface from the current DEM to estimate the valley depth and derive valley volume.

Luo et al. (2015) improved this method, developing a Progressive Black Top Hat (PBTH) algorithm that iteratively increases the windows size into the traditional BTH algorithm to capture the smaller and shallower features such as tributaries and thus improves the accuracy of the estimation of valleys' volume (Luo et al., 2015). Using this technique, Luo et al. (2017) evaluated the eroded volumes of the same valleys studied by Hoke et al. (2011). Their values (see **Table 3.2**) and those obtained in our work are generally consistent within one order of magnitude. The observed differences can be due to the distinct approaches used and how the valleys' areas were defined. In the present work, like in those of Hynek et al. (2010) those areas were outlined using a manual drawing and this can be the primary difference between these values. This approach can be more accurate even though it requires more time and more human intervention to outline the boundaries of a valley than the PBTH method which can generate this area automatically. Overall, as we will see, this would not result in significant differences in global estimate of VN volumes (Luo et al., 2017).

I then compared the volumes obtained for 16 valleys in our sample for which the volume has been obtained also by Matsubara et al. (2013). The results are shown in **Table 3.3**.

Matsubara et al. (2013) used a different approach for the volume determination. They applied a hydrologic routing model which uses MOLA DEMs from which post-Noachian craters have been replaced by a smooth surface (Matsubara et al., 2013) to determine the flow direction. The drainage network was automatically delineated by steepest-descent pathways, with overflow at the lowest point of enclosed depressions (Matsubara et al., 2013). They estimated valleys depths of selected reaches of Noachian highlands by searching MOLA data along the valley network within a 4 km × 4 km window and taking the difference between the 75$^{th}$ percentile value within the window (as elevation of valley floor). The local valley volume was measured by overlaying a 16 km × 16 km box centered on a valley and calculating the valley volume from the detrended 75$^{th}$ percentile elevation (Matsubara et al., 2013). With this method Matsubara et al. (2013) have evaluated in an indirect way the volume of 16 valleys that are also included in our sample. In this case the differences with our data are much larger than in the previous one, but generally the order of magnitude is the same, except for some systems, such as *Verde Vallis*, which shows the maximum discrepancy reaching a factor of 9 (see below). This situation is not surprising, however, since the method of Matsubara and colleagues is not addressed to a precise determination of the eroded volume.

**Table 3.3**
Comparison between the volumes obtained in this work ($V_1$) and those obtained by Matsubara et al. (2013) in two different cases: 1) wet environment ($V_4$); 2) dryer conditions ($V_5$).

| Valley ID | Valley name | Lat. (°N) | Long. (°E) | $V_1$ (m$^3$) | $V_4$ (m$^3$) | $V_5$ (m$^3$) |
|---|---|---|---|---|---|---|
| 1 | *Parana Valles* | 22.9 | 349.0 | 1.33E+13 | 2.96E+12 | 2.40E+12 |
| 2 | *Naktong Vallis* | 5.0 | 33.0 | 7.80E+12 | 1.98E+12 | 1.07E+12 |
| 3 | *Zarqa Vallis* | 1.6 | 80.8 | 3.16E+12 | 6.34E+12 | 6.06E+12 |
| 5 | *Licus Vallis* | -2.4 | 126.1 | 8.31E+11 | 5.12E+12 | 3.98E+12 |
| 9 | *Durius Vallis* | -17.7 | 172.1 | 7.04E+11 | 5.04E+12 | 4.24E+12 |
| 10 | *Al-Qahira Vallis* | -18.0 | 165.5 | 3.31E+12 | 6.61E+12 | 5.10E+12 |
| 11 | *Samara Vallis* | -23.8 | 340.9 | 1.14E+13 | 1.10E+13 | 9.58E+12 |
| 26 | *Verde Vallis* | -0.7 | 30.1 | 1.09E+11 | 9.50E+11 | 4.11E+11 |
| 29 | *Naro Vallis* | -4.0 | 60.6 | 8.91E+10 | 7.37E+11 | 6.74E+11 |
| 31 | *Tinto Vallis* | -5.1 | 111.4 | 7.42E+11 | 1.12E+12 | 7.92E+11 |
| 36 | *Tagus Vallis* | -6.5 | 114.4 | 2.52E+11 | 1.76E+12 | 1.31E+12 |
| 46 | *Evros Vallis* | -12.0 | 13.9 | 5.37E+12 | 4.15E+12 | 3.35E+12 |
| 54 | *Loire Vallis* | -18.1 | 343.3 | 8.82E+11 | 2.41E+12 | 1.90E+12 |
| 55 | *Vichada Valles* | -18.6 | 88.1 | 1.82E+11 | 9.01E+11 | 8.27E+11 |
| 56 | *Marikh Vallis* | -19.2 | 3.9 | 4.63E+12 | 3.51E+12 | 2.24E+12 |
| 58 | *Ma'adim Vallis* | -22.0 | 177.3 | 1.41E+13 | 1.13E+13 | 1.06E+13 |

For two of these 16 valleys, *Verde Vallis* and *Ma'adim Vallis* we have other volume estimations reported in literature using different approaches (Luo et al., 2015; Luo et al., 2011; Jung et al., 2012; Luo et al., 2015; Goldspiel and Squyres, 1991; Cabrol et al., 1996).

The eroded volume of *Verde Vallis* has been evaluated by Jung et al. (2012) applying an algorithm, known as Axelsson algorithm, based on progressive Triangular Irregular Network (TIN) densifications to detect valley networks. For comparison, the authors evaluated the volume of the same valley applying BTH technique (Rodriguez et al., 2002; Luo et al., 2011).

Luo et al. (2015) also applied the PBTH algorithm, to evaluate the eroded volume of the *Ma'adim Vallis*.

Goldspiel and Squyres (1991) estimated the volume of *Ma'adim Vallis* approximating its geometry and then breaking it into segments along its length and treating each segment as a trapezoidal prism. The volumes of the individual valley segments are summed to approximate the total volume of material eroded.

Cabrol et al. (1996) estimated the *Ma'adim Vallis* eroded volume evaluating a mean cross section of the valley and multiplying that for the total length of the main branch of the system. They hypothesize that a V-shape geometry represent the form of the valley before infilling by sediment deposition created the observed flat floor.

Unlike the other estimations, Goldspiel and Squyres (1991) and Cabrol et al. (1996) evaluated the eroded volume using the Viking data instead of the MOLA data which have a much better resolution.

**Table 3.4** summarizes the results concerning *Ma'adim* and *Verde Vallis* obtained by these different approaches. As it can be seen, the various evaluations of the volume eroded from the two valleys are in good agreement each other, apart from those obtained by Matsubara et al. (2013), especially for *Verde Vallis* (for the reasons outlined before).

**Table 3.4**
Comparison between the volumes of *Verde Vallis* and *Ma'adim Vallis* obtained in the present and in previous works, using different methods. Matsubara et al. (2013) have obtained for the two valleys different values adopting the same procedure but in two different environmental conditions (see text); Luo et al. (2015) have derived different volumes for *Ma'adim Vallis* using the same algorithm (BTH or PBTH), but with two different input data (HRSC or MOLA).

| **Verde Vallis - (Valley ID 26)** | | |
|---|---|---|
| This work | | $1.09\text{E}+11\ m^3$ |
| Jung et al. (2012) – BTH | | $1.22\text{E}+11\ m^3$ |
| Jung et al. (2012) – Alexsson algorithm | | $1.41\text{E}+11\ m^3$ |
| Matsubara et al. (2013) | wet | $9.50\text{E}+11\ m^3$ |
| | dry | $4.11\text{E}+11\ m^3$ |
| **Ma'adim Vallis - (Valley ID 58)** | | |
| This work | | $1.41\text{E}+13\ m^3$ |
| Luo et al. (2015) – BTH | HRSC | $1.35\text{E}+13\ m^3$ |
| | MOLA | $1.25\text{E}+13\ m^3$ |
| Luo et al. (2015) – PBTH | HRSC | $1.46\text{E}+13\ m^3$ |
| | MOLA | $1.32\text{E}+13\ m^3$ |
| Goldspiel and Squyres (1991); Cabrol et al. (1996) | | $1.40\text{E}+13\ m^3$ |
| Matsubara et al. (2013) | wet | $1.13\text{E}+13\ m^3$ |
| | dry | $1.06\text{E}+13\ m^3$ |

It is also evident that the method of the mean cross section produces sometimes an overestimation and other time an underestimation of the eroded volume of a valley.

I tested in fact the method of the mean cross section applying it to our sample to some of our valleys. The results are shown in **Table 3.5** in comparison with QGIS estimations.

**Table 3.5**
Comparison between the volumes obtained using QGIS algorithm plus correction ($V_1$) and those ($V_6$) evaluated using the mean cross section (S) and the total length ($L_T$) for each valley.

| Valley ID | Valley name | Lat. (°N) | Long. (°E) | S (m²) | $L_T$ (m) | $V_1$ (m³) | $V_6$ (m3) |
|---|---|---|---|---|---|---|---|
| 2 | *Naktong Vallis* | 5.0 | 33.0 | 2.56E+05 | 1.06E+07 | 7.80E+12 | 2.71E+12 |
| 18 | | 12.0 | 43.0 | 4.40E+04 | 2.42E+06 | 1.36E+12 | 1.06E+11 |
| 24 | | 0.0 | 23.0 | 5.60E+04 | 2.28E+06 | 1.46E+12 | 1.27E+11 |
| 28 | | -3.0 | 5.0 | 1.25E+05 | 5.82E+06 | 3.53E+11 | 7.27E+11 |
| 34 | | -6.0 | 45.0 | 9.80E+04 | 2.32E+06 | 6.22E+11 | 2.27E+11 |
| 37 | | -7.0 | 3.0 | 2.50E+05 | 1.15E+07 | 7.17E+11 | 2.87E+12 |
| 46 | *Evros Vallis* | -12.0 | 13.9 | 2.67E+06 | 9.05E+06 | 5.37E+12 | 2.42E+13 |
| 58 | *Ma'adim Vallis* | -22.0 | 177.3 | 9.30E+06 | 7.16E+06 | 1.41E+13 | 6.66E+13 |
| 61 | *Warrego Valles* | -42.2 | -93.0 | 1.60E+05 | 3.59E+06 | 2.77E+12 | 5.75E+11 |

### *3.2.2 Total eroded volume estimation*

Using the PBTH method, Luo et al. (2017) also estimated the total volume of the Martian valley networks obtaining a value of $2.96 \times 10^{14}$ m³.

For all the valleys in our sample, I obtained a total volume around $2.99 \times 10^{14}$ m³. If we consider that our sample contains all the greatest valleys of the Martian surface, we can say that this volume is representative, in terms of order of magnitude, of the total volume.

A rapid test can prove this statement. In order to estimate the contribution to the total eroded volume of the remaining valley networks, I, at first, analyzed the distribution of the total length of these valleys using the data from the global map (Chapter II). From this distribution (**Fig. 3.6**), I obtained a mean value of 190 km and a median of 150 km.

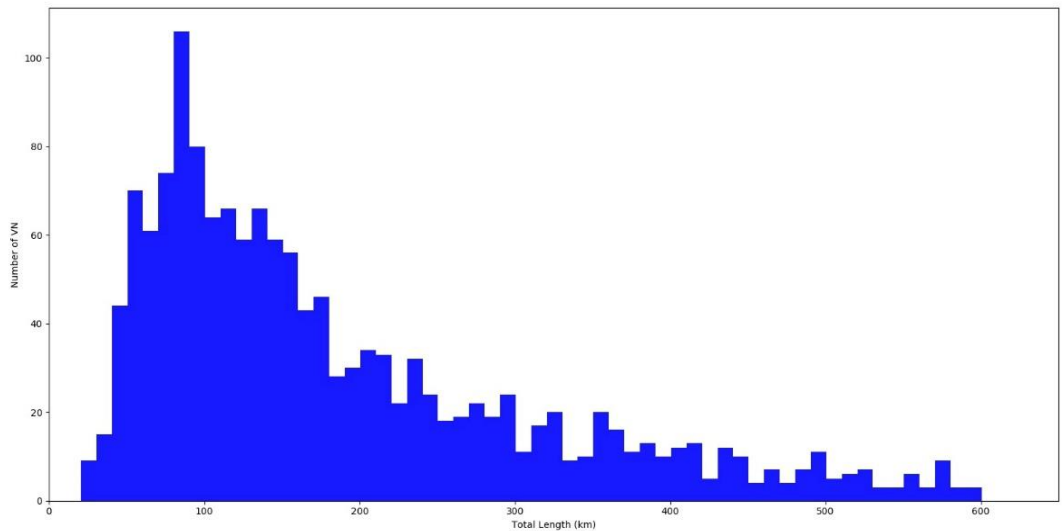

**Fig. 3.6** Distribution of the total length of the mapped Martian valley networks with a total length smaller than 600 km.

Considering a sample of valleys with a total length around the average value (in particular between 180 and 200 km) I obtained a mean volume of $1.66 \times 10^{10}$ m$^3$. By multiplying this volume for the number of the remaining 1575 valley networks (assuming that this dimension is representative of all the remaining valleys), I estimated a total volume $2.62 \times 10^{13}$ m$^3$. We can conclude that the volume of the rest of valley networks mapped is negligible respect to the volume of the 63 longer fluvial systems. Therefore, the volume of the latter valleys can be considered as representative, in terms of order of magnitude, of the total eroded volume since the remaining valleys give only a small contribution.

It is now possible to compare the above reported value with those obtained by previous works. In a recent study, Rosenberg and Head (2015) estimated the cumulative volume of water needed to carve the valley networks based on a fluid/sediment flux ratio function derived from terrestrial empirical data. To do that they used the estimated volume of only the seven largest valleys previously studied by Hoke et al. (2011) and assumed that the rest of the valley networks have a negligible contribution to the total global volume. They thus obtained a sediment volume of $1 \times 10^{13}$ m$^3$.

As widely discussed above, with the development of many morphological image processing techniques, other estimations of valleys' volume have been obtained (Rodriguez et al., 2002; Jung et al., 2012; Luo et al., 2011, 2015, 2017). The value obtained in the present work is, in terms of order of magnitude, in good agreement with that obtained by Luo et al. (2017): the observed discrepancies can be due to the different approaches and, in particular, to the procedure of evaluation of the valley area. In our approach, as well as in that by Hoke et al. (2011), this area was outlined using a manual drawing. As acknowledged by Luo et al. (2017), the manual procedure can be more accurate, even though it is time-consuming and requires a direct human intervention to outline the boundaries of the valley, while the PBTH method evaluates the valley area and the eroded volume automatically. Both volume estimates, the one evaluated in the present work and that obtained by Luo et al. (2017), are one order of magnitude larger than that obtained the Rosenberg and Head (2015). These authors, in fact, based their estimation on the eroded volume of eight large systems previously studied by Hoke et al. (2011), considering negligible the contribution of the rest of valleys. For this reason, their value underestimate the total volume.

### *3.2.3 Width and depth of the interior channel*

To determine the width ($w$) and depth ($h$) of the interior channels, I analyzed cross section profiles obtained using the THEMIS and MOLA data with the Profile QGIS tool. To create the profiles the following rules were adopted:

1) draw the profiles perpendicular to the course of the valley;
2) intersect the valley where it appears well defined, regular and not where it is degraded by impact craters or landslides;
3) trace the profile along the visible parts of the interior channel.

For each valley, at least 3 cross section profiles were drawn at the visible interior channel. For each section, a value of width and depth was evaluated. The final values were obtained as the average of those estimated from these cross-sectional profiles. Two examples of width and depth measurements are shown in **Fig. 3.7** and **Fig. 3.8**. The first example (see **Fig. 3.7**) is that of *Parana Valles* (Valley ID 1), a well-integrated valley network located around 22.9°S 349.0°E. The image is centered around 24.1°S 350.0°E where a portion of the interior channel of the valley is visible. Combining the photographic CTX data and MOLA topographic data we estimated width and depth of the interior channel. It is important to note that this procedure is made difficult by many factors: data resolution, lighting conditions of the surface, preservation of the interior channel. **Fig. 3.8** shows a portion of a valley centered around 12.0°S 161.8°E (Valley ID 7). In the image centered around 11.0°S 161.8°E it is possible to note that, in this case, the channel seems to cover the whole valley floor. This is a likely indication of two different evolution stages: in the first case, we can see a less developed channel in the valley, in the second case the flow has grown to fulfill the whole valley.

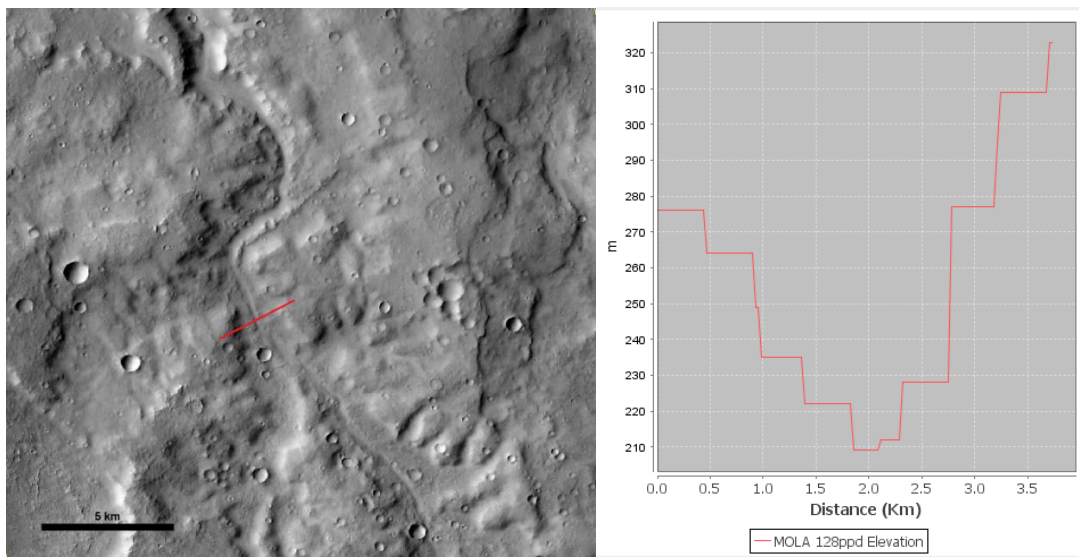

**Fig. 3.7** Left: portion of *Parana Valles* centered around 24.1°S 350.0°E. Gray scale CTX image (P02_002008_1558_XI_24S009W). Right: MOLA channel section obtained drawing lines on the valleys course as shown in the figure on the left (red line).

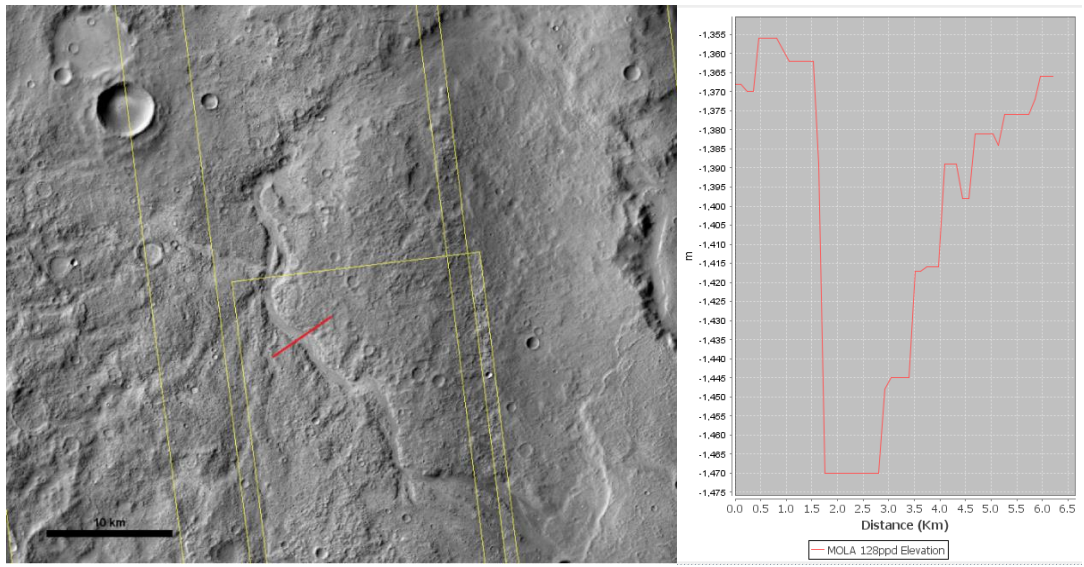

**Fig. 3.8** Left: portion of valley 7 centered around 11.0°S 161.8°E. Gray scale CTX images (P17_007512_1682_XI_11S162W - P01_001565_1685_XN_11S162W). Right: MOLA channel section obtained drawing lines on the valleys course as shown in the figure on the left (red line).

Also for the channel width, as for the volumes, it is possible to compare the present results with those obtained by other authors:

- Irwin et al. (2005a) evaluated the widths of some channels using VIKING and/or THEMIS data;
- Jaumann et al. (2005) estimated the channel dimensions of *Zarqa Vallis* using MOLA and THEMIS data.
- Hoke et al. (2011) studied a sample of Martian valleys including the *Naktong Vallis*, the only one that match with the first group of our samples (i.e. valleys with visible interior channel). These authors used MOLA and THEMIS data too.

The results of all these works are compared in **Table 3.6**.

**Table 3.6**
Comparison between the channel width obtained in the present work (w) and those obtained by Irwin et al. (2005a) ($w_1$), Jaumann et al. (2005) ($w_2$) and Hoke et al. (2011) ($w_3$).

| Valley ID | Valley Name | Lat. (°N) | Long. (°E) | w (m) | $w_1$ (m) | $w_2$ (m) | $w_3$ (m) |
|---|---|---|---|---|---|---|---|
| 1 | *Parana Valles* | 22.9 | 349.0 | 533 | | 180 | |
| 2 | *Naktong Vallis* | 5.0 | 33.0 | 600 | | | 1400 |
| 3 | *Zarqa Vallis* | 1.6 | 80.8 | 464 | 450 | | |
| 4 | | -0.9 | 30.0 | 810 | | 360 | |
| 5 | *Licus Vallis* | -2.4 | 126.1 | 529 | | 380 | |
| 6 | | -9.8 | -14.0 | 1074 | | 700 | |
| 7 | | -12.0 | -161.8 | 1244 | | 1000 | |
| 8 | | -13.6 | -174.3 | 1320 | | | |
| 9 | *Durius Vallis* | -17.7 | 172.1 | 780 | | 460 | |
| 10 | *Al-Qahira Vallis* | -18.0 | 165.5 | 840 | | | |
| 11 | *Samara Vallis* | -23.8 | 340.9 | 936 | | 400 | |
| 12 | | -25.2 | -12.4 | 882 | | 700 | |
| 13 | | -25.2 | -3.1 | 546 | | 310 | |

Considering the difficulties, previously discussed, in evaluating the interior channels dimensions, we are not surprised by this disagreement. Overall, we think that our data are more reliable of those obtained by Irwin et al. (2005a) because these authors used just photographic VIKING and THEMIS data with a lower resolution than the photographic data used in this work. In addition, we used topographic MOLA data which allow a better identification of the interior channel dimensions.

As regards the width of *Naktong Vallis*, the value obtained from Hoke et al. (2011) is higher than ours. This is probably due to the fact that they estimated the channel dimension at the end of the valley course, where it is usually wider. In addition, they seem to confuse the entire floor of the valley with the true interior channel. In contrast, we evaluated a mean value on different section profiles of the interior channel. For *Zarqa Vallis*, we are in good agreement with the value obtained by Jaumann et al. (2005). The authors, in fact, estimated a mean value like us.

We can also compare the width-to-depth ratios (*w/h*) obtained for the valleys of our sample with a typical mean terrestrial value. For our valleys *w/h* ranges from minimum of 15 to a maximum of 50 with a mean value of 32. Typical terrestrial valleys have a mean *w/h* ratio of 58. So, the *w/h* ratio obtained for the Martian channels here analyzed is less than what we would expect for bankfull discharge in terrestrial alluvial rivers (Jaumann et al., 2005).

### *3.2.4 Valley slopes*

The slope of the valleys was obtained, using the MOLA data, by analyzing the longitudinal profiles made by drawing a line along the valley course through the profile tool QGIS.

The slope measured for each valley was obtained by making three longitudinal profiles along three different sections of the valley system: one at the beginning of the system, one intermediate and one at the end of the course. To obtain these profiles, we used MOLA topographic data, processed in the form of false color images, and derived from the polygon shapefile of the valleys previously mapped.

As can be seen in **Fig. 3.9** these data show the different units of the valley along its longitudinal development. The valleys in the image are, in fact, false color ranging from red for higher areas to the blue for the most depressed. It was therefore chosen for each valley: a profile in correspondence with a higher stretch (dark orange); one in correspondence of portion at intermediate height (teal); and a last close to a more depressed part (blue). The slope values used in the calculation of the valley formation timescales are the average of the three values obtained from these parts. Each value was obtained from longitudinal profiles of about 20 - 25 km in length, considering the maximum and the minimum height of the portion concerned. This length value was chosen because it allows a good evaluation of the slope, instead of lower values for which more local behaviors could emerge. In addition, estimating the valley slope by making different longitudinal profiles no longer than 25 km instead of a longitudinal profile of the entire valley, allows us to follow with higher accuracy the valley floor.

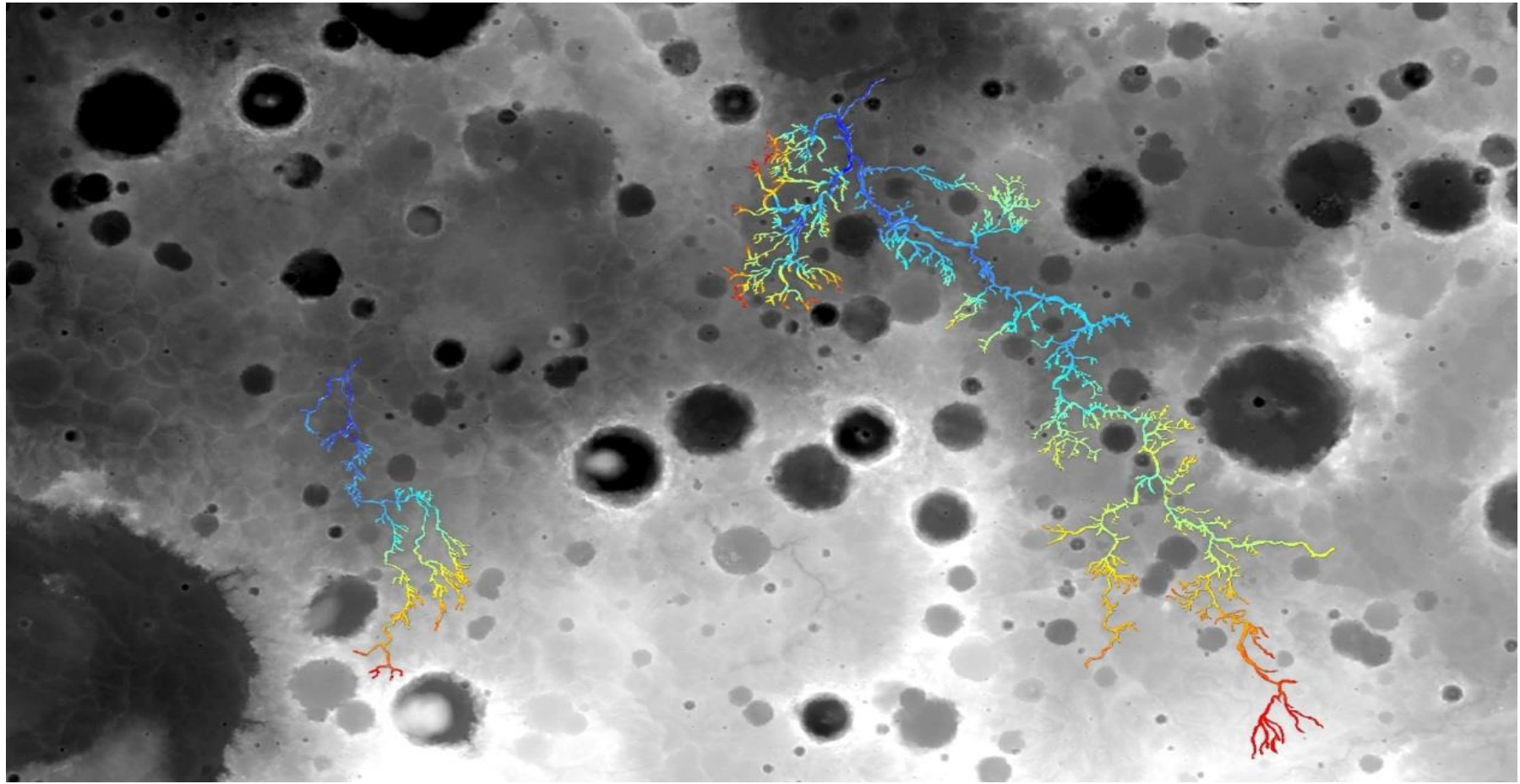

**Fig. 3.9** Image centered around 1.0°N and 28.0°E that shows *Naktong Vallis* (located at 5.0°N 33.0°E – Valley ID #2) and the valley at the coordinates 0.0°N 23.0°E (Valley ID #24). The valleys are in false color ranging from red for the highest areas to blue for the most depressed. The background image is a MOLA altimetry mosaic in false colors that range from black to white.

**Fig. 3.10** is an example of slope measurement for *Natkong Vallis*. It can be noted that, along the 25 km length, the valley height varies from 1.18 to 1.10 km. Clearly, the slope is given by the difference between these two values divided by the length of the portion considered (25 km in our case).

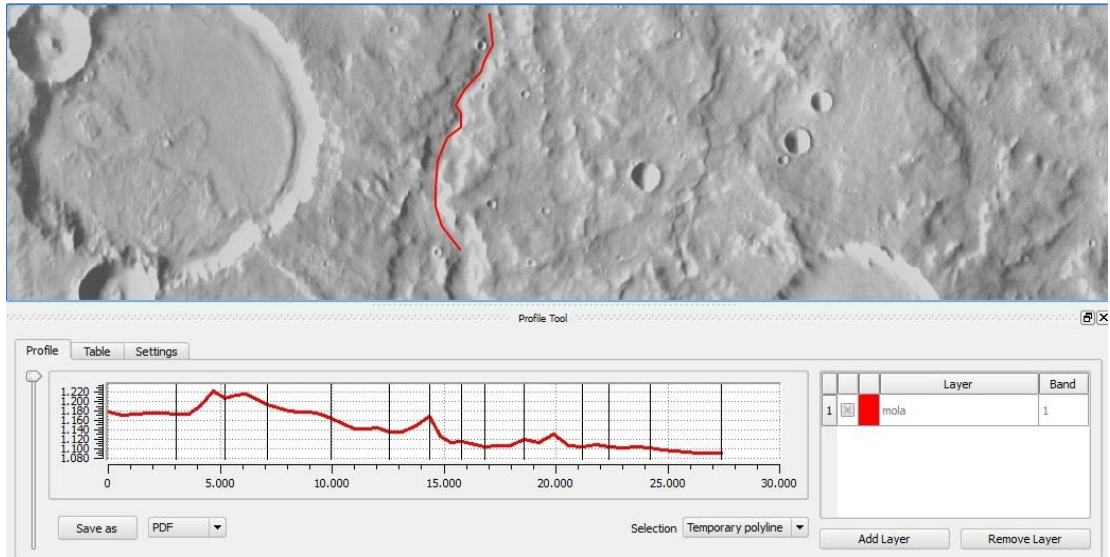

**Fig. 3.10** Example of longitudinal profile obtained through QGIS. The image, centered around 37.0°E 2.1°S, shows the initial slope of the *Naktong Vallis*. The profile obtained by drawing a line along the bottom of the valley (from top to bottom) is shown in the box below. The unit of measurement shown in the graph is in meters.

### *3.2.5 Summary of the obtained values*

For the valleys with a visible interior channel we obtained the values shown in **Table 3.7**. Those relative to valleys that do not show an interior channel are reported in **Table 3.8**.

For the valleys of the first group (**Table 3.7**) we have obtained a mean area of $5.56 \times 10^9$ m$^2$ and a mean eroded volume of $9.35 \times 10^{12}$ m$^3$; for those of the second group (**Table 3.8**) the mean area is $2.78 \times 10^9$ m$^2$ with an average eroded volume of $3.56 \times 10^{12}$ m$^3$. So, it is possible to conclude that the valleys that show an interior channel have on average a larger volume and a larger area than those in which inner channels are not visible.

**Table 3.7**
Geometrical parameters relative to fluvial systems with an inner channel. The values of width (*w*) and depth (*h*) of the channel, along with of slope (*S*), area (*A*) and volume (*V*) of the valley are reported.

| ID | Valley Name | Lat. (°N) | Long. (°E) | w (m) | h (m) | S | A (m²) | V (m³) |
|---|---|---|---|---|---|---|---|---|
| 1 | *Parana Valles* | 22.9 | 349.0 | 533 | 12 | 0.007 | 3.85E+09 | 1.33E+13 |
| 2 | *Naktong Vallis* | 5.0 | 33.0 | 600 | 23 | 0.004 | 1.16E+10 | 7.80E+12 |
| 3 | *Zarqa Vallis* | 1.6 | 80.8 | 464 | 32 | 0.010 | 1.82E+09 | 3.16E+12 |
| 4 | | -0.9 | 30.0 | 810 | 27 | 0.002 | 1.19E+09 | 1.19E+11 |
| 5 | *Licus Vallis* | -2.4 | 126.1 | 529 | 32 | 0.004 | 2.95E+09 | 8.31E+11 |
| 6 | | -9.8 | -14.0 | 1074 | 25 | 0.002 | 1.01E+10 | 5.76E+11 |
| 7 | | -12.0 | -161.8 | 1244 | 25 | 0.003 | 1.67E+09 | 1.71E+11 |
| 8 | | -13.6 | -174.3 | 1320 | 30 | 0.009 | 1.29E+09 | 1.61E+11 |
| 9 | *Durius Vallis* | -17.7 | 172.1 | 780 | 20 | 0.006 | 2.47E+09 | 7.04E+11 |
| 10 | *Al-Qahira Vallis* | -18.0 | 165.5 | 840 | 42 | 0.004 | 1.00E+10 | 3.31E+12 |
| 11 | *Samara Vallis* | -23.8 | 340.9 | 936 | 26 | 0.002 | 1.74E+10 | 1.14E+13 |
| 12 | | -25.2 | -12.4 | 882 | 19 | 0.004 | 2.23E+09 | 2.66E+13 |
| 13 | | -25.2 | -3.1 | 546 | 13 | 0.005 | 4.19E+09 | 5.25E+13 |

**Table 3.8**
Same as in **Table 3.7**, but for valleys that do not show an interior channel.

| Valley ID | Valley Name | Lat. (°N) | Long. (°E) | Area (m$^2$) | Volume (m$^3$) |
|---|---|---|---|---|---|
| 14 | *Indus Vallis* | 19.2 | 38.7 | 1.66E+09 | 1.43E+11 |
| 15 | *Scamander Vallis* | 15.8 | 28.5 | 4.20E+09 | 1.78E+11 |
| 16 | | 14.7 | 51.8 | 7.62E+08 | 1.89E+10 |
| 17 | *Cuscus Valles* | 13.6 | 50.5 | 2.18E+09 | 1.47E+11 |
| 18 | | 12.0 | 43.0 | 2.09E+09 | 1.36E+12 |
| 19 | | 9.4 | 46.6 | 3.10E+09 | 4.14E+11 |
| 20 | *Locras Valles* | 8.6 | 47.7 | 3.05E+09 | 3.08E+11 |
| 21 | | 4.5 | 17.4 | 1.30E+09 | 1.30E+11 |
| 22 | | 3.3 | 115.7 | 7.65E+08 | 4.94E+10 |
| 23 | | 1.8 | 90.0 | 4.41E+08 | 1.71E+11 |
| 24 | | 0.0 | 23.0 | 1.87E+09 | 1.46E+12 |
| 25 | | -0.5 | 123.2 | 8.32E+08 | 1.23E+11 |
| 26 | *Verde Vallis* | -0.7 | 30.1 | 9.81E+08 | 1.09E+11 |
| 27 | | -1.6 | 124.0 | 1.13E+09 | 3.49E+13 |
| 28 | | -3.0 | 5.0 | 5.09E+09 | 3.53E+11 |
| 29 | *Naro Vallis* | -4.0 | 60.6 | 2.39E+09 | 8.91E+10 |
| 30 | | -4.9 | 131.8 | 8.49E+08 | 2.29E+09 |
| 31 | *Tinto Vallis* | -5.1 | 111.4 | 1.78E+09 | 7.42E+11 |
| 32 | | -5.6 | 120.8 | 1.44E+09 | 2.19E+11 |
| 33 | | -5.8 | 128.1 | 3.59E+09 | 2.50E+13 |
| 34 | | -6.0 | 45.0 | 2.63E+09 | 6.22E+11 |
| 35 | | -6.2 | 31.0 | 2.76E+09 | 4.45E+11 |
| 36 | *Tagus Valles* | -6.5 | 114.4 | 7.06E+08 | 2.52E+11 |
| 37 | | -7.0 | 3.0 | 9.11E+09 | 7.17E+11 |

**Table 3.8** – continued

| 38 |                | -7.3  | 131.6  | 2.20E+09 | 6.14E+13 |
|----|----------------|-------|--------|----------|----------|
| 39 |                | -7.4  | -7.6   | 3.70E+09 | 6.61E+11 |
| 40 | *Brazos Valles* | -7.6  | 18.9   | 4.05E+09 | 3.96E+11 |
| 41 |                | -8.3  | -167.1 | 1.88E+09 | 1.91E+11 |
| 42 |                | -9.2  | 117.1  | 4.61E+09 | 2.55E+11 |
| 43 |                | -9.7  | 127.0  | 6.13E+09 | 5.85E+11 |
| 44 |                | -10.5 | 98.8   | 8.62E+08 | 6.30E+10 |
| 45 |                | -10.8 | 94.1   | 1.16E+09 | 1.13E+10 |
| 46 | *Evros Vallis* | -12.0 | 13.9   | 7.08E+09 | 5.37E+12 |
| 47 |                | -12.4 | 155.4  | 5.97E+09 | 1.59E+13 |
| 48 |                | -12.6 | 23.9   | 2.66E+09 | 8.50E+10 |
| 49 |                | -12.7 | 96.4   | 5.75E+08 | 2.21E+10 |
| 50 |                | -13.5 | -159.6 | 7.57E+08 | 5.78E+10 |
| 51 |                | -15.0 | 149.0  | 1.20E+09 | 1.72E+11 |
| 52 |                | -17.5 | -7.9   | 9.96E+08 | 8.69E+10 |
| 53 |                | -18.0 | 78.4   | 1.27E+09 | 1.83E+11 |
| 54 | *Loire Vallis* | -18.1 | 343.3  | 8.44E+09 | 8.82E+11 |
| 55 | *Vichada Valles* | -18.6 | 88.1 | 5.11E+09 | 1.82E+11 |
| 56 | *Marikh Vallis* | -19.2 | 3.9   | 7.17E+09 | 4.63E+12 |
| 57 |                | -20.0 | 167.0  | 9.85E+08 | 2.87E+11 |
| 58 | *Ma'adim Vallis* | -22.0 | 177.3 | 7.12E+09 | 1.41E+13 |
| 59 |                | -34.6 | -71.3  | 2.03E+09 | 4.76E+11 |
| 60 |                | -40.6 | -74.6  | 1.77E+09 | 4.33E+10 |
| 61 | *Warrego Valles* | -42.2 | -93.0 | 2.56E+09 | 2.77E+12 |
| 62 |                | -43.8 | -78.4  | 1.79E+09 | 4.22E+11 |
| 63 |                | -48.0 | -74.9  | 2.13E+09 | 4.90E+11 |

Comparing **Table 3.1** with **Tables 3.7** and **3.8**, we can note that the valleys with the largest eroded volumes are not the ones with the widest areas and the longest lengths. This can be better seen in **Fig. 3.11**. We analyzed the volume trend in correlation with the total length and the area of the valleys in our sample. For example, the valley #38, which has the largest volume in the sample, is characterized by relatively small values of $A$ and $L_T$, indicating the decisive importance of the depth of the valley in generating the large eroded volume.

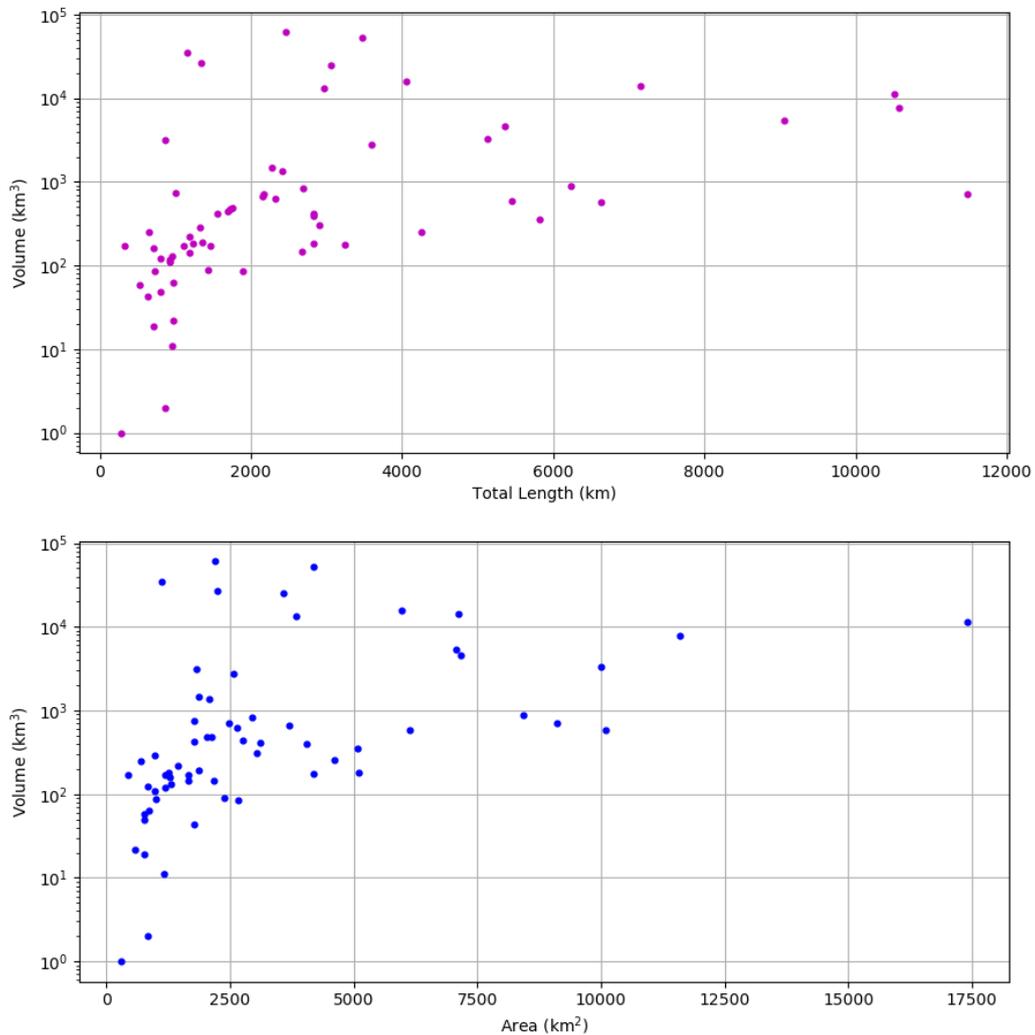

**Fig. 3.11.** Upper: volume (km$^3$) vs total length (km) for each valley. Lower: volume (km$^3$) vs area (km$^2$) of our valleys.

## *3.3 Calculation method and results*

For the first group of valleys I obtained the formation time using a method based on the evaluation of water and sediment discharges (Jaumann et al., 2005). In particular we used a modified Manning's equation for steady, uniform flow considering the lower gravitational acceleration on Mars (Komar, 1979; Manning, 1981; Gioia and Bombardelli; 2002; Wilson et al., 2004; Jaumann et al., 2005). This equation allows one to express the volumetric flow rate $Q_w$ (volume transported per unit of time) of the water in a channel, as it follows:

$$Q_w = \frac{\Sigma \sqrt{S} \, R^{4/3}}{\eta} \quad (1)$$

where $\Sigma$ is the flow cross-sectional area, $S$ is the local slope, $R$ is the hydraulic radius and $\eta$ is the so-called Manning roughness coefficient (Komar, 1979; Manning, 1981). An analysis performed by Gioia and Bombardelli (2002) demonstrated that an analogue modified Manning's equation can be derived from the turbulence theory in a way that links the Manning roughness coefficient $\eta$ with a dimensionless constant $k$:

$$\eta = r^{1/6} g^{-1/2} k^{-1} \quad (2)$$

where $r$ is the bed roughness scale and $g$ is the gravity acceleration. Later, Wilson et al. (2004) used the results of the analysis of Gioia and Bombardelli (2002) to derive a value of $\eta$ applicable to the Martian case. To do that, these authors started from an analysis of terrestrial river data. Comparing the water flow velocities predicted by the new version of the Manning's equation with those obtained using instead the Darcy-Weisbach relationships, they optimized the values of both k and $\eta$ for Mars obtaining: $k = 6.01$ and $r = 0.064$ m (Wilson et al., 2004). Using the gravity acceleration $g = 3.74$ m/s² So $\eta = 0.0545$ s m$^{-1/3}$. The value of $r$, defined by Gioia and Bombardelli (2002) as the "typical size" of the bed roughness elements, was obtained by Wilson et al. (2004) on the basis of the analysis of the median Martian rock size. The latter was evaluated using the data of the Viking and Pathfinder lander (Wilson et al., 2004). Comparing their value of $\eta$ with those used in previous works on Martian channel systems, Wilson et al. (2004) infer that previous works both overestimate and underestimate flow velocities by a factor 2.

In order to obtain, through eq. (1), the volumetric flow rate $Q_w$ of the water in the channel (of width $w$ and depth $h$) visible on the bottom of a Martian valley, the slope $S$ can be reasonably well approximated with that of whole valley. The slope is an important parameter for the determination of valleys' formation timescales. Since steeper slopes facilitate the transport of material downstream carving out the valleys in shorter time (Hynek et al., 2011), it has a strong influence on formation timescales. However, one cannot directly measure the slope of the interior channel of a valley because it is only partly visible. We assumed that the channel slope is the same as the valley slope and this is a reasonable assumption considering the uncertainty of our calculation (Hynek et al., 2011).

Moreover, assuming a rectangular-shaped cross section of the channel, the flow cross-sectional area is $\Sigma = w\,h'$ (where $h'$ is the flow depth), while the hydraulic radius is $R = h'$ (see for example Hoke et al., 2011).

Since the valley-incising flow depths of ancient rivers cannot be directly measured, they must be assumed or evaluated from analysis of the interior channel morphology (Hoke et al., 2011). A possibility is simply that $h' = h$, assuming that the dominant discharge filled the channel to the top of its banks (bankfull discharge). However, according to Jaumann et al. (2005), for one of our channels (the one in *Zarqa Vallis* ID #3 with $w/h$=15), the width to depth ratio is about an order of magnitude less than what we would expect for bankfull discharge in terrestrial alluvial rivers and, mainly for this reason, these authors assumed an active flow depth of the dominant discharge equal to 10% of bankfull ($h' = h/10$). Even if the $w/h$ ratios of the other interior channels present in the valleys of **Table 3.7** are greater than 15, all of them remain still smaller than the ratios expected for bankfull discharge. So, generalizing the conclusion of Jaumann et al. (2005), it seems unlikely that the interior channels of the valleys in **Table 3.7** experienced a bankfull discharge for the entire duration of the hydrological activity of the system.

Alternatively, and more plausibly the flow depth $h'$ can be evaluated following the method used by Hoke et al. (2011). According these authors we have:

$$h' = \frac{\theta D}{S}\frac{\rho_s - \rho}{\rho} \tag{3}$$

where $\rho_s$ and $\rho$ are respectively the density of the sediment grains and of the water, $D$ is the median grain size, while $\theta$ is the so-called Shields parameter, given by $\theta = \alpha\,\theta_{cr}$; here $\theta_{cr}$ is the critical Shields number, a dimensionless quantity which depends on the nature of the channel bed, while $\alpha$ is the flow strength (Rosenberg and Head, 2015), another dimensionless parameter which measures how much the shear stress of the flow exceeds the critical shear stress required for sediment transport. Again according to Hoke et al. (2011), in Eq. (3) we can put $\rho_s = 3400$ kg/m$^3$ (assuming a sediment basaltic composition), $\rho_s = 1000$ kg/m$^3$, $D = 0.1$ m (Kleinhans, 2005), and $\theta = 0.036$, corresponding to $\alpha = 1.2$ (Hoke et al., 2011; Parker, 1978) and $\theta_{cr} = 0.03$, typical of gravel bed rivers (e.g. Kleinhans and Van den Berg, 2011).

In this way, for our channels we obtain flow depths between 0.9 and 4.3 m, with an average value of 2.4 m. These flow depths are consistent with those obtained by Hoke et al. (2011) with the same formula for their valleys, considering the wider widths of the inner channels. Furthermore, if we calculate for each channel of the valleys listed in **Table 3.7** the flow depth using the simple relation $h' = h/10$ (suggested by Jaumann et al., 2005) instead of Eq. (3), then we obtain values of $h'$ between 1.2 and 4.2 m, with an average value of 2.5 m. So, even if for a few valleys differences exist between the flow depths derived with the two equations, the results of the two methods are in general agreement, supporting the Jaumann and colleagues' assumption of an active flow depth of the dominant discharge equal to 10% of bankfull.

We note that recently, Rosenberg and Head (2015) have quantified the minimum cumulative water volume that was required to carve the Late Noachian valley networks on Mars, using in their calculations a grain size $D$ between 1 and 6 mm and $\alpha$ in the range 35 - 60, with the latter quite distant from the values quoted in the literature (Parker, 1978; Paola and Mohring, 1996; Hoke et al., 2011). We will return to this possibility at the end of this section. For the moment we note that, even assuming values of grain size and flow strength consistent with those used by Rosenberg and Head (2015) ($D$ = 3 mm and $\alpha$ = 45), for each channel one obtains a value of $h'$ in very good agreement with what was previously calculated with a more traditional choice of $D$ and $\alpha$.

In order to determine the time needed to erode a valley network, the sediment load $q$ of the interior channel (sediment mass transported per unit of water volume) has to be known. This parameter is linked to $Q_w$ by the equation:

$$q = \frac{Q'_s}{Q_w} = \frac{\rho'_s Q_s}{Q_w} \qquad (4)$$

where $\rho_s'$ is the sediment bulk density, while $Q_s'$ (expressed in kg/s) and $Q_s$ (expressed in m$^3$/s) are respectively the sediment mass and the sediment volume transported per unit of time.

For terrestrial rivers the water/sediment flow ratio $Q_w/Q_s$ can vary by some orders of magnitude (due to differences in grain size of the sediment, as well as slope and hydraulic radius of the channel and so on). Analyzing a sample of 67 terrestrial rivers spread around the world, Milliman and Meade (1983) have measured some important fluvial parameters, among which the water and the sediment discharges which allow one to calculate for each river the water/sediment flow ratio. From these ratios it is possible to derive for the whole sample of rivers an average value of about $1 \times 10^4$, which seems quite significant, if we consider not only the number of rivers but also their widespread geographical distribution.

Taking advantage of this result, we can reasonably assume for our Martian channels the same value $q = 0.40$ kg m$^{-3}$. In fact, as discussed by Rosenberg and Head (2015), the use of terrestrial values of the water/sediment flow ratio (and by consequence of the sediment load – see Eq. (4)) for the Martian case is fully correct because the dependence of $Q_w/Q_s$ and hence of $q$ on gravity is very weak.

Starting from $q$, the formation time $T$ of a valley can be evaluated through the equation:

$$T = \frac{V_s'}{Q_s} = \frac{V_s'}{q\, Q_w}\, \rho_s' \qquad (5)$$

Here we assume a sediment bulk density $\rho_s' = 1500$ kg/m$^3$, equal to that of the sedimentary deposits at the Mars Pathfinder landing site (Matijevic et al., 1997), while the volume $V_s'$ is linked to the one evaluated in Section 3.2.1, by means of the relation (Kleinhans, 2005; Hoke et al., 2011):

$$V_s' = \frac{V_s}{(1-\lambda)}, \qquad (6)$$

where $\lambda = 0.3$ (Kleinhans, 2005) is a coefficient which takes into account the porosity of the transported material.

The formation timescales thus obtained are representative of a continuous flow. This is possible only in environments that invoke an essentially continuous source of supply for liquid water. Even on Earth channel forming conditions occur rarely owing to the variability of the climatic phenomena. According to Barnhart et al. (2009), there is no reason to think this would be different for Martian valley networks. So, we assume three different possible values of intermittency:

a) 5%, if conditions on early Mars were humid or sub-humid (Barnhart et al., 2009; Matsubara et al., 2014);

b) 1% if Martian environments were analogues to semiarid/arid terrestrial ones (Barnhart et al., 2009);

c) 0.1% for hyperarid environments (Fassett and Head, 2008).

For each of these three cases we obtained the formation timescales *T'*, *T''* and *T'''*, respectively. They are reported, along with the timescale *T* for continuous flux in **Table 3.9**, where the values of $Q_w$, $Q_s$, $V'_s$ are also shown.

Once the formation times for each valley are obtained, one can evaluate the rate of vertical erosion (incision) of the terrain *E* (expressed in m/yr) which represents the increase in depth of the valley per unit of time:

$$E = \frac{V'_s}{TA} \tag{7}$$

In **Table 3.10**, I report the values of *E* obtained for the valleys of the first group, in the case of both continuous or intermittent fluxes (assuming the above reported values of intermittency). This parameter is linked to the superficial erosion rate (expressed in t km$^{-2}$ yr$^{-1}$), defined as:

$$E_0 = \frac{M_S}{TA} \tag{8}$$

(where *A* is the superficial area of the valley obtained mapping each valley with polygonal shapefiles in QGIS and $M_S$ is the mass of the sediment), by the equation $E_0 = \rho'_s E$.

The mean erosion rates obtained for the 13 valleys with a visible inner channel can be used to calculate, by inverting Eq. (9), the formation times for the other 50 valleys in our sample. In this way, values in the range $8 \times 10^2$ - $9 \times 10^6$ yr were obtained for a continuous flux. Considering, instead, an intermittency of 5% the formation times of these 50 valleys range from $2 \times 10^4$ to $2 \times 10^8$ yr; finally, in the last two cases (1% and 0.1% of intermittency), I found values in the ranges $8\times10^4$ - $9\times10^8$ yr and $8\times10^5$ - $9\times10^9$ yr, respectively.

      Summarizing the results obtained (see **Fig. 3.12**) for the whole sample of the fluvial systems studied in this work, I can conclude that, assuming a continuous flow, the formation timescales are in the range $8\times10^2$ - $1\times10^7$ yr (with an average of $8\times10^5$ yr), while with an intermittency of 5% they range from $2\times10^4$ to $3\times10^8$ yr (mean value $2\times10^7$ yr). Instead, with an intermittency of 1%, the formation timescales are between $8\times10^4$ and $1\times10^9$ yr (mean value $8\times10^7$ yr) and between $8\times10^5$ and $1\times10^{10}$ yr (mean value $8\times10^8$ yr) with a 0.1% intermittence.

**Table 3.9**

Values of Q, $Q_s$, $V'_s$ for the valleys with a visible interior channel studied in this work. Formation times were obtained for continuous flow (T) and assuming different possible values of intermittency: 5% (T'), 1% (T'') and 0.1% (T''').

| Valley ID | Valley Name | Lat. (°N) | Long. (°E) | Q (m³/s) | Qs (kg/s) | V' (m³) | T (yr) | T' (yr) | T''(yr) | T''' (yr) |
|---|---|---|---|---|---|---|---|---|---|---|
| 1 | *Parana Valles* | 22.9 | 349.0 | 1.11E+03 | 0.30 | 1.90E+13 | 2.04E+06 | 4.09E+07 | 2.04E+08 | 2.04E+09 |
| 2 | *Naktong Vallis* | 5.0 | 33.0 | 2.51E+03 | 0.67 | 1.11E+13 | 5.28E+05 | 1.06E+07 | 5.28E+07 | 5.28E+08 |
| 3 | *Zarqa Vallis* | 1.6 | 80.8 | 6.62E+02 | 0.18 | 4.52E+12 | 8.11E+05 | 1.62E+07 | 8.11E+07 | 8.11E+08 |
| 4 |  | -0.9 | 30.0 | 7.54E+03 | 2.01 | 1.70E+11 | 2.68E+03 | 5.36E+04 | 2.68E+05 | 2.68E+06 |
| 5 | *Licus Vallis* | -2.4 | 126.1 | 2.28E+03 | 0.61 | 1.19E+12 | 6.18E+04 | 1.24E+06 | 6.18E+06 | 6.18E+07 |
| 6 |  | -9.8 | -14.0 | 1.00E+04 | 2.67 | 8.22E+11 | 9.78E+03 | 1.96E+05 | 9.78E+05 | 9.78E+06 |
| 7 |  | -12.0 | -161.8 | 7.31E+03 | 1.95 | 2.44E+11 | 3.97E+03 | 7.94E+04 | 3.97E+05 | 3.97E+06 |
| 8 |  | -13.6 | -174.3 | 2.15E+03 | 0.57 | 2.30E+11 | 1.27E+04 | 2.54E+05 | 1.27E+06 | 1.27E+07 |
| 9 | *Durius Vallis* | -17.7 | 172.1 | 2.03E+03 | 0.54 | 1.01E+12 | 5.88E+04 | 1.18E+06 | 5.88E+06 | 5.88E+07 |
| 10 | *Al-Qahira Vallis* | -18.0 | 165.5 | 3.62E+03 | 0.97 | 4.74E+12 | 1.55E+05 | 3.11E+06 | 1.55E+07 | 1.55E+08 |
| 11 | *Samara Vallis* | -23.8 | 340.9 | 8.71E+03 | 2.32 | 1.63E+13 | 2.22E+05 | 4.44E+06 | 2.22E+07 | 2.22E+08 |
| 12 |  | -25.2 | -12.4 | 3.80E+03 | 1.01 | 3.80E+13 | 1.19E+06 | 2.37E+07 | 1.19E+08 | 1.19E+09 |
| 13 |  | -25.2 | -3.1 | 1.76E+03 | 0.47 | 7.49E+13 | 5.05E+06 | 1.01E+08 | 5.05E+08 | 5.05E+09 |

**Table 3.10**

Values of erosion rates E, E', E'' and E''' corresponding to the formation times obtained for continuous flow (T) and assuming different possible values of intermittency: 5% (T'), 1% (T'') and 0.1% (T''').

| Valley ID | Valley Name | Lat. (°N) | Long. (°E) | E(m/yr) | E'(m/yr) | E''(m/yr) | E'''(m/yr) |
|---|---|---|---|---|---|---|---|
| 1 | *Parana Valles* | 22.9 | 349.0 | 2.42E-03 | 1.21E-04 | 2.42E-05 | 2.42E-06 |
| 2 | *Naktong Vallis* | 5.0 | 33.0 | 1.82E-03 | 9.10E-05 | 1.82E-05 | 1.82E-06 |
| 3 | *Zarqa Vallis* | 1.6 | 80.8 | 3.06E-03 | 1.53E-04 | 3.06E-05 | 3.06E-06 |
| 4 | | -0.9 | 30.0 | 5.35E-02 | 2.67E-03 | 5.35E-04 | 5.35E-05 |
| 5 | *Licus Vallis* | -2.4 | 126.1 | 6.50E-03 | 3.25E-04 | 6.50E-05 | 6.50E-06 |
| 6 | | -9.8 | -14.0 | 8.32E-03 | 4.16E-04 | 8.32E-05 | 8.32E-06 |
| 7 | | -12.0 | -161.8 | 3.68E-02 | 1.84E-03 | 3.68E-04 | 3.68E-05 |
| 8 | | -13.6 | -174.3 | 1.40E-02 | 7.01E-04 | 1.40E-04 | 1.40E-05 |
| 9 | *Durius Vallis* | -17.7 | 172.1 | 6.93E-03 | 3.46E-04 | 6.93E-05 | 6.93E-06 |
| 10 | *Al-Qahira Vallis* | -18.0 | 165.5 | 3.05E-03 | 1.52E-04 | 3.05E-05 | 3.05E-06 |
| 11 | *Samara Vallis* | -23.8 | 340.9 | 4.21E-03 | 2.11E-04 | 4.21E-05 | 4.21E-06 |
| 12 | | -25.2 | -12.4 | 1.43E-02 | 7.16E-04 | 1.43E-04 | 1.43E-05 |
| 13 | | -25.2 | -3.1 | 3.54E-03 | 1.77E-04 | 3.54E-05 | 3.54E-06 |
| | | | *Average values* | 1.22E-02 | 6.10E-04 | 1.22E-04 | 1.22E-05 |

As discussed above, owing to the variability in climatic phenomena, a continuous flow is unlikely on Mars, as well as on Earth. For this reason, we will not take in further consideration this case. On the other hand, using an intermittency of 0.1% we obtained for eight valleys formation timescales greater than $10^9$ yr, with a maximum of 10 billion years, and these values are unreasonable or even impossible when compared with the age of the planet (4.5 billion years). So there are two plausible scenarios for the formation of the Martian valleys: a humid and temperate environment or a semiarid/arid one. We think that the latter is preferable according to the scenario proposed by several authors which suggest that, during the period of valley formation, the climatic conditions on Mars were arid or semiarid (Barnhart et al., 2009; Matsubara et al., 2014) even if we cannot completely rule out the possibility of humid climatic conditions. In any case, the distribution of the formation timescales for the two scenarios outlined above is shown in **Fig. 3.12**.

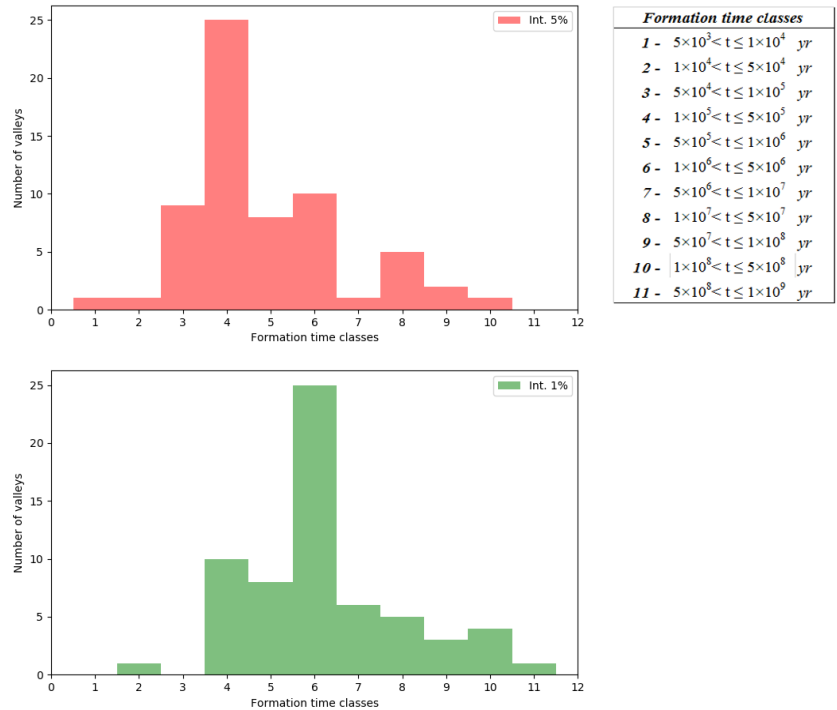

**Fig. 3.12.** Histograms of the formation timescales obtained for the 63 valleys, here studied, in two different possible situations: intermittency of 5% (upper), and of 1% (lower). The timescales obtained were divided in 10 formation time classes: class 1 - $5\times10^3 \leq t < 1\times10^4$ yr, class 2 - $1\times10^4 \leq t < 5\times10^4$ yr and so on. The last class (1) has $5\times10^8 \leq t < 1\times10^9$ yr.

## 3.4 Comparison with the terrestrial case and with previous works on Martian valleys

A key parameter in our model is the incision rate of the terrain; its average values were obtained for the valleys of the first group (reported in **Table 3.10**) and then used to calculate the formation timescales of the remaining valleys of our sample (second group). It is, therefore, important to verify the plausibility of these average values by comparing them with those measured on Earth for rivers located in different regions. The fluvial incision rate depends on a large number of factors such as the size of the river, the nature of the load, the speed of the current and the gradient of the river valley and the climatic conditions. As a consequence, for the Earth the values of incision rates vary by some order of magnitude.

Vigier et al. (2006) analyzed a sample of 20 Icelandic rivers that flow on basaltic rocks obtaining values of the erosion rate $E_0$ between a minimum of 21 t km$^{-2}$ yr$^{-1}$ and a maximum of 4864 t km$^{-2}$ yr$^{-1}$, implying an incision rate between 6.2 m/Myr and 1430 m/Myr (using the density of basalt).

Hoke and Jordan (2010) evaluated the rate of vertical incision $E$ for 28 individual catchments in the Atacama Desert, Chile, one of the driest regions on Earth, obtaining values between 5 m/Myr and 30 m/Myr.

On the other hand, Duxbury (2009) measured the rate of fluvial incision in the Blue Ridge province in and around Shenandoah National Park, USA; he obtained values of $E$ in the range 3.8 -24 m/Myr for various lithologies that crop out in the Park (granite, metabasalt, quartzite, and siliciclastic rocks).

Finally, Hubacz (2012) analyzed the ground erosion produced by the Bluff's Parcel Tributary, a second order system of Strahler located in the Coastal Plain of Delaware, USA. This author found, for the loose material in which the river is engraved, a mean value of the incision rate E = 22000 m/Myr.

According to Irwin et al. (2004), the fluvial incision rate on a given planet depends on gravity by the law: $E \propto g^{0.7}$, where $g$ is the gravity of that planet. Therefore, the incision rate on Mars ($E_M$) if the values of the other parameters are the same, is linked to terrestrial value by the following equation:

$$E_M = \left(\frac{g_M}{g_T}\right)^{0.7} E_T \tag{10}$$

where $E_T$ is the rate of incision on Earth, while $g_M$ and $g_T$ are the values of gravity on Mars and on Earth, respectively.

When scaled to the Martian gravity by Eq. (10), the above reported values of vertical incision rate measured on Earth, vary from 1.9 m/Myr, corresponding to the minimum value of $E_T$ measured for the rocky terrain in Shenandoah National Park (Duxbury, 2009), to 11000 m/Myr corresponding to the loose material in the Coastal Plain of Delaware (Hubacz, 2012).

By comparison, for the valleys with visible interior channels we obtained (see from **Table 3.10**):

a) 12200 m/Myr for continuous flow;
b) 610 m/Myr assuming 5% of intermittency typical of humid or sub-humid climate;
c) 122 m/Myr for arid environments corresponding to 1% of fluvial intermittency;
d) 12 m/Myr considering 0.1% of intermittency typical of hyperarid environments.

We conclude that the incision rates derived for our Martian valleys, are substantially within the range of terrestrial values scaled to Martian gravity. Even the value obtained for 0.1% of intermittency (12 m/Myr) is greater than that corresponding to the terrestrial minimum (1.9 m/Myr).

As already mentioned in Section 3.1.2, Hoke et al. (2011) estimated the formation timescales of seven Martian valley networks using three different models of sediments transport: Ribberink, (1998); Meyer-Peter and Muller (1948) and Van Rijn, (1984). For each valley, Hoke and colleagues obtained the formation time, assuming an intermittency of 5%, using the value of the flow depth calculated in two ways: the first was obtained by the equation $h' = w/58$, which gives flow depths very close to those derivable in the case of bankfull discharge, the second applying exactly the same equation, Eq. (3), used here. In the latter case they obtained mean timescales of $6\times10^6$ yr, $8\times10^6$ yr and $8\times10^7$ yr using the Ribberink, Meyer-Peter and Muller, and Van Rijin methods, respectively. Overall, considering the three methods, Hoke et al. (2011) obtained a mean value of $3\times10^7$ yr. If we compare this result with the mean value obtained for our sample ($2\times10^7$ yr), for the same value of intermittency (5%), the agreement is very good.

We also note that the sediment load $q = 0.40$ kg m$^{-3}$ assumed in our calculations is associated by Eq. (4) with a water/sediment flow ratio $Q_w/Q_s = 3600$. This represent a conservative estimate with respect to that by Hoke et al. (2011) who in the case of $h'$ obtained from Eq. (3) (that is in the case different from that of the bankfull discharge) have derived for their valley networks an average value $Q_w/Q_s = 4 \times 10^4$.

Another possible comparison concerns the formation time of the *Ma'adim Vallis*. This valley is one of the most studied fluvial structures on Mars and many theories have been proposed to explain its origin (Sharp e Malin, 1975; Masursky et al., 1980; Carr, 1981; Baker, 1982; Cabrol, 1991; Cabrol et al., 1997; Landheim, 1995; Irwin et al. 2002). Some features of the valley are suggestive of an origin due to runoff, while others seem, instead, to indicate groundwater sapping. Anyway, multiple episodes of flows have been proposed to explain the presence of several terraces (Cabrol et al., 1998a). Based on the study of these terraces, Cabrol et al. (1998a) identified in the valley three distinct levels: level 1 (ancient), level 2 (intermediate) and level 3 (recent). Assuming a hydraulic radius $h'$=10 m, they also evaluated a water discharge rate of $1.2\times10^3$ m$^3$/s and from this they obtained a duration of water flow of only 200 yr and 50 yr, necessary to erode the valley up to levels 1 and 2, respectively. Since *Ma'adim Vallis* has no visible interior channels, I indirectly evaluated its formation timescale, as reported in the previous section, starting from the total eroded volume of the valley ($1.4\times10^{13}$ m$^3$ - see **Table 3.4**) and using the mean incision rate obtained for the 13 valleys of the first group. In this way, we obtained for a continuous flow $T = 2.0\times10^5$ yr, which is very different from the value reported by Cabrol et al. (1998a), although their value of $Q$ ($1.2\times10^3$ m$^3$/s) is not too far from those derived for the valley of the first group (see **Table 3.9**). This high discrepancy is due to the water/sediment flow ratio assumed by those authors. In fact, they adopted $Q/Q_s = 4$, obtained by Goldspiel and Squyres (1991) starting from the value relative to the *Mississipi* river, extrapolated to Mars using some assumptions about the dependence of this ratio on gravity, slope and grain size. Instead, the sediment load $q = 0.40$ kg m$^{-3}$ assumed in our calculations (representative of the sample of 67 rivers analyzed by Milliman and Meade (1983) is associated by Eq. (4) to a water/sediment flow ratio $Q/Q_s = 3600$, that is about 1000 times greater than the previous one. If we scale by this factor their time of 200 yr, we obtain $2 \times 10^5$ yr, in perfect agreement with our result. In other words, the very short formation time obtained by Cabrol et al. (1998a) is due to a value of $Q/Q_s$ which seems very far from the terrestrial standards.

Another case of very different evaluations, which in any case can be ultimately reconciled, is that of *Zarqa Vallis* for which Jaumann et al. (2005) obtained for a continuous flow a formation time of 1800 yr, while in the same conditions I found $2.2 \times 10^6$ yr. Actually, this large discrepancy, can be easily explained if we consider three main factors in the calculations of Jaumann et al. (2005): the flow depth, assumed exactly equal to 10% of the channel depth, instead of calculated by Eq. (3); the eroded volume, crudely evaluated multiplying the correctly measured area of the valley by an assumed average depth of 250 m; and the sediment load *q*, assumed much greater than that relative to the terrestrial rivers of the sample studied by Milliman and Meade (1983) (see previous section).

In particular, h' used by these authors is three times the typical value here adopted. This results in a Q eight times greater, a q 12 times greater and $M_s$ 5 times smaller than those here obtained starting from V'. Adopting these values, it is possible to obtain a formation timescale 480 ($8 \times 12 \times 5$) times longer than our value; in fact: $1800 \times 480 = 8.6 \times 10^5$ yr is in good agreement with the value obtained in this work which is equal to $8.1 \times 10^5$ yr.

As discussed above, Rosenberg and Head (2015) evaluated the minimum water volume required to carve the valley networks on Mars, and, by consequence, the minimum formation timescale of these structures. To achieve this goal, they performed their calculations using a grain size *D* between 1 and 6 mm and a flow strength α between 35 and 60. By means of the same equations they used in their calculations, the minimum formation timescale of our valleys can be obtained in the following way. First of all, one can evaluate for each Martian channel the new volumetric flow rate $Q_w$, using the empirical relationship (Irwin et al. 2008; Rosenberg and Head, 2015):

$Q_w = \frac{a\,w^2}{k_w}$, where the constants $a$ and $k_w$ have values respectively of about 0.62 and 3, in order to obtain $Q_w$ in cubic meters/seconds when $w$ is expressed in meters (Rosenberg and Head, 2015). Then, using the other empirical relationship (Rosenberg and Head, 2015): $\frac{Q_w}{Q_s} = \exp(11.95\,\alpha^{-0.1778})$, it is possible to obtain that in the range $35 \leq \alpha \leq 60$ used by Rosenberg and Head (2015), the water/sediment flux ratio is in the range 320–570. Assuming a typical value $Q_w/Q_s = 450$ and a sediment bulk density $\rho_s' = 1500$ kg/m$^3$, as before, Eq. (4) gives $q = 3.3$ kg/m$^3$. Finally, using the new values of $q$ and $Q_w$ coupled with the geometric parameters of the channels and valleys, one can derive for the first group of our valleys, in the case of a continuous flux, an average formation time of $7\times10^3$ yr, corresponding to a mean incision rate of $1\times10^6$ m/Myr. Using this rate, one finds for the remaining 50 valleys and for the whole sample an average formation timescale of $1\times10^3$ and $3\times10^3$ yr, respectively.

The comparison of these results with those previously obtained with a much smaller value of the flow strength clearly shows that the formation times here obtained using the approach of Rosenberg and Head (2015) are much shorter than those obtained with our method. So that, as expected, the former represents a lower limit for the formation timescales of the Martian valleys. This is also supported by the mean incision rates, which are well above the terrestrial values.

## *3.5 Valley ages and formation times*

Because of their key role as climate indicators, it is also useful to evaluate the time of onset of these fluvial systems, in addition to the duration of their activity. Using size-frequency crater distribution methods Fassett and Head (2008) and Hoke and Hynek (2009) evaluated the absolute ages of a sample of Martian valleys.

To obtain an estimation of the ages of the selected valley networks, Fassett and Head (2008) applied a crater counting technique. Their mapping of both valleys and superposed craters rely on Viking MDIM 2.1 (Mars Digital Image Mosaic) and THEMIS IR global mosaic (both of which are ~200 m/pixel), MOLA grid topography (~463 m/pixel) and HRSC image data (where available up to 13 m/pixel). Fassett and Head's (2008) crater-size frequency distributions were fitted to the Hartmann production function (Hartmann, 2005) and the Neukum production function (Neukum and Ivanov, 2001; Ivanov, 2001; Werner, 2005).

Hoke and Hynek (2009) used THEMIS mosaic and MOLA data in ArcGIS and followed the method outlined by Tanaka et al. (1982). Their method includes only craters whose rims overlap the actual valley network and an area that is a function of the actual valley dimensions rather than an average width for the entire system (Hoke and Hynek., 2009).

Both groups of authors tried to analyze the crater population since the end of valley formation. In this way, they obtained an indication of the end of the fluvial activity of a valley.

Using our formation times, we can go back to the start of the fluvial activity of these structures; in fact, knowing the formation timescales and the cessation times of these fluvial systems, we can determine their onset time.

In **Table 3.11** I report the age obtained by Fassett and Head (2008) (using both the Hartmann and Neukum production functions) and by Hoke and Hynek (2009) (using Neukum chronology only) for all the valleys that are in common between our sample and theirs.

Adding to these ages the formation times $T'$ and $T''$ here evaluated in the cases of 5% and 1% of intermittency, respectively, I obtained the results shown in **Fig. 3.12**. Note that, for those valleys for which we have two different Neukum ages (see **End 1** and **End 3** in **Table 3.11**) we used the mean value of them.

**Table 3.11**
Valleys studied in this work for which we have an age indication obtained by Fassett and Head (2008) using the crater size distribution technique and the Neukum (**End 1**) and Hartmann production functions (**End 2**). And also, the age obtained by Hoke and Hynek (2009) considering only Neukum absolute ages (**End 3**)

| Lat (°N) | Long (°E) | End 1 (Gyr) | End 2 (Gyr) | End 3 (Gyr) |
|---|---|---|---|---|
| -42.0 | -93.0 | 3.77 | 3.62 | |
| -23.8 | 340.9 | 3.73 | 3.49 | |
| -22.9 | 349.0 | 3.73 | 3.49 | 3.68 |
| -19.2 | 3.9 | 3.85 | 3.69 | |
| -18.6 | 88.1 | 3.80 | 3.65 | |
| -18.1 | 343.3 | 3.73 | 3.49 | |
| -18.0 | 78.4 | 3.75 | 3.58 | |
| -18.0 | 165.5 | 3.86 | 3.71 | |
| -17.5 | -7.9 | 3.85 | 3.67 | |
| -12.7 | 96.4 | 3.74 | 3.54 | |
| -12.4 | 155.4 | 3.72 | 3.50 | |
| -12.0 | 13.9 | 3.76 | 3.62 | 3.63 |
| -9.8 | -14.0 | 3.79 | 3.61 | |
| -7.6 | 18.9 | 3.71 | **3.46** | |
| -7.4 | -7.6 | 3.79 | 3.67 | |
| -7.3 | 131.6 | 3.74 | 3.50 | |
| -7.0 | 3.0 | 3.71 | 3.54 | 3.65 |
| -6.8 | 114.4 | 3.74 | 3.56 | |
| -6.0 | 45.0 | | | 3.70 |
| -4.0 | 60.6 | 3.90 | 3.81 | |
| -3.0 | 5.0 | | | **3.62** |
| -2.4 | 126.1 | 3.83 | 3.73 | |
| -0.7 | 30.1 | 3.79 | 3.63 | |
| -0.5 | 123.2 | 3.93 | 3.77 | |
| 0.0 | 23.0 | 3.69 | 3.50 | 3.68 |
| 5.0 | 33.0 | 3.69 | 3.50 | 3.72 |
| 12.0 | 43.0 | | | 3.73 |
| 13.6 | 50.5 | 3.81 | 3.62 | |
| 15.0 | 28.5 | 3.69 | 3.50 | 3.72 |

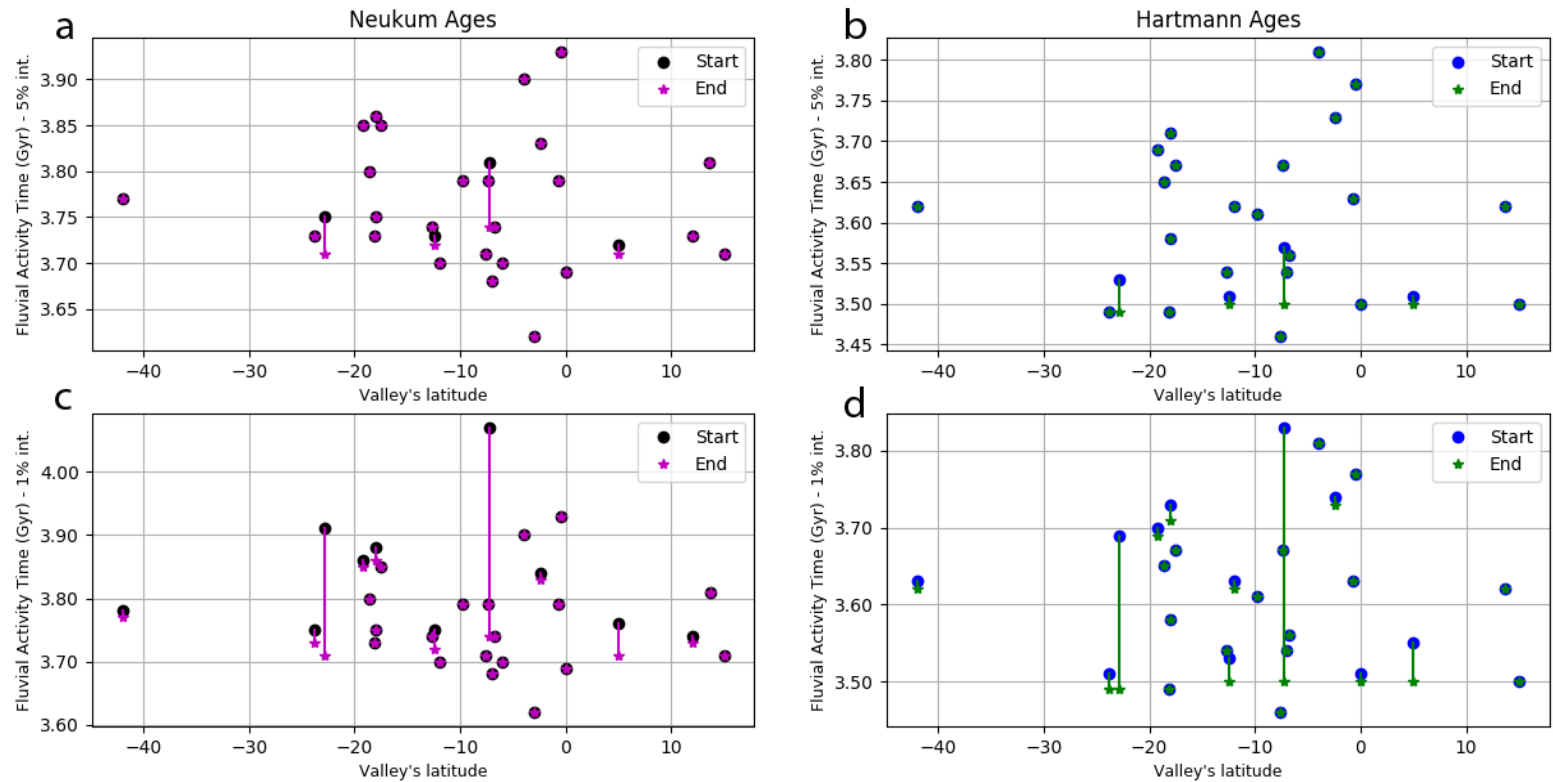

**Fig. 3.13.** Fluvial activity duration of the valleys in **Table 3.11** in relation with their age. The evaluation of the time of onset of these structure has been made in four different cases: a) considering Neukum absolute ages and 5% of intermittency; b) for Hartmann absolute ages and 5% of intermittency; c) considering Neukum absolute ages and 1% of intermittency; d) for Hartmann absolute ages and 1% of intermittency.

Considering Hartmann absolute ages, I obtained for the subset of valleys reported in **Table 3.11** that the period of the fluvial activity is about 350 Myr for 5% of intermittency and 370 Myr for 1%, respectively. Considering Neukum absolute ages I obtained ~ 310 Myr and ~ 450 Myr for 5% and 1% of intermittency.

As one can see in **Fig. 3.13**, the valleys for which the ages are known are not equally distributed in the whole temporal range of hydrological activity. In particular, focusing our attention on the case of 1% of intermittency and considering Hartmann absolute ages (**Fig. 3.13d**), we can see that before 3.8 Gyr ago only two of these valleys were active. Instead between 3.6 and 3.7 Gyr, 11 valleys were active. Although the numbers are a little different, a similar situation still holds in the case of Neukum absolute ages with the same intermittency (**Fig. 3.13c**).

I analyzed these data in more detail to see if there is a trend in the number of valleys simultaneously active during the period of hydrological activity. Considering Hartmann and Neukum absolute ages and 1% of intermittency, I obtained the result shown in **Fig. 3.14a** and **b**: the number of active valleys increases towards the end of the period of hydrological activity.

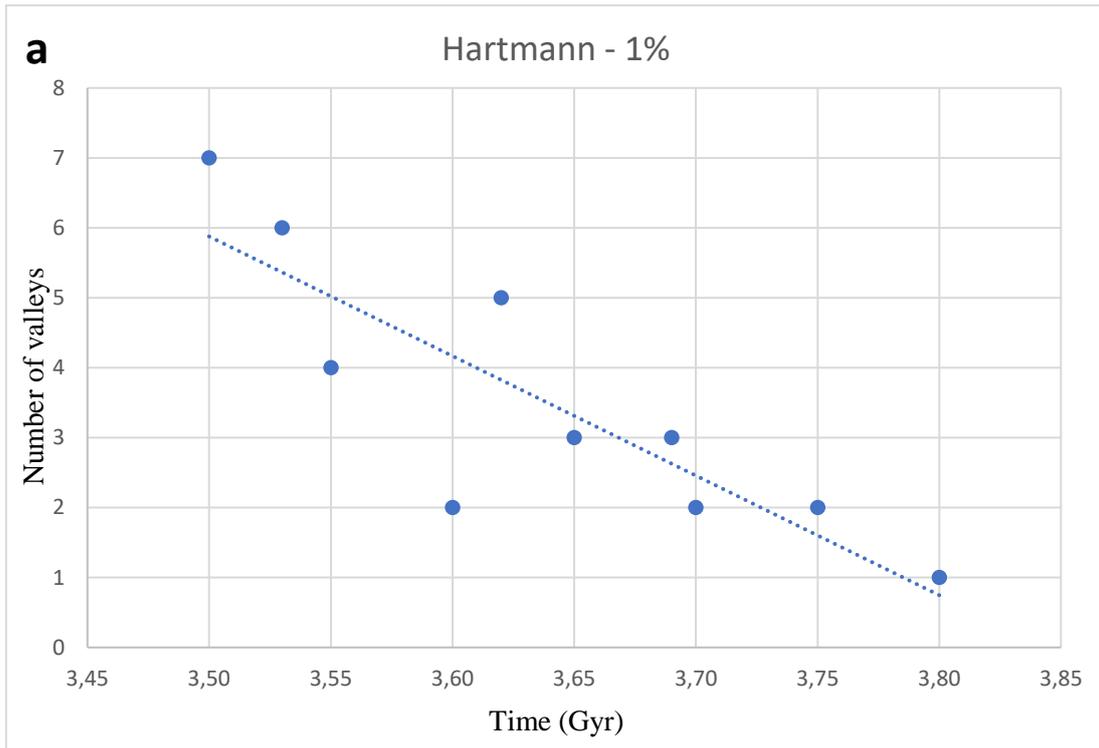

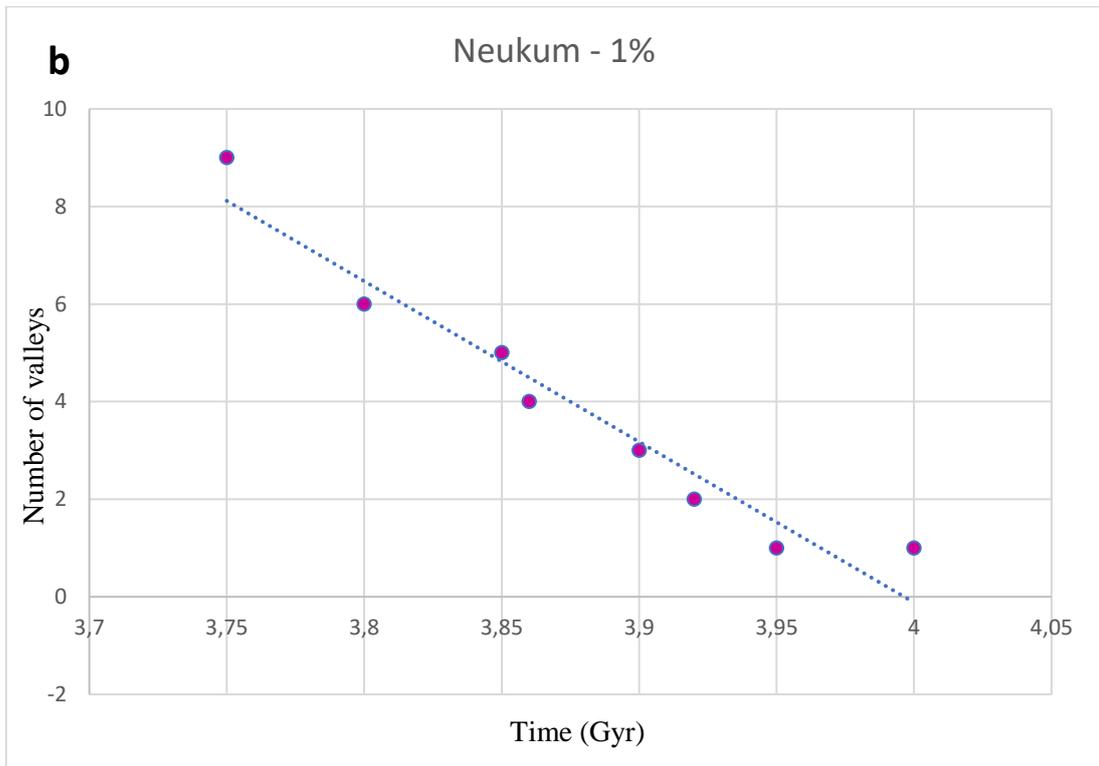

**Fig. 3.14.** Trend in the number of valleys simultaneously active during the period of hydrological activity: a) considering Hartmann Ages; b) considering Neukum Ages.

## *3.6 Discussion and conclusions*

In the present work I evaluated the duration of water flow in a sample of large Martian valley networks to better understand their formation mechanisms. I applied a method based on the evaluation of water and sediment discharge using equations of transport models (Jaumann et al., 2005) to a first group of valleys in our sample (13 fluvial systems with visible interior channels). Sediment discharges were evaluated from water discharges using a median value obtained from an analysis of terrestrial rivers (Milliman and Meade, 1983). The erosion rates obtained for the first group of valleys (those with visible interior channels) were used to evaluate the formation times of the other group (50 valleys).

In this work, for the first time, a model of sediment transport has been applied to a considerable large sample of Martian valleys. In fact, so far sediment transport models (e.g. Jaumann et al., 2005; Hoke et al., 2011) have been applied to a limited number of valleys, as discussed in Section 3.1. Due to the widespread distribution of the analyzed valleys, our results give a more global vision of the duration of the hydrological cycle on Mars and can be useful to constrain the ancient climate of the planet. In addition, the identification of the interior channels for the valleys of the first group and the evaluation of its dimensions, thanks to data at higher resolution, allowed us to better estimate the formation time for these valleys. The use of the interior channel certainly does give more definite geometrical dimensions than using the entire valley. Obviously, it is possible that in some cases the forming events of the interior channel and of the whole valley could be unrelated. For example, larger events could have carved the valley while smaller subsequent events could have formed the interior channel later. In this regard, very recently, some authors (Goudge and Fassett, 2016; 2017) suggested that the interior channel present in *Licus Vallis* may not be a preserved example of the formative channel for the larger valley, but a smaller valley that originated from a lake overflow flood at *Licus*'s head. However, we still cannot exclude that this channel formed together with the valley, and in any case, it is unlikely that the scenario proposed for *Licus Vallis* also applies to all the valleys with interior channels here studied.

Another important result is related to the eroded volume estimations. We evaluated this volume for each of the 63 valleys by means of an extracted DEM, obtained starting from a manual mapping procedure. In this way, we obtained, for each valley an accurate map and a good estimation of the eroded volume.

The results support the hypothesis of an ancient arid or semiarid Mars with episodic warmer and wetter periods (Barnhart et al., 2009; Matsubara et al., 2013; Matsubara et al., 2014). However, we cannot entirely exclude the scenario of a long warm and wet Mars during the whole Noachian period (Hoke et al., 2011).

An accurate analysis of our method and the comparison with other approaches, revealed that the main cause of uncertainty in the application of the models of sediment transport to the Martian case is the sediment load $q$, which, in turn, is linked to the ratio $Q/Q_s$ by Eq. (4). This is in fact the main reason of discrepancy between the results of different sediment transport models applied to Mars. The most difficult part of these methods is in fact the evaluation of the sediment discharge $Q_s$. This is a fundamental parameter for timescale determinations through models of sediment transport and depends on a large number of variables that are very difficult to evaluate such as flow velocity, friction factor, grain shear stress. In addition, measuring sediment transport is quite complex even on Earth where we know much more compared to Mars. $Q_s$ can be estimated using sediment transport predictors such as those used by Meyer-Peter and Mueller (1948), Van Rijin (1984) and Ribberink (1998). However, the differences between various predictors are considerable (Kleinhans, 2005). Instead of using those predictors to evaluate $Q_s$ directly for the Martian case, we decided to use the sediment load (and consequently the ratio $Q/Q_s$) obtained on a large sample of terrestrial valleys distributed all over the world (Milliman and Meade, 1983) in order to minimize the uncertainties due to the large variations that we observe on Earth because of local and time-limited behaviors. Although this sediment concentration is a rough estimate, it can be used to constrain the duration of the valley formation based on an order-of-magnitude calculation.

The adopted values of $Q/Q_s$ are much higher than those used by other authors, such as Jaumann et al. (2005) and Cabrol et al. (1996). However, the values adopted by these authors are related to specific rivers and/or specific flow conditions characterized by large flows of sediment. This occurs rarely even on Earth. In particular, the value of $Q/Q_s$ equal to 4, adopted by Cabrol et al. (1996), is not measured in any of the 67 rivers of Milliman and Meade (1983) (in this sample the lowest value is 37). In addition, the value $q = 5$ kg/m$^3$ adopted by Jaumann et al. (2005), that, according to these authors, seems to be a reasonable choice, is actually quite high. In fact, in the sample of Milliman and Meade (1983) only 6 rivers out of 67 have a sediment load equal or greater than that. In other words, the values assumed by these authors appear atypical for terrestrial rivers.

On the other hand, our value of $Q/Q_s$ is lower than those obtained by Hoke et al. (2011 and 2014). These authors did not make direct assumptions on this ratio, but they used sediment transport models (Meyer-Peter and Mueller, 1948; Van Rijin, 1984; Ribberink, 1998) to directly evaluate $Q_s$. Probably neither of these approaches are reliable, due to the already mentioned large uncertainties about the various parameters of the sediment transport model they used. On the contrary, we are rather confident about the values $q$ and $Q/Q_s$ adopted in our calculations, because the used sample of terrestrial rivers is widespread and large enough to cover a large spectrum of sediments and bedrock applicable to Mars, and, in addition, the comparison with Earth can be considered significant and no extrapolation to the Martian case is needed (Rosenberg and Head, 2015). Obviously, it is unlikely that for all the valleys the sediment load was exactly equal to the assumed value ($q = 0.40$ kg m$^{-3}$). However, although this is a rough estimate, it can be used to evaluate the valley formation time, at least in terms of order of magnitude.

It is also important to note that, as normally done in this kind of calculation (Kleinhans, 2005; Jaumann et al. 2005; Hoke et al., 2011; Rosenberg and Head, 2015), the formation time of a valley has been obtained eith the hypothesis that this would be exactly equal to the time needed for the removal and transport of the sediment. However, the latter assumption is valid only in the hypothesis of an incision of the valley through incoherent terrains. This implies an underestimation of the real values since we do not take into account the time required to carve the rocks and to break it up into transportable pieces. Actually, there is morphological evidence that these valley networks formed on a substrate that had experienced heavy bombardment by impactors. Therefore, at the time of their formation the surface material was in the form of bedrock under a layer of regolith debris (Baker and Partridge, 1986; Carr and Malin, 2000; Hartmann and Neukum, 2001). We must acknowledge, however, that we do not know the depth and the distribution of such a regolith in the first part of the Martian history. If this mantling of unconsolidated terrain was not widespread and thick enough, then the bedrock would have been rapidly exposed in large areas of the planet. In this case, the material transported through bedload processes would have acted to break up the surface into additional transportable pieces, depending on sediment supply and transport capacity (Sklar and Dietrich, 2004). But this process would have been much slower than the simple sediment removal and transport. In fact, the comparison with the Earth (as in the cases of the rivers in Iceland and in the Shenandoah Park – Vigier et al., 2006; Duxubury, 2009) suggests a lengthening of the formation timescales by a factor of ten or more in this case.

Another relevant consideration is about the above discussed intermittency of the flow. In fact, we shall not necessarily think about an intermittent water flow but rather about a discontinuous sediment transport. In this scenario, like in many terrestrial rivers, the water flow could have been, at least in principle, continuous over the whole period of valley formation while the sediment transport may have occurred only during a much shorter span of time. It is likely, however, that in some of these Martian valleys the hydrological activity experienced periods of true quiescence. This is suggested by the fact that for some valleys the time required to produce the eroded volume we observe now is much less than the total period necessary for the valley formation, as deduced by stratigraphic analyses.

A typical example of this situation is *Ma'adim Vallis* (Cabrol et al., 1996; 1998a, 1998b) for which, even in the most favorable case (0.1% of intermittency), the formation time obtained in the present work is $2\times10^8$ yr, while the overall duration of the *Ma'adim-Gusev* system, estimated from stratigraphic analyses, should have lasted up to 2 billion years (Cabrol et al., 1998b). Actually, despite this long span of time, the valley has several features suggesting that the river never reached the full stage of maturity. Among them, the most indicative are: the linear trend of the longitudinal profile of the valley and the lack of a well-developed tributary system; in addition, as observed by Cabrol et al. (1998b), some tributaries seem to get into the main branch through waterfalls, rather than directly entering the bottom of the river course, as in the case of mature systems. This apparent contradiction between the long duration of the valley system and its poor maturity could be explained by assuming that the total duration of water flow in the valley was much less than the overall duration of the system, derived by stratigraphic analyses. In other words, *Ma'adim Vallis* could have experienced several and successive flow episodes, interrupted by relatively long periods of inactivity (Cabrol et al., 1996). This was probably caused by global (such as cyclical variations in the inclination of the axis of rotation and the eccentricity of the orbit of Mars – Carr, 1996) and/or local effects (such as the displacement of high precipitation areas - Hoke and Hynek, 2009). Such effects could have affected the hydrological cycle of this system, interrupting and reactivating it periodically (Cabrol et al., 1998b).

Analyzing **Fig. 3.13** (i.e. **3.13d**), it is possible to note that, for any given instant of the period of hydrological activity on Mars, there were only a few active valleys at the same time. In addition, there is a trend to a more pronounced temporal overlapping of the fluvial activity for the younger valleys, in the sense that towards the end of the period of the hydrological activity (a few hundred million years long - see Section 3.5) the number of valleys that were active at the same time seems have been larger than at the beginning of that period. And this conclusion agrees with the findings of other authors (Howard et al., 2005; Irwin et al., 2005a).

The previous result is very important for its paleoclimatic implications, since it gives an indication of the intensity of the global hydrological activity on Mars, which would seem to have been somewhat localized and sporadic. This is because, unlike Earth (where now a few hundred large rivers are active at the same time), Mars would have hosted only a few large fluvial valleys that were active at the same time for any given instant of the whole temporal range of hydrological activity.

This conclusion, however, is valid only with the hypothesis, adopted in this work, of valley incision within incoherent terrain. As discussed before, in the case of a rocky terrain, the formation timescales would be much longer and the differences in the fluvial activity (and climatic conditions) on present and ancient Mars could have been less accentuated than those found in the present work.

# CHAPTER IV
# Spectral analysis of Martian fluvio-lacustrine structures

## 4.1 State of the art

As discussed Section 1.2.1 (Chapter I), several candidate paleolake basins have been observed on Mars (De Hon, 1992; Cabrol and Grin, 1999, 2001; Fassett and Head, 2008) and catalogued in detail on the basis of their morphologic features into two major categories: closed-basin and open-basin lakes (Cabrol and Grin, 1999; Fassett and Head, 2008). Unlike closed-basin lakes that have inlet valleys but lack outlets, open-basin lakes show also the presence of outlet valleys. This feature makes open-basin lakes more important from a paleoclimatic point of view because their formation seems to require long lasting fluvial activity (Fassett and Head, 2008). In fact, the presence of an outlet means that water within the basin must have ponded to approximately the level of the surface adjacent to the outlet valley before breaching and overflowing the basin (Fassett and Head, 2008; Goudge et al., 2012).

A catalogue of 210 open-basin lakes fed by valley networks has been produced by Fassett and Head (2008). Later, Goudge et al. (2012) added 16 more open-basin lakes, and analyzed the morphologies and mineralogies associated with the basins of the entire catalogue. They found that 34 of 79 open basin lakes with observed sedimentary deposits have targeted CRISM observation that cover these deposits (Goudge et al., 2012). Only 10 of these 34 CRISM data have spectral signatures consistent with the presence of aqueous alteration minerals and one of them contains the only observed signature consistent with evaporite minerals (Bandfield, 2006; Ehlmann et al., 2008a, 2008b, 2009; Wray et al., 2009; Dehouck et al., 2010; Ansan et al., 2011; Hughes et al., 2011; Goudge et al., 2011; 2012).

As far as the closed-basin lakes are concerned, Goudge et al. (2015) produced a catalogue that includes 205 of these structures. Same as in the case of the study on the open-basin lakes, they studied the morphology and spectroscopy of these structures. Their study shows that 55 candidate closed-basin lakes have sedimentary fan deposits in their interiors, 22 of the 55 have overlapping CRISM images. Of these 22, 4 shows the indication of the presence of water alteration minerals (Goudge et al., 2015).

In the following section I describe the sample of open- and closed-basin lakes analyzed in this work and the dataset used for the spectral analysis. In this context, the structure of the used data is discussed in detail. The methodology adopted for the data analysis and the instruments used are described in detail in Section 4.3. The results obtained from this study are presented in Section 4.4, focusing the attention on four basin lakes for which we found some interesting spectral evidences. Finally, in Section 4.5 I discuss our results, the implications of our findings and the possible future developments.

## *4.2   Dataset description*

### *4.2.1 Analyzed open and closed-basin lakes*

During the manual mapping procedure of Martian valleys (discussed in Chapter II), in addition to the ones previously reported in the literature, 17 closed-basins and 41 open-basins were found (see **Tables 4.1** and **4.2**). We decided to look for CRISM observations of these new basins as well as for observations of previously catalogued open- and/or closed-basin lakes showing sedimentary deposits. **Tables 4.1** and **4.2** report the open and closed-basin lakes mapped in this work plus those of Fassett and Head (2005) and Goudge et al. (2015) that I analyzed in search for more recent CRISM observations.

**Table 4.1**

List of closed-basin lakes analyzed in this work. For each basin the central coordinates, the ID of the CRISM data, where present, and eventually the alteration minerals detected are reported. For those basins where CRISM observation were not found, and consequently no minerals were detected, the corresponding slots in the table are marked with a "-". Finally, in the case of presence of the CRISM data but no mineral observed, the slots are marked with "None".

| ID | Lat (°N) | Long (°E) | CRISM OBS. | Alteration mineral detec. |
|---|---|---|---|---|
| 1 | 29.9 | -16.7 | - | - |
| 2 | 28.1 | 27.9 | FRT00019CB5 | None |
| | | | FRT00019894 | None |
| | | | FRT00019F17 | None |
| | | | HRL00020BCD | None |
| | | | FRT0001FBB0 | None |
| 3 | 22.3 | 53.1 | HRL00012522 | None |
| 4 | 10.2 | -16.6 | - | - |
| 5* | 8.4 | -56.9 | HRL000116DF | None |
| 6* | 5.4 | -58.6 | FRT000199E0 | None |
| | | | FRT000128F3 | None |
| 7 | -4.3 | 46.8 | - | - |
| 8 | -5.5 | 52.6 | FRT000106C3 | None |
| 9 | -5.6 | 27.1 | - | - |
| 10 | -9.5 | 98.2 | HRL000203C4 | None |
| | | | FRT0001127E | None |
| | | | FRT00009717 | None |
| 11 | -12.3 | 60.6 | FRT00011718 | None |
| 12 | -16.5 | 64.7 | FRT0000A26B | None |
| | | | FRT000137E9 | None |
| 13 | -17.9 | 178.7 | - | - |
| 14 | -19.2 | 176.6 | - | - |
| 15* | -19.4 | 52.1 | FRT00011D18 | None |
| | | | FRT0001192C | None |
| **16** | **-24.4** | **61.4** | **FRT0001689A** | **Fe/Mg-smectites** |
| 17 | -28.0 | -23.1 | - | - |
| 18 | -30.0 | 21.4 | - | - |
| 19 | -30.1 | -167.8 | - | - |
| 20 | -36.6 | -72.8 | FRT0001D9B7 | None |

Note:

(*) closed-basin lake mapped by Goudge et al. (2015) for which subsequent observations are available

**Table 4.2**
Same as in Table 4.1 but for open-basin lakes analyzed in this work.

| ID | Lat (°N) | Long (°E) | CRISM OBS. | Alteration mineral detec. |
|---|---|---|---|---|
| 1 | 35.1 | -2.7 | - | - |
| 2 | 34.4 | 49.1 | - | - |
| 3 | 33.6 | -10.1 | - | - |
| 4 | 27.9 | 26.6 | FRT0001648B | None |
| 5 | 21.4 | 151.9 | - | - |
| 6 | 18.5 | -54.5 | - | - |
| 7 | 8.6 | 37.6 | - | - |
| 8 | 4.6 | -58.6 | - | - |
| 9 | 1.0 | 43.0 | - | - |
| **10*** | **-4.0** | **109.2** | FRT0001BBA7** | None |
| | | | FRT0001A267 | None |
| | | | HRL0001851C | None |
| | | | **FRT0000D34E** | **Fe/Mg-smectites** |
| 11 | -5.5 | 124.0 | - | - |
| 12 | -7.8 | 82.0 | - | - |
| 13 | -10.4 | 148.5 | - | - |
| 14 | -10.6 | 147.9 | - | - |
| 15 | -11.3 | 133.0 | FRT0001F5DB | None |
| 16 | -12.3 | -25.6 | - | - |
| 17 | -12.9 | -161.4 | - | - |
| 18 | -14.1 | 24.8 | - | - |
| 19 | -15.2 | 61.3 | - | - |
| 20 | -16.2 | -22.5 | - | - |
| 21 | -17.6 | 173.1 | - | - |
| 22 | -19.6 | -9.6 | - | - |
| 23 | -20.1 | 57.0 | - | - |
| 24 | -24.5 | 62.9 | - | - |
| **25** | **-24.7** | **59.2** | **FRT0000C7A1** | **Fe/Mg-smectites** |
| | | | **FRT0001CA9D** | **Fe/Mg-smectites** |
| 26 | -25.3 | 138.3 | - | - |
| 27 | -29.0 | -174.3 | - | - |
| **28*** | **-30.1** | **73.4** | HRL000A153** | None |
| | | | **FRT00009720** | **Fe/Mg-smectites** |
| 29 | -30.5 | 82.1 | - | - |

**Table 4.2** – continued

| | | | | |
|---|---|---|---|---|
| 30 | -31.3 | -35.8 | - | - |
| 31 | -32.3 | -167.8 | - | - |
| 32 | -33.9 | -169.9 | - | - |
| 33 | -35.6 | -157.9 | - | - |
| 34 | -35.9 | -40.0 | - | - |
| 35 | -36.8 | -118.0 | - | - |
| 36 | -37.3 | -152.5 | - | - |
| 37 | -39.9 | 162.3 | - | - |
| 39 | -40.1 | 162.5 | - | - |
| 40 | -48.2 | -136.8 | - | - |
| 41 | -48.4 | -136.7 | - | - |
| 42 | -54.1 | -86.7 | - | - |
| 43 | -61.6 | -141.3 | - | - |

Note:
(*) Open-basin lake mapped by Fassett and Head (2005) for which subsequent observations are available.
(**) Observations studied by Goudge et al. (2012).

### *4.2.2 CRISM instrument*

In order to study from a spectroscopic point of view the structures described in the previous section, I looked for CRISM observations located in and/or close to them and in association with sedimentary deposits (see **Tables 4.1** and **4.2**). The CRISM instrument is a visible-infrared hyperspectral imaging spectrometer onboard Mars Reconnaissance Orbiter (MRO) that maps the geology, composition, and layering of surface features on Mars. Each readout of the detector is a line of a spatial image built as MRO moves along its orbit. In addition, each pixel has a spectrum whose absorptions can be compared with those of the various minerals (Murchie et al., 2007).

CRISM has two operational modes for observing the martian surface and atmosphere: gimbaled and pushbroom imaging. In the first mode, a gimbal is used to target a location and take multiple images in a sequence. Those images are then combined to produce one image of the chosen area (target). In the second mode, pushbroom (or mapping) observations are taken by turning CRISM on and letting the spacecraft do the pointing. As MRO flies, CRISM acquires data to take long image strips.

CRISM uses three main methods to collect data:

- Targeted observations - The instrument's gimbal tracks a point on the surface, and superimposes a scan to cover a region approximately 10 km x 10 km at about 18 meters per pixel, in 544 wavelengths.

- Mapping observations – The instrument collects fewer wavelengths (between 72 and 262) to provide a global map of Mars at 5 to 10 times lower resolution.

- Atmospheric observations – There are several types of these. In one type, the MRO spacecraft pitches forward or backward to view Mars' limb (viewing the atmosphere edge-on) and CRISM's gimbal scans back and forth to capture the vertical structure of the atmosphere in 544 wavelengths.

In **Figs. 4.1** and **4.2** are shown and described in detail all the CRISM observation modes, how the measurements are acquired and their primary science functions.

| Mode | Description | Surface Tracking | | Spectral Sampling (# channels) | | Spatial Resolution (m/pix) | | | | Footprint Size (km) | Primary Science Function | | | Acquisition Dates (DOY) | ID Range (hex) |
|---|---|---|---|---|---|---|---|---|---|---|---|---|---|---|---|
| | | Gimbaled | Push-broom | VNIR | IR | 20 | 40 | 100 | 200 | | Targeted | Mapping | Atmospheric | | |
| FRT | Full Resolution Targeted | ● | | 107 | 438* | ● | | | | ~10x10 | ● | | ● | start: 2006_270 end: 2012_146 | 27B2 252AB |
| HRL | Half Resolution Targeted | ● | | 107 | 438* | | ● | | | ~10x20 | ● | | ● | start: 2006_273 end: 2012_146 | 282D 252A3 |
| HRS | Half Resolution Short | ● | | 107 | 438* | | ● | | | ~10x8 | ● | | ● | start: 2006_273 end: 2012_144 | 2847 25248 |
| EPF | Emission Phase Function | ● | | 107 | 438* | | | | ● | variable | | | ● | start: 2006_273 end: 2012_146 (restricted > 2010_280) | 27FF 252AB |
| FRS | Full Resolution Short | ● | | 107 | 438* | ● | | | | ~10x3 | ● | | | start: 2012_275 | 26D50 |
| MSW | Multispectral Window | | ● | 19 | 55 | | | ● | | ~10x45, 10x180, 10x540 | | ● | | start: 2006_272 end: 2008_072 | 27F6 A659 |
| MSV | Multispectral VNIR | | ● | 90 | 0* | | | ● | | ~10x45, 10x180, 10x540 | | ● | | start: 2012_013 | 2267B |
| MSP | Multispectral Survey | | ● | 19 | 55 | | | | ● | ~10x45, 10x180, 10x540 | | ● | | start: 2006_270 | 27DC |
| HSP | Hyperspectral Mapping | | ● | 107 | 154 | | | | ● | ~10x45, 10x180, 10x540 | | ● | | start: 2011_142 | 1E328 |
| HSV | Hyperspectral VNIR | | ● | 107 | 0* | | | | ● | ~10x45, 10x180, 10x540 | | ● | | start: 2009_174 | 135BA |
| ATO† | Along-Track Oversampled | ● | | 107 | 438* | non-square pixels: up to ~3 m/pix downtrack, 20 m/pix crosstrack | | | | ~10x3 or 10x1† | ● | | | start: 2011_079 end: 2012_142† start: 2012_293 | 1D678 251D4 27155 |
| ATU | Along-Track Undersampled | ● | | 107 | 438* | non-square pixels: ~40 m/pix downtrack, 20 m/pix crosstrack | | | | ~10x6 | ● | | | start: 2012_293 | 2715A |
| LMB | Limb Scan | ● | | 107 | 438 | non-square pixels | | | | N/A | | | ● | start: 2009_191 | 2B5C |
| TOD | Tracking Optical Depth | | ● | 107 | 438 | non-square pixels | | | | ~10x540 | | | ● | start: 2007_217 | 6F2B |
| FFC | Flat Field Calibration | | ● | varies by type | varies by type | non-square pixels | | | | ~10x540 | N/A | | | start: 2006_274 | 2856 |

*Most gimbaled observing modes including FRT, HRL, HRS, EPF, FRS, ATO, and ATU can be commanded as VNIR-only; MSV and HSV are VNIR-only by definition. LMB, TOD, and FFCs cannot be commanded as VNIR-only.

†From DOY 2011_079 to 2012_142, ATO mode was a variation of FRT and labeled as "FRT" in the PDS archive. Beginning again on 2012_293 and ongoing, ATOs are labeled as "ATO" in the archive. These two periods also correspond to a change in the way that ATOs were acquired, resulting in a difference in footprint size and shape.

**Fig. 4.1.** Summary of CRISM observing modes and characteristics. Image credits: APL (Applied Physics Laboratory) - The Johns Hopkins University.

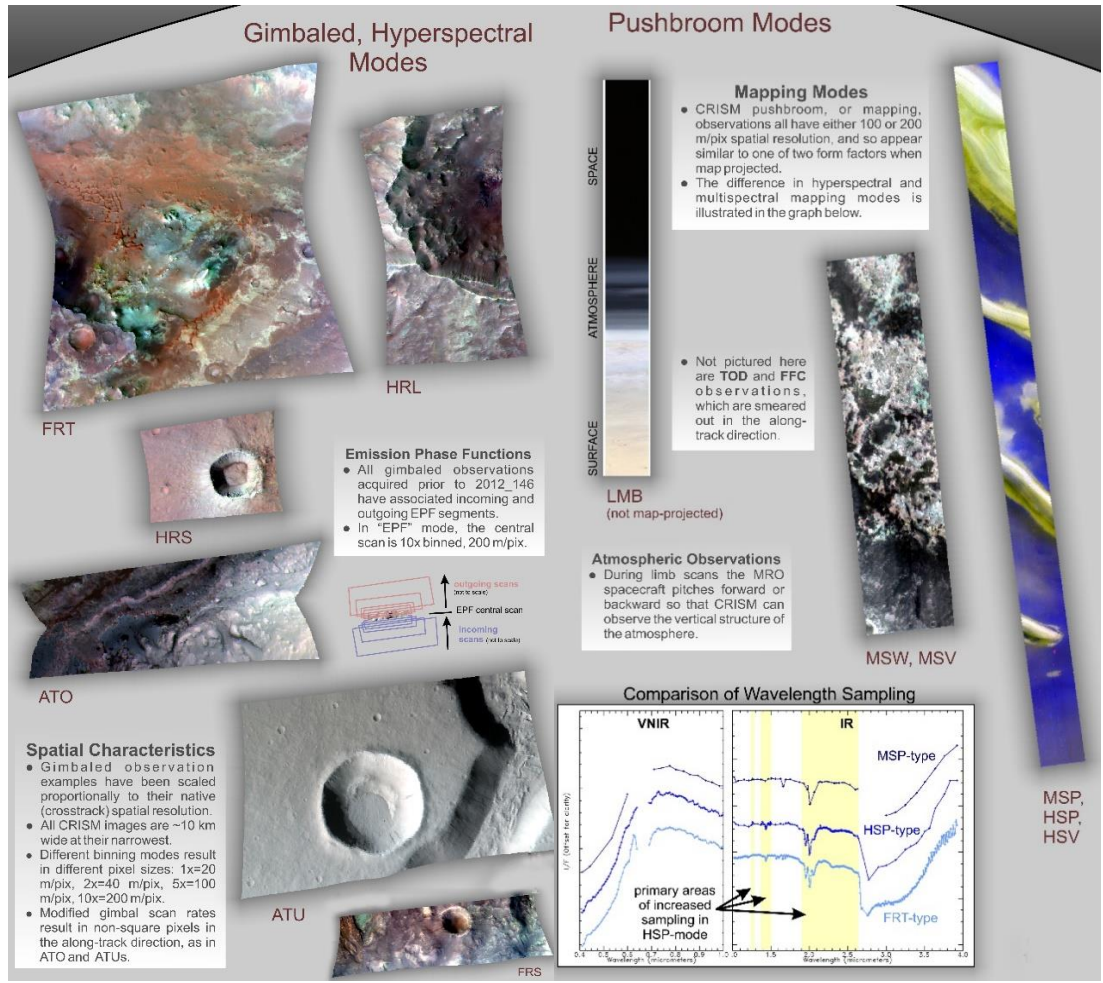

**Fig. 4.2.** Examples of CRISM observations in the two operational modes: gimbaled and pushbroom. Image credits: APL (Applied Physics Laboratory) - The Johns Hopkins University.

In the present work, I analyzed targeted observations (gimbal tracks surface with superimposed scan for each image) with a spatial resolution ranging from 18 m/pixel (Full Resolution Targeted, FRT, observations) to 36 m/pixel (Half Resolution Long channel, HRL, observations). In both modes, the spectrum, in the VNIR (Visible and Near InfraRed) spectral region from 0.4 to 4.0 µm is sampled in 544 channels at 6.55 nm using two different detectors (Viviano-Beck et al., 2014). The short wavelength channel covers the range between 0.4 and 1.0 μm, while the long wavelength channel ranges from 1.0–4.0 μm (Murchie et al., 2007). Since I paid attention to the possible detection of aqueous alteration minerals that have spectral features located in the infrared range, I analyzed CRISM targeted long wavelength observations.

CRISM can also work in a survey mode at 100 or 200 m/pixel with a variable spectral resolution (covering the whole range from 72 to 261 bands depending on the observation mode), with sampling density varying from 53.4 nm between channels at sparsely sampled wavelengths to contiguous sampling at 6.55 nm/channel (Viviano-Beck et al., 2014).

### *4.2.3 Structure of a CRISM data*

Each CRISM image is structured as a cube (see **Fig. 4.3**) with two spatial dimensions (one given by the along track motion and the other by the slit length) and one spectral. In other words, each pixel in a two-dimensional CRISM image has a corresponding spectrum of the reflected sunlight covering CRISM wavelength range, from the visible to the infrared: this spectral data can be thought of as the third dimension of the cube. There is one image for the VNIR focal plane and one image for the IR focal plane. CRISM builds this cube by taking images one line at a time. A two-dimensional spatial image of a target is built up by taking successive lines as the spectrometer is swept across the targeted region. This sweeping motion can be performed by either the movement of CRISM's gimbal or by MRO's along-track motion over the Martian surface.

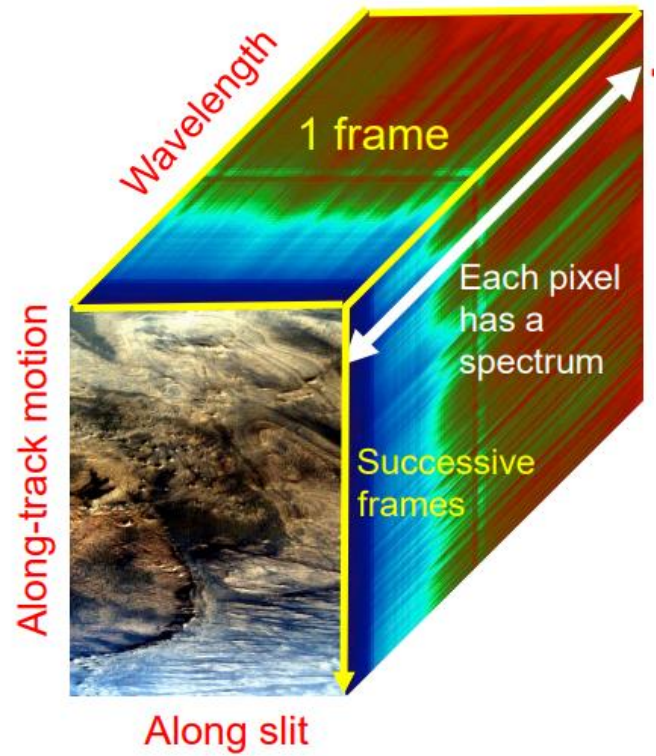

**Fig. 4.3**. 3D representation of a CRISM 'cube'. Each readout of the detector is 1 line of a spatial image. The whole image is built as MRO moves along its orbit. Each pixel has a spectrum that can be compared with those of candidate minerals. Image credits: JPL/NASA

The image from each focal plane has a header with 220 housekeeping items containing full status of the instrument hardware, including data configuration, lamp and shutter status, gimbal position, a time stamp, the target ID and macro within which the frame of data was taken.

The raw data from all observations are stored in a CRISM product called Experimental Data Record (EDR), but the main data of interest are the so-called Targeted Reduced Data Record (TRDR). In a TRDR the image data has been converted to units of radiance or I/F. Each observation is accompanied by a Derived Data Record (DDR) that contains: - geometric information such as latitude, longitude, incidence, emission and phase angles; - information on surface physical properties, such as slope, azimuth and thermal inertia. The combination of TRDR and DDR images produces the map-projected data of the same scene.

Each CRISM file has a name that fully describes the type of data, the used detector, the version of the processing, and gives the unique ID and counter. In detail, the file name structure is the following (let us consider FRT00003F12_07_SC166L_TTR3.LBL for example):

- Class Type (e.g. FRT, HRL, HRS);
- 8-digit hexadecimal observation ID (e.g 00003F12);
- Hexadecimal counter for image within observation (e.g. 07);
- Processing (RA for radiance, IF for I/F in TRDRs or DE for derived information in DDRs) and internal command used (e.g. SC166);
- Sensor ID (S for VNIR or L for IR);
- Software version of calibration (TTR3 for TRDRs or DDR1 for DDRs);
- File extension (IMG for binary data, LBL for detached ASCII PDS label and TAB for detached ASCII table of housekeeping).

## 4.3 Data analysis and methodology

### 4.3.1 CRISM Analysis Toolkit (CAT)

In order to analyze the CRISM data I used the CRISM Analysis Toolkit (CAT) which is a collection of ENVI and IDL procedures (initiated by Pelkey S. and colleagues at Brown University and then developed by the CRISM Science Team) for reading, displaying and analyzing CRISM data.

A CRISM cube is stored in a PDS (Planetary Data System) standard format and it is converted in a file easily readable by ENVI. The Planetary Data System is an archive of data products from NASA planetary missions, which is sponsored by NASA's Science Mission Directorate. All PDF-produced products are peer-reviewed, well-documented and easily accessible through a system of online catalogs that are organized by planetary disciplines.

## *4.3.2 Spectral atmospheric and photometric correction*

Before the spectral analysis, atmospheric and photometric corrections were applied to the cubes. The photometric correction is a first order correction to radiance for non-normal solar incidence, i.e. a division by the cosine of the incidence angle. The atmospheric correction, required only for IR cubes, allows correction of the spectrum for the absorption of $CO_2$ around 2 μm, along with other atmospheric gases. This contribution of the atmosphere is removed by applying to the CRISM data-spectra the Volcano Scan method of McGuire et al. (2009), adapted from earlier methods (Erard and Calvin, 1997; Murchie et al., 2007; Pelkey et al., 2007). Overall, the Volcano Scan is a simple and efficient method to separate the spectral contributions of the surface and the atmosphere, but small residual absorptions near 2 μm can be still present. After the atmospheric correction, an algorithm removes along track column–oriented stripes and also isolated spikes (Seelos et al., 2011).

In order to minimize residual atmospheric and instrumental artifacts in the reduced spectra, average spectra from regions of interest were compared to the average spectra from a spectrally unremarkable region, called 'neutral', in the same scene. When possible, these two regions are composed by spectra belonging to the same lines of the un–projected image to reduce the detector–dependent noise. Nevertheless, some artifacts could still be present, as, for example, near 1.65 μm owing to a detector filter boundary (Murchie et al., 2009).

## *4.3.3 Spectral parameters*

Once the corrected spectra are obtained by applying the methodology described in the previous section, there are three fundamental steps to be accomplished for a good spectral analysis:

1) Locate interesting material(s) (by means of the use of spectral parameters);
2) Collect best possible spectra (pixel average, Region Of Interest -ROI);
3) Interpret endmember spectra (comparison with laboratory data).

Being interested in the detection of aqueous alteration minerals, I focused my analysis on the data from the long wavelength detector. Those minerals are, in fact, detectable and mappable in remote sensing data because of distinctive absorptions that occur at infrared wavelengths due to vibrations of molecules within the mineral structure. Among them: fundamental H − O stretching and bending modes (near 3.0 μm and 6.0 μm) and their overtones and combinations of overtones (near 1.4 μm and 1.9 μm). In the NIR (Near-InfraRed) wavelength range we have, also, Metal-OH stretching (near 2.8 μm), overtone (near 1.4 μm) and combination absorptions (2.2–2.5 μm). Those are particularly useful in identifying the precise mineral because the position and shape of the absorption features depend on the composition of the cations and the configuration of the cation site (e.g. Michalski et al. 2005; Bishop et al. 2008). For example, in the case of smectite clays, the NIR overtones and combination tones are located at 1.41 μm and 2.21 μm for Al-rich montmorillonites, at 1.43 μm and 2.28 μm for Fe-rich nontronites, and at 1.38 μm and 2.32 μm for Mg-rich saponites (Bishop et al. 2008; Ehlmann et al. 2009).

For the identification of regions where hydrated minerals are located I applied to the CRISM observations the so-called spectral parameters, i.e. parameters directly linked to spectral signatures of the minerals of interest. The idea of utilizing parameters to analyze spectral data was used with success in past analyses of spectral data (see for example Bell et al., 2000; Murchie et al., 2009).

The idea is that a given spectral feature, can be captured by a single parameter value, which is calculated by applying an algorithm to the data. Each parameter is designed with the precise intent to capture spectral features diagnostic of a specific mineralogy. Typically, the parameter value is then mapped across a region to assess spatial variations of the spectral feature, which can in turn be interpreted as spatial variations of the associated mineralogy.

For example, absorption features can be parametrized by calculating the band depth defined as $\Delta = 1 - R_\lambda/R_\lambda^*$, where $R_\lambda$ is the reflectance at the center wavelength of the band and $R_\lambda^*$ is the interpolated continuum reflectance at the same wavelength (Clark and Roush, 1984). The continuum level is created from a linear fit between two wavelengths from either absorption sides that are part of the local continuum levels.

The original spectral parameters used for the analysis of CRISM data were developed by Pelkey et al. (2007) and designed for the CRISM 72-band multispectral mapping mode based on the surface composition derived by the OMEGA data (Poulet et al., 2007). In this work, I used the revised set of spectral parameters described by Viviano-Beck et al. (2014) and designed to improve the definition of a given spectral feature in the hyperspectral mode of CRISM. In fact, in this new description a value of reflectance of a particular wavelength is replaced with the median of the reflectance values corresponding to a certain number of wavelengths around it; this reduces the effect of residual noise.

I took into consideration all the possible spectral parameters able to detect the presence of phyllosilicates, sulfates and carbonates. Where I found water alteration minerals, I looked also for the presence of mafic material which could have been the starting material of the alteration process. For this reason, I used the spectral parameters useful to identify mafic minerals, such as olivine (OLINDEX) and pyroxenes, both with a low (LCP) and high (HCP) content of Ca (LCPINDEX and HCPINDEX, respectively), as well as hydrated minerals, such as phyllosilicates and carbonates (for example: BD1900, BD2210, BD2350, BD2500, D2300 and CINDEX). The definition of each of these parameters is reported in Viviano-Beck et al. (2014).

In addition, a mineral can be detected using more than just one spectral parameter. In this way, checking if these parameters are positive in the same region of interest allows a more reliable identification of that mineral.

Parameter maps, calculated using the CRISM Analysis Toolkit (CAT) version 7.3.1, were analyzed alongside image cubes covering the same spatial regions.

For each cube, it is possible to extract by means of CAT a file containing some spectral parameters. Each spectral parameter can be visualized as a single greyscale image. Alternatively, three different images can be merged creating a three level RGB image useful to detect mineral deposits. RGB combinations of CRISM spectral parameters summary products by Viviano-Beck et al. (2014) useful for the detection of mafic and hydrated minerals are shown in **Table 4.3**.

**Table 4.3**
RGB combination of spectral parameter summary products. Extract from the Table 3 of Viviano-Beck et al. (2014).

| Abbreviation | RGB association | Interpretation |
| --- | --- | --- |
| **FAL** | R2529<br>R1506<br>R1080 | From "false color." An enhanced infrared false color representation of the scene. The wavelengths chosen highlight differences between key mineral groups. Red/orange colors are usually characteristic of olivine-rich material, blue/green colors often indicate clay, green colors may indicate carbonate, and gray/brown colors often indicate basaltic material. |
| **MAF** | OLINDEX3<br>LCPINDEX2<br>HCPINDEX2 | From "mafic mineralogy." Shows information related to mafic mineralogy. Olivine and Fe-phyllosilicate share a 1.0–1.7 μm bowl-shaped absorption and will appear red in the MAF browse product. Low and high-Ca pyroxene display additional ~2.0 μm absorptions and appear green/cyan and blue/magenta, respectively. |
| **HYD** | SINDEX2<br>BD2100_2<br>BD1900_2 | From "hydrated mineralogy." Shows information related to bound water in minerals. Polyhydrated sulfates have strong 1.9 μm and 2.4 μm absorption bands, and thus appear magenta in the HYD browse product. Monohydrated sulfates have a strong 2.1 μm absorption and a weak 2.4 μm absorption band, and thus appear yellow/green in the HYD browse product. Blue colors are indicative of other hydrated minerals (such as clays, hydrated silica, carbonate, or zeolite). |
| **PHY** | D2300<br>D2200<br>BD1900r2 | From "phyllosilicates." Shows information related to hydroxylated minerals including phyllosilicates. Fe/Mg-OH bearing minerals (e.g., Fe/Mg-phyllosilicates) will appear red, or magenta when hydrated. Al/Si-OH bearing minerals (e.g., Al-phyllosilicates or hydrated silica) will appear green, or cyan when hydrated. Blue colors are indicative of other hydrated minerals (such as sulfates, hydrated silica, carbonate, or water ice). |
| **PFM** | BD2355<br>D2300<br>BD2290 | From "phyllosilicates with Fe and Mg." Shows information related to cation composition of hydroxylated minerals including Fe/Mg-phyllosilicate. Red/yellow colors indicate the presence of prehnite, chlorite, epidote, or Ca/Fe carbonate, while cyan colors indicate the presence of Fe/Mg smectites or Mg carbonate. |

**Table 4.3** - continued

| | | |
|---|---|---|
| **PAL** | BD2210_2<br>BD2190<br>BD2165 | From "phyllosilicates with Al." Shows information related to cation composition of hydroxylated minerals including Al-phyllosilicate and hydrated silica. Red/yellow colors indicate the presence of Al smectites or hydrated silica, cyan colors may indicate the alunite, and light/white colors indicate the presence of kaolinite group minerals. |
| **HYS** | MIN2250<br>BD2250<br>BD1900r2 | From "hydrated silica." Shows information related to Si/Al-hydroxylated minerals that can be used to differentiate between hydrated silica and Al-phyllosilicates. Light red/yellow colors indicate the presence of hydrated silica, whereas cyan colors indicate Al-OH minerals. Additionally, jarosite will appear yellow. Blue colors are indicative of other hydrated minerals (such as sulfates, clays, hydrated silica, carbonate, or water ice). |
| **ICE** | BD1900_2<br>BD1500_2<br>BD1432 | From "ices." Shows information related to water or carbon dioxide frost or ice. CO2 frost or ice displays a sharp 1.435 μm absorption and thus appears blue in the ICE browse product. Water ice or frost has a strong 1.5 μm absorption and thus appears green in the ICE browse product. Red colors are indicative of hydrated minerals (such as sulfates, clays, hydrated silica, carbonate, or water ice). |
| **IC2** | R3920<br>BD1500_2<br>BD1435 | From "ices, version 2." Shows complementary information related to water or carbon dioxide frost or ice. $CO_2$ frost or ice displays a sharp 1.435 μm absorption and thus appears blue in the IC2 browse product. Water ice or frost has a strong 1.5 μm absorption and thus appears green in the IC2 browse product. The reflectance at 3920nm is a proxy for silicates, which are more reflecting than ices at 3.9 μm, so red colors represent ice-free surfaces. |
| **CHL** | ISLOPE<br>BD3000<br>IRR2 | From "chloride." Shows information related to inferred chloride deposits detected from Thermal Emission Imaging System (THEMIS) data and spatially associated to hydrated mineral deposits. Of the THEMIS-based chloride detections studied to date, these surfaces have a relatively positive near infrared spectral slope and are comparatively desiccated, so chlorides appear blue in the CHL browse product. Yellow/green colors are indicative of hydrated minerals, especially phyllosilicates. |

| **Table 4.3** – continued | | |
|---|---|---|
| **CAR** | D2300<br>BD2500H2<br>BD1900_2 | From "carbonates." Shows information related to Mg carbonate minerals. Blueish- or yellowish-white colors indicate Mg carbonates, while red/magenta colors indicate Fe/Mg-phyllosilicates. Blue colors are indicative of other hydrated minerals (such as sulfates, clays, hydrated silica, or carbonate). |
| **CR2** | MIN2295_2480<br>MIN2345_2537<br>CINDEX2 | From "carbonates, version 2." Shows information distinguishing carbonate minerals. Red/magenta colors indicate Mg carbonates, while green/cyan colors indicate Fe/Ca carbonates. |

We have stretched the summary products applied to our observations using the limits fixed by Viviano-Beck et al. (2014). The lower stretch limit is fixed at a value equal to 0, usually corresponding with a band depth of zero, while the upper stretch limit is either set to the 99th percentile of the cumulative histogram, or the 99.9th percentile of the local histogram. Within those limits, it is possible to be sure that the resulting stretched summary product will not over-enhance a small tail of the distribution that may be subject to outliers (caused by uncontrolled noise) (Viviano-Beck et al., 2014).

For a better identification of minerals, for each data cube, appropriate RGBs are created, associating different colors to the different compositions. Once I obtained the RGBs that were chosen on the basis of the selected minerals, I collected the spectra from some areas of interest in order to compare them to laboratory spectra of minerals extracted from the RELAB database (Pieters, 1983; Pieters and Hiroi, 2004 - http://www.planetary.brown.edu./relab/). To do that, it is possible, using CAT, to create a so-called Region Of Interest (ROI) on the CRISM observation composed of pixel with high values of a spectral parameter.

With the use of a ROI TOOL it is then possible to obtain statistical results on the selected ROI, such as average spectrum and standard deviation. After that, spectra in the original CRISM cube corresponding to accumulation of pixels with the same color in the RGBs are averaged and compared to that of neutral region obtained in the same way. The advantage of this method is to greatly reduce the detector-dependent noise.

However, even with the use of this expedient, sometimes artifacts still remain, such as those centered around 1.65 µm and 3.18 µm. In particular, the former is present in our spectra and it is indicated by a yellow stripe.

The resulting ratioed spectra are then compared to laboratory RELAB spectra that are available in CAT.

In the next section we will review in detail the CRISM data analyzed from those basins where we suggest the possible presence of hydrated minerals and the comparison with RELAB spectra.

## *4.4 Results*

As reported in **Tables 4.1** and **4.2** interesting CRISM observations were found in one of the analyzed closed-basin lakes and in three of the open ones. The morphology of the basin lakes was analyzed using a combination of CTX data (with a resolution up to 6 m/pixel) and THEMIS data (resolution of ~ 100 m/pixel). These interesting basin lakes, along with their CRISM observations, are discussed in the following subsections.

### *4.4.1 Closed basin #16*

This basin, shown in **Fig. 4.4**, is centered around 24.4°S 61.4°E and it is located in *Terra Sabaea*. The crater has a diameter of ~ 15 km and it was also observed by HiRISE (High Resolution Imaging Science Experiment) camera (**Fig. 4.5**). HiRISE is a camera, onboard the Mars Reconnaissance Orbiter, operating in the visible range with a telescopic lens that allows it to obtain images at resolution never seen before in the other planetary exploration missions. Thanks to this high resolution it is possible to discern 1-meter-size objects on Mars and to analyze the morphology in a more detailed way with respect to what has been done so far.

In the upper part the basin shows an exposure of light-toned bedrock (as reported by the HiRISE team at the link - https://www.uahirise.org/ESP_016668_1555).

The observed exposure extends ~ 3.5 km horizontally and ~4 km vertically. The rest of the basin seems to be resurfaced. All the valleys visible in **Fig. 4.4** are inlets of the basin. This crater was not previously catalogued among the closed-basin lakes of Goudge et al. (2015). I found this potential closed paleolake during the manual mapping procedure of Martian valleys (see Chapter II).

For this basin I found a CRISM observation (FRT0001689A) in the region corresponding to this light-toned bedrock exposure (**Fig. 4.4**). The CRISM data were processed with the application of CAT, as described in Section 4.4. After the photometric and atmospheric corrections and data filtering, I applied all the spectral parameters listed in **Table 4.3**.

**Fig. 4.4b** shows the map-projected RGBs, while the spectral parameter maps **Fig. 4.4c**, **d** and **e**, correspond respectively to the MAF, PHY and PFM groups of spectral parameters (for the definition of these parameters see **Table 4.3**).

The RGBs and the parameters maps shown in **Fig. 4.4** were obtained after conveniently stretching the histogram of each spectral parameter values as described in Section 4.3.3.

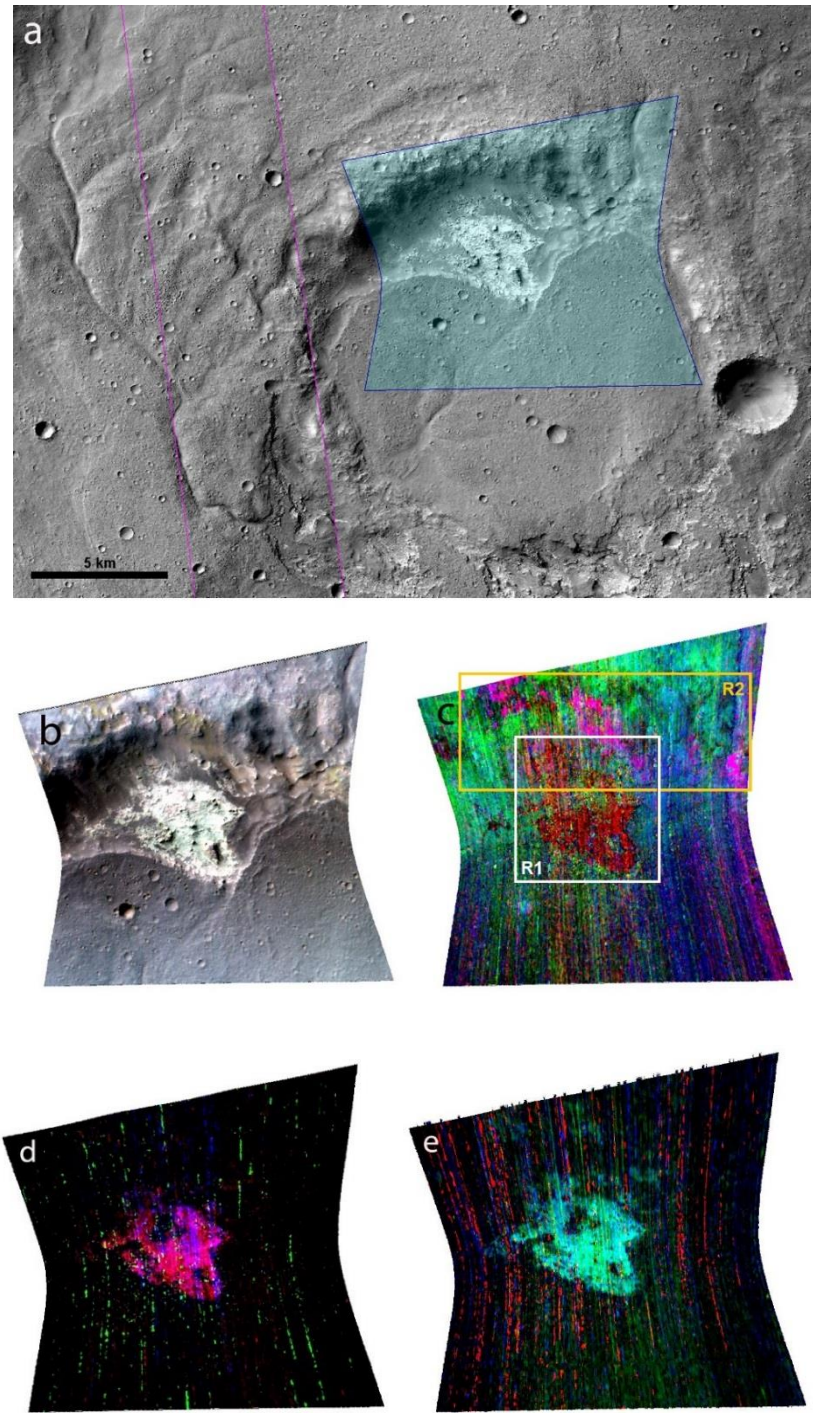

**Fig. 4.4.** a) Closed basin #16 shown in a mosaic of CTX images (B18_016668_1536_XN and F02_036632_1537_XN, whose boundaries are indicated by the pink lines) with superimposed the CRISM observation footprint FRT0001689A, blue swath; b) Map-projected RGB of FRT0001689A observation showing the enhanced IR false color (red: 2.52; green: 1.50; blue: 1.08 µm); c) MAF parameter map (red: OLINDEX3; green: LCPINDEX2; blue: HCPINDEX); d) PHY parameter map (red:D2300; green:D2200; blue: BD1900r2); e) PFM parameter map (red: BD2355; green: D2300; blue: BD2290). White and yellow boxes in c) indicate the two regions where pixels were collected for creating the studied ROIs (R1 and R2, respectively). Here as well as in all the following figures North is up.

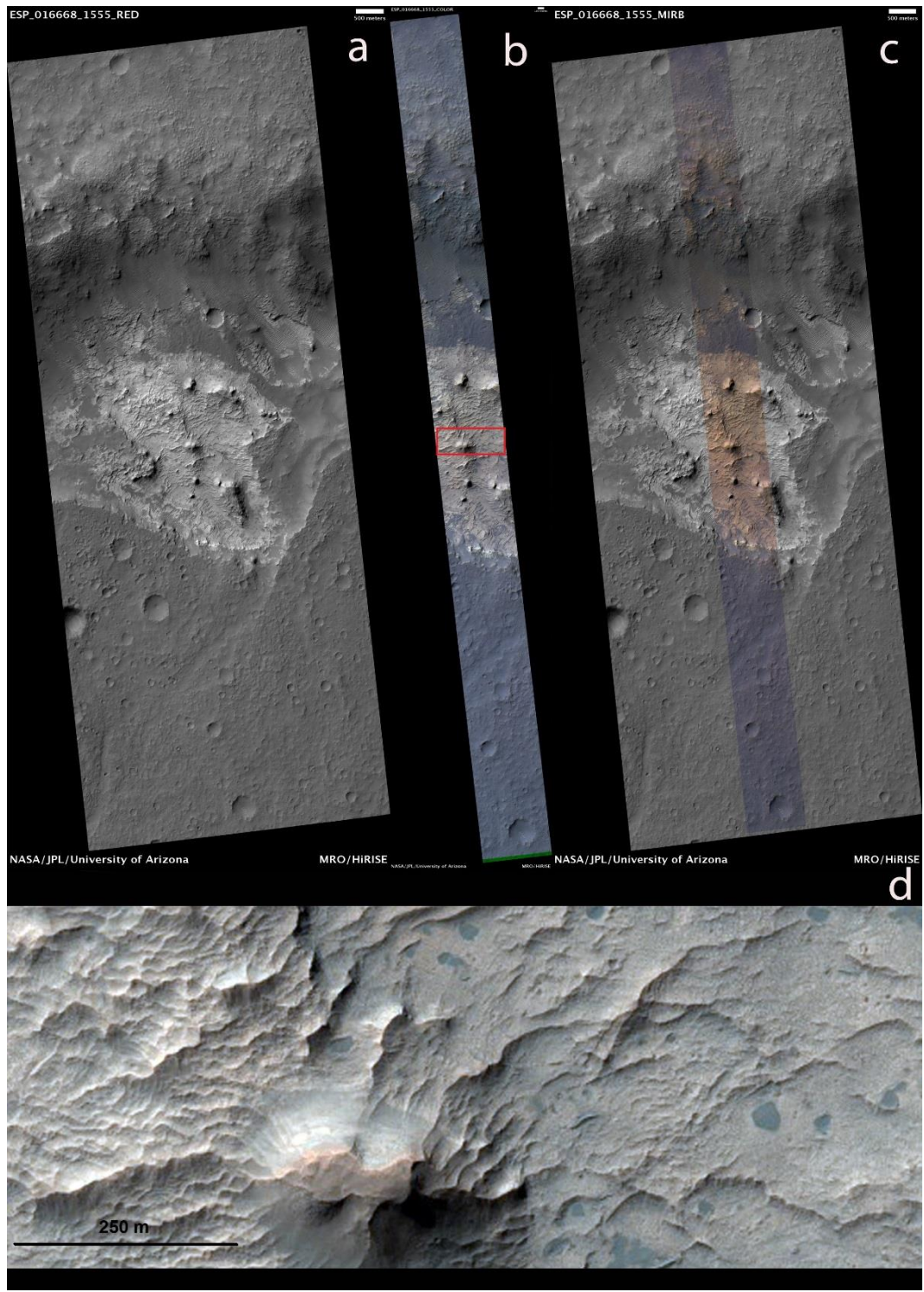

**Fig. 4.5.** Set of HiRISE images showing the region of the closed basin #16 where light toned exposed bedrock is visible: a) Black and white projected map; b) IRB (infrared-red-blue) color projected map; c) merged IRB color projected map; d) close-up of the central target zone shown in the HiRISE observations (red box in b). The resolution of the projected map is 25 cm/pixel and the map projection is equirectangular. Credits: NASA/JPL/University of Arizona.

The spectral parameter maps indicate the presence of mafic material (red and magenta pixels in **Fig.4.4c**) along with water related minerals (magenta pixels in **Fig. 4.4d**; cyan pixels in **Fig. 4.4e**). As reported in **Table 4.3**, in the mafic (MAF) parameter map, red pixels indicate the presence of Olivine and/or Fe-phyllosilicate, while green/cyan and blue/magenta indicate respectively the presence of HCP and LCP. In the phyllosilicates (PHY) parameter map, magenta pixels correspond, instead, to Fe/Mg-phyllosilicates and in the PFM map cyan pixels corresponds to Fe/Mg-smectites, a class of the Fe/Mg-phyllosilicates. It is possible to observe that the same points of the surface which are positive to the spectral parameters that indicate the presence of Olivine and/or Fe-phyllosilicates in the MAF parameter map (red pixels in **Fig.4.4c**) are positive to the PHY (magenta pixels in **Fig.4.4d**) and to the PFM spectral parameters (cyan pixels in **Fig. 4.4e**). This correspondence suggests that in these points of the surface, the original mafic materials were converted into Fe/Mg phyllosilicates and in particular, into Fe/Mg-smectites. In all these cases, the distribution of the previous mentioned pixels is consistent with the light-toned exposure visible in the CTX and HIRISE images (**Figs. 4.4a** and **4.5**).

In addition, the region on the crater rim North of the dune field is positive to the MAF spectral parameters (magenta pixels in **Fig. 4.4c**). This suggests the presence of unaltered mafic material on the rim.

For this CRISM observation I created two ROIs:

1) R1 – collecting pixels from the exposed region, where the spectral indexes related to the identification of Fe/Mg phyllosilicates are positive (white box in **Fig. 4.4c**);
2) R2 – selecting pixels from the region above the light-toned exposure where the spectral parameters corresponding to the presence of olivine are positive (yellow box in **Fig. 4.4c**).

The average spectra obtained from these ROIs were then normalized to the average spectra of neutral spectral regions, where the spectral parameters are not positive to the identification of the searched minerals. The obtained spectra are shown in the top paned of **Fig. 4.6**.

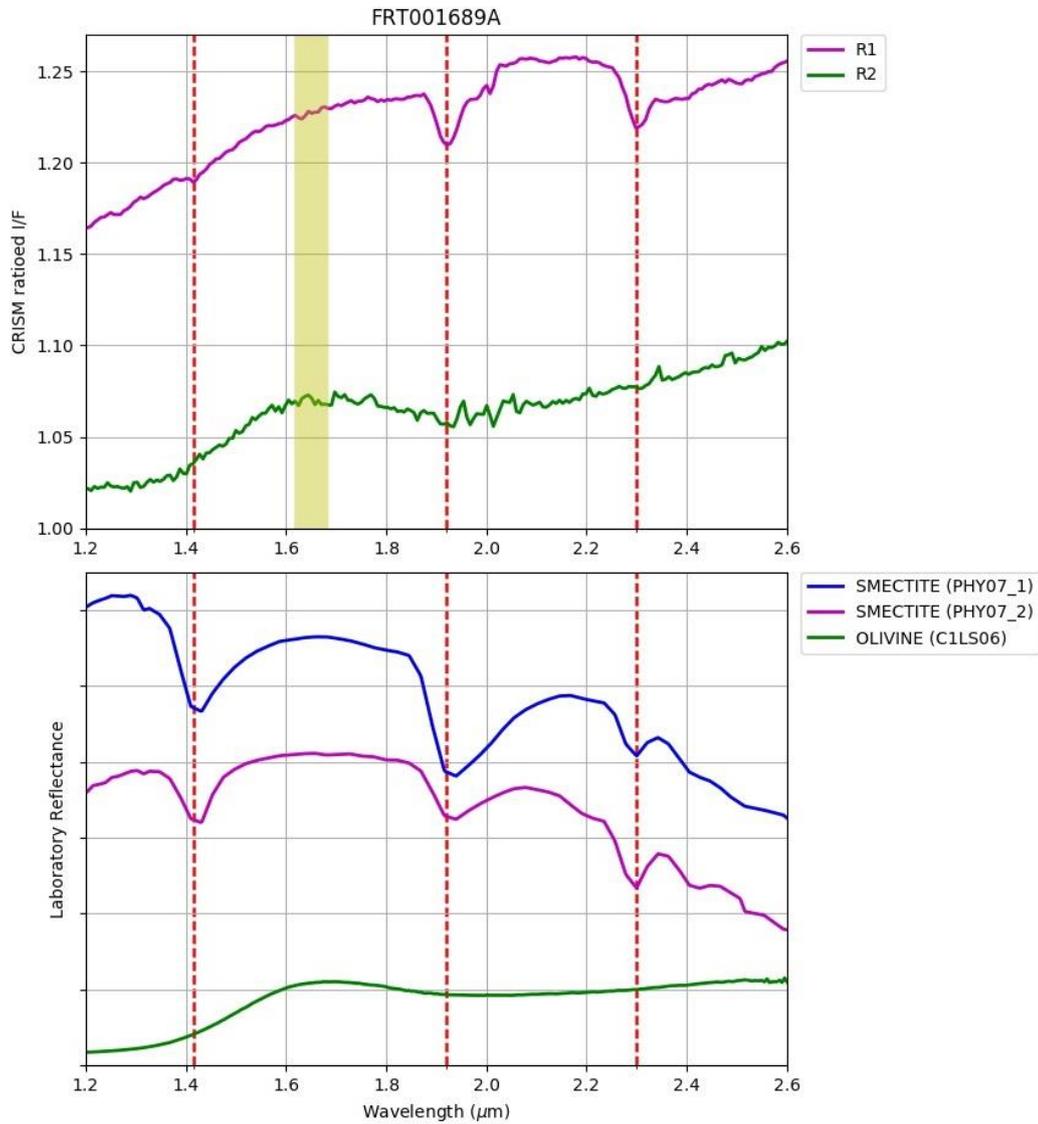

**Fig. 4.6.** Top panel: ratioed CRISM spectra obtained from R1 and R2. Bottom panel: RELAB spectra of nontronite and olivine (shifted for clarity). Vertical red dashed lines at 1.42, 1.92, 2.30 µm indicate the central positions of the bands observed in the CRISM spectra. The yellow stripe indicates the spectral region around 1.65 µm where an instrumental artifact is present.

The spectrum of the region R1 shows clearly the presence of the bands at 1.92 µm and 2.30 µm. It is also possible to catch a glimpse of a band centered around 1.42 µm. Instead the spectrum of the region R2 presents a broad minimum around 2 µm.

For a reliable interpretation of these spectra it is necessary to make a direct comparison with laboratory spectra. To confirm the identification of mafic and hydrated minerals obtained on the basis of the spectral parameters I compared band center positions of the spectra obtained from CRISM observations with laboratory spectra from RELAB database. Among all the RELAB spectra analyzed for comparison the best agreement in band locations was found with the spectra of a nontronite (PHY_07_1 and PHY_07_2 - Fe-Phyllosilicate) and an olivine (C1LS06). Those spectral are showed in the bottom panel of **Fig. 4.6**.

### *4.3.2 Open basin #10*

This basin, shown in **Fig. 4.7**, is centered around 4.0°S 109.2°E and has a diameter of ~ 100 km. It has been catalogued as a potential open-basin lake by Fassett and Head (2005). The inlet valley is NE of the crater rim. In contrast the outlet valley is not well-defined, but my analysis suggests that it is located South of the crater center, as indicated by the white arrow in **Fig**. **4.7a**.

Later Goudge et al. (2012), studying the basin from a geologic and spectroscopic point of view, observed that the basin shows exposed floor materials (**Fig. 4.7**) and analyzed the CRISM observation FRT0001BBA7 (1 - located more or less in the central part of the basin). But the analysis of this observation did not display the presence of hydrated minerals.

We found three more CRISM targeted observations: FRT0000D34E (2); HRL0001851C (3); FRT0001A267 (4). These three subsequent observations are, instead, on a minor crater located SE of the center of the basin and close to its outlet.

For the observation FRT0001A267 there are only the short wavelength data that were therefore not studied for the reasons explained in Section 4.2.2.

We focused our attention on the observations FRT0000D34E and HRL0001851C. As for the previous case, I applied photometric and atmospheric corrections and data filtering. The subsequent application of spectral parameters (**Table 4.3**) in both observations did not show any significant presence of hydrated minerals and/or mafic materials. We found only some indicative pixels (magenta) of the possible presence of hydrated minerals in the observation FRT0000D34E (**Fig. 4.7c**) after the application of the PHY spectral parameters (**Fig. 4.7d**).

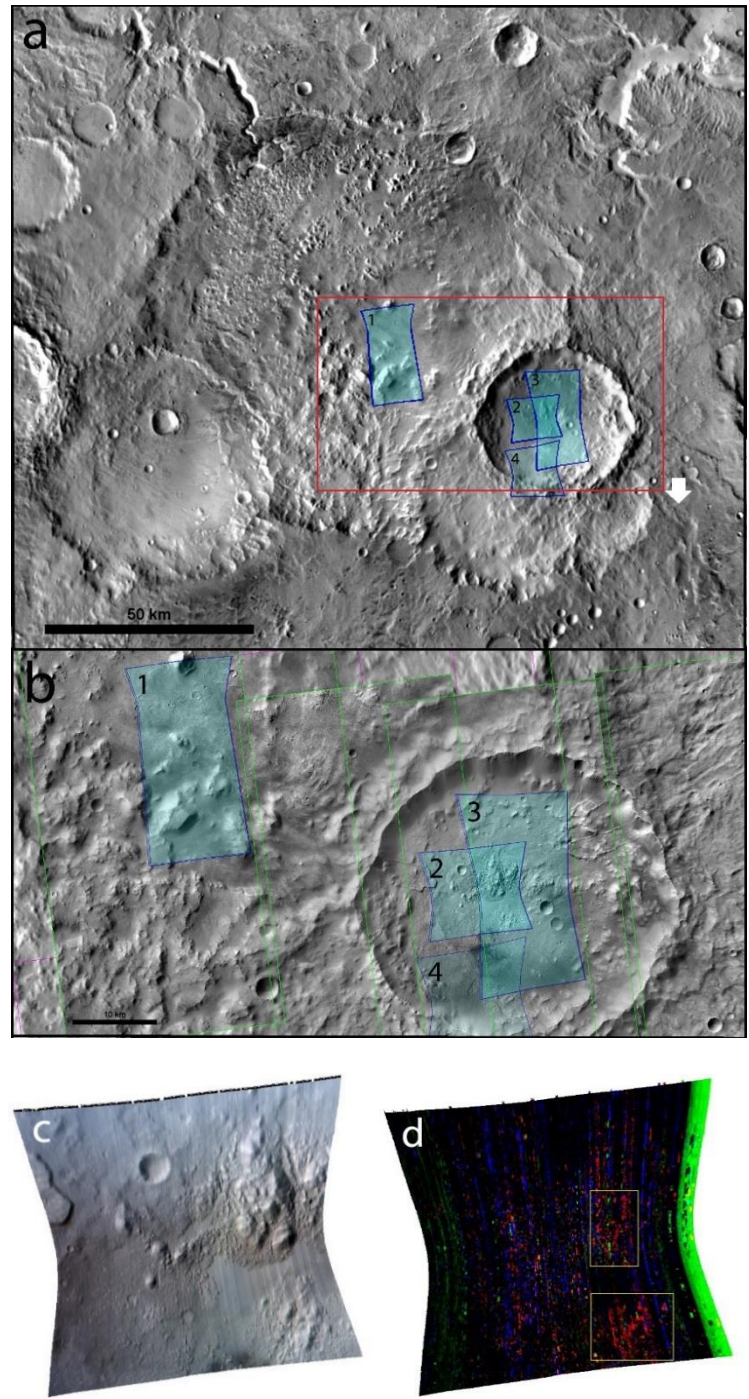

**Fig. 4.7.** a) Open basin #10 centered at 4.0°S 109.2°E, shown in a THEMIS IR daytime mosaic (~ 100 m/pixel) with superimposed the CRISM observation footprints (blue swaths): FRT0001BBA7 (1); FRT0000D34E (2); HRL0001851C (3); FRT0001A267 (4); b) close-up of the secondary crater area (red box in a), shown in a mosaic of five CTX images (P14_006645_1740_XN, G23_027347_1754_XN, B20_017576_1754_XN, B03_010640_1752_XN and B01_010073_1761_XN, whose boundaries are indicated by the green lines); c) Map-projected RGB of FRT0000D34E observation showing the enhanced IR false color (red: 2.52; green: 1.50; blue: 1.08 µm); d) PHY parameter map (red: D2300; green:D2200; blue: BD1900r2). Yellow boxes in d) indicate the two regions where pixels were collected for creating the studied ROI.

However, also in this case, the observation is very noisy, and the useful pixels are very few (yellow boxes **Fig. 4.7d**). I created a ROI (R1) selecting those pixels (magenta ones) and obtained the average spectrum reported in the upper panel of **Fig. 4.8**. This spectrum has been normalized to a spectrum from a neutral region containing more or less the same number of pixel.

Owing to the high noise it is difficult to assess the presence of some bands. However, it is possible to recognize the presence of the water bands centered at 1.91 µm and 2.31 µm and it is also possible to see an indication of a band centered around 1.40 µm, but being almost within the noise it cannot be considered diagnostic.

I tried to compare the spectrum with some RELAB spectra of hydrated minerals (bottom panel - **Fig. 4.8**) and we found a possible match with the band positions of some smectite spectra (BKR1JB172, C1JB172, PHY07 and CJB170).

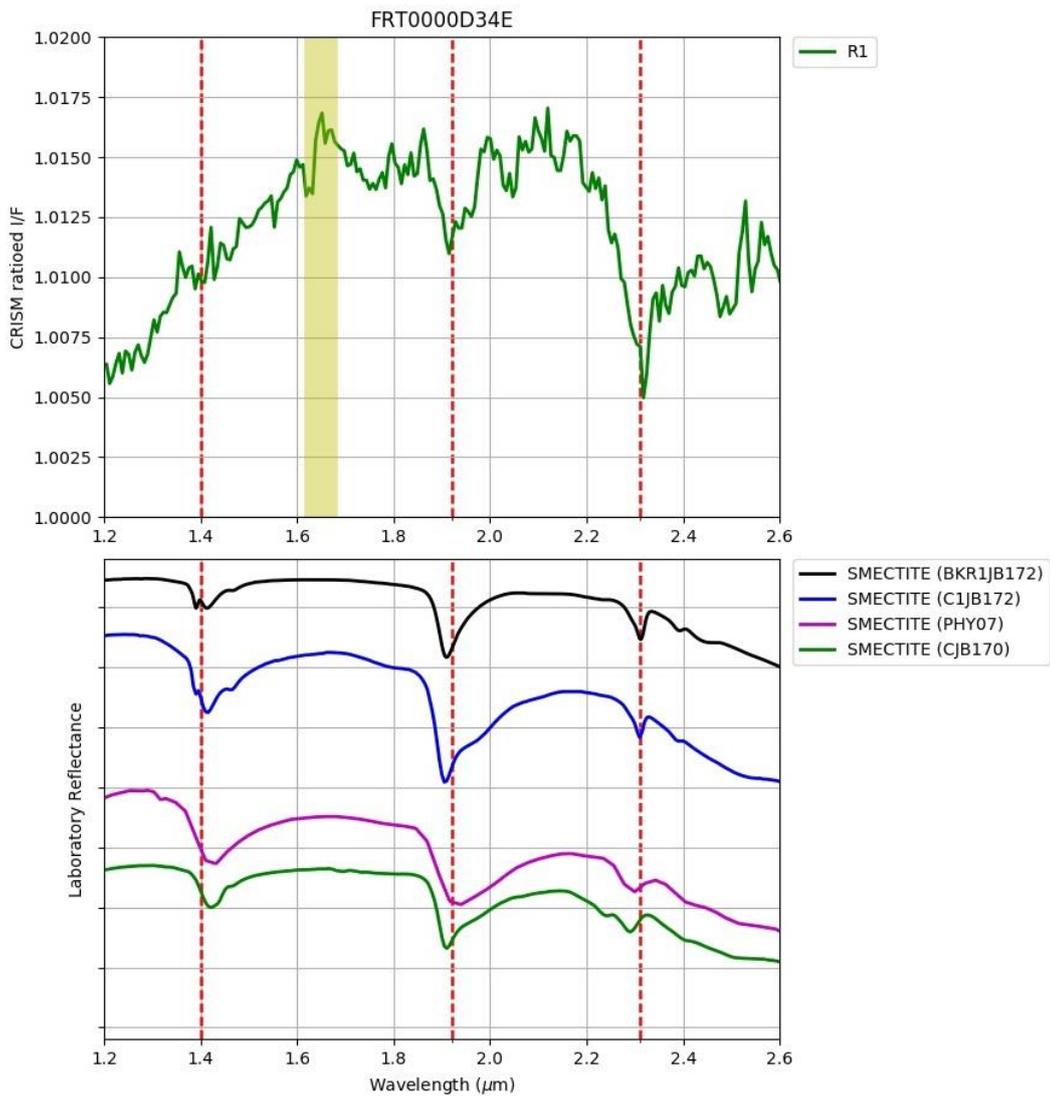

**Fig. 4.8.** Top panel: ratioed CRISM spectrum obtained from R1. Bottom panel: RELAB spectra (shifted for clarity). Vertical red dashed lines at 1.40, 1.91, 2.32 µm indicate the central positions of the bands observed in the CRISM spectra. The yellow stripe indicates the spectral region around 1.65 µm where an instrumental artifact is present.

### *4.3.3 Open basin #25*

The open basin #25, is centered at 24.7°S 59.2°E, and it is located in *Terra Sabaea* close to the closed-basin 16.

I found this potential open paleolake during the mapping of fluvial valleys discussed in Chapter II. Checking the two-previous works on open-basin lakes (Fassett and Head, 2008; Goudge et al., 2012), I noticed that this basin was not previously catalogued by these authors.

The crater has a diameter of ~ 35 km and the inlet valley is located NE of the crater, while the outlet is SW (**Fig. 4.9a**)

For this basin I found two FRT CRISM observations (FRT0001CA9D and FRT0000C7A1) at the crater rim close to the outlet valley (**Fig. 4.9**). As in the case of the other CRISM data so far studied, I applied to these two observations the photometric and atmospheric corrections. Then the noise was reduced through a process of data filtering.

I applied several combinations of spectral parameters reported in **Table 4.3**, but the best information was obtained from a combination of spectral indexes previously suggested by Bishop et al. (2013). Adopting their approach, I generated a map with the following association between colors and spectral indexes: red - D2300; green - OLINDEX3; blue - BD2210. In this color configuration, green pixels indicate the presence of olivine while red/orange ones indicate the presence of Fe/Mg smectite.

The spectral parameter maps so obtained indicate the presence of olivine (green pixels in **Fig. 4.9c**, along with water related minerals (red/orange pixels in **Fig. 4.9c** and in **Fig. 4.9e**).

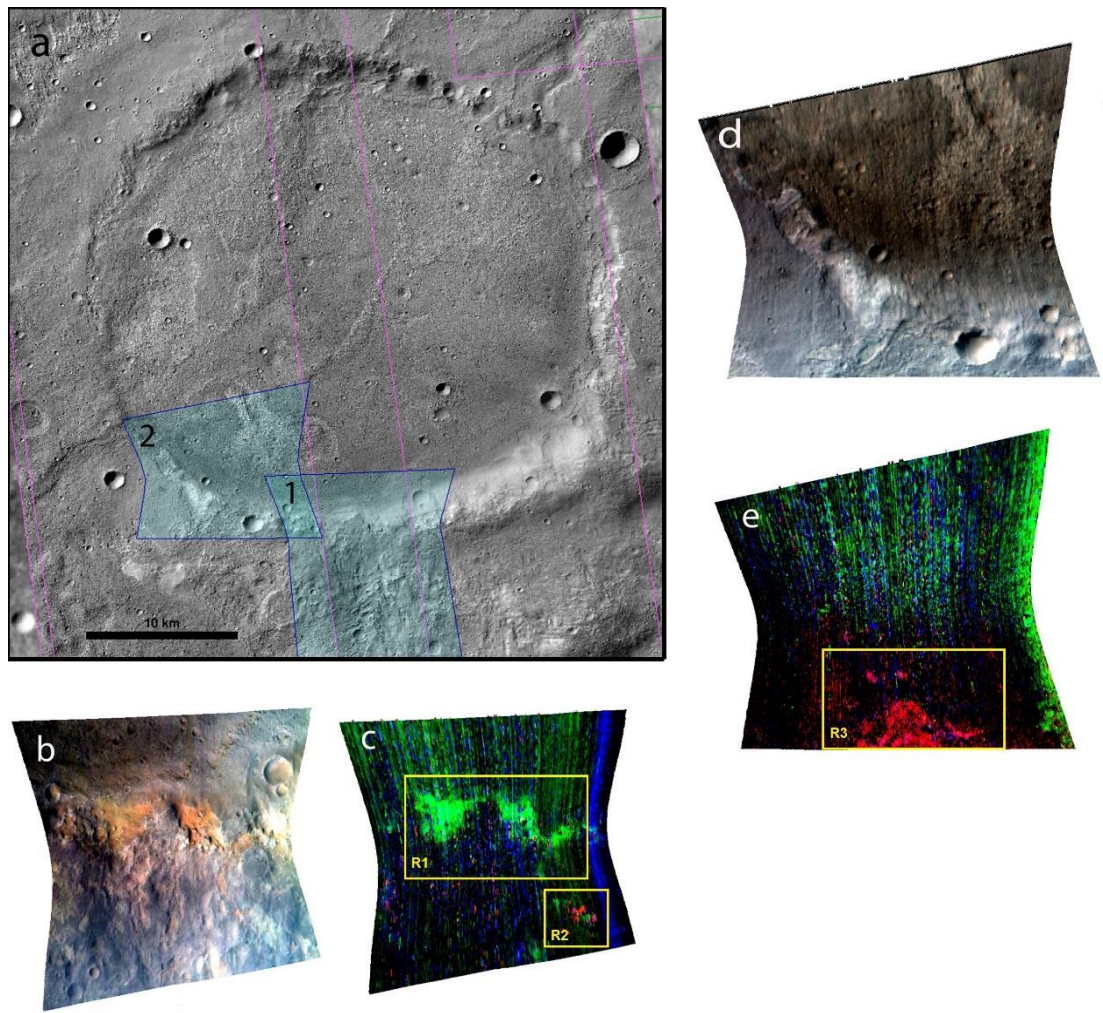

**Fig. 4.9**. a) Crater centered at 24.7°S 59.2°E shown in a CTX mosaic (F01_036276_1546_XN, J03_046074_1551_XN, B01_010009_1552_XI, B02_010431_1551_XI, whose boundaries are indicated by the pink lines). The CRISM observation footprints FRT0001CA9D (1) and FRT0000C7A1 (2) are superimposed and shown in blue swaths. b) and d) Map-projected RGB of FRT0001CA9D and FRT0000C7A1 observations (respectively) showing the enhanced IR false color (red: 2.52; green: 1.50; blue: 1.08 µm). c) and e) Parameter map (red: D2300; green: OLINDEX3; blue: BD2210) where Fe/Mg-smectite appear red/orange while olivine appears green. Yellow boxes in c) and e) indicates the three regions where pixels were collected for creating the studied ROIs (R1, R2 and R3).

For this CRISM observation I created three ROIs:

1) R1 – collecting green pixels in **Fig. 4.9c**;
2) R2 – selecting red/orange pixels from **Fig. 4.9c**;
3) R3 – selecting red/orange pixels in **Fig. 4.9e**.

As usual the spectra obtained from these ROI were then normalized to average spectra of neutral spectral regions (where the spectral parameters are not positive to the identification of the searched minerals) and are shown in the top panel of **Fig. 4.10**. As for the previous cases, I compared them with laboratory spectra. From this comparison (bottom panel **Fig. 4.10**) we observed a similarity between the spectrum of the region R1 and that of an olivine (C1LS06), while the spectra of regions R2 and R3 are similar, because of the presence of the water bands, to the spectra of phyllosilicates and in particular smectite (C1JB172 and PHY07_2) and saponite (LASA59).

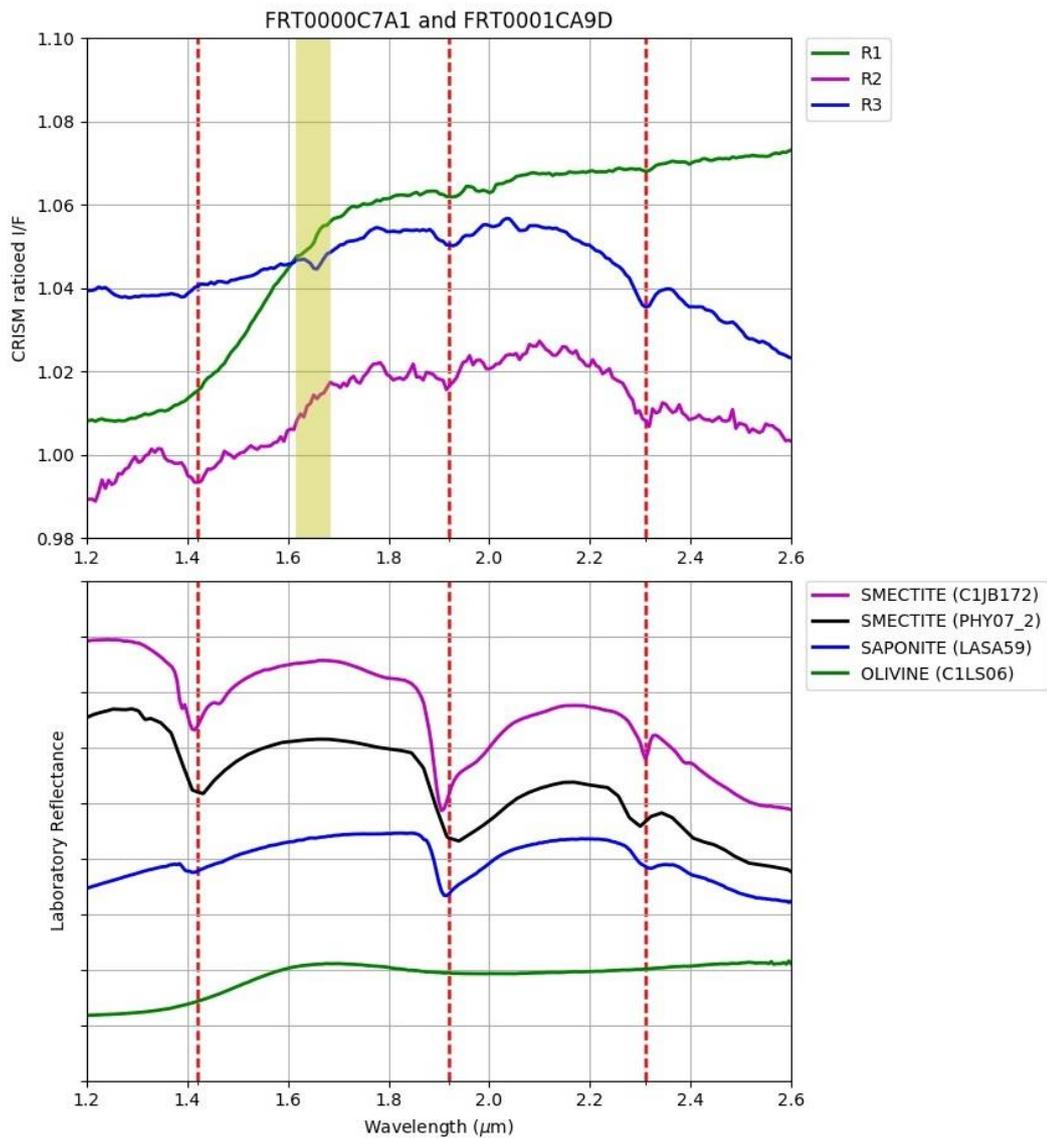

**Fig. 4.10.** Top panel: ratioed CRISM spectra obtained from R1, R2 and R3. Bottom panel: RELAB spectra (shifted for clarity). Vertical red dashed lines at 1.42, 1.92, 2.31 µm indicate the central positions of the bands observed in the CRISM spectra. The yellow area indicates the spectral region around 1.65 µm where an instrumental artifact is present.

### *4.3.4 Open basin #28*

Catalogued as an open-basin by Fassett and Head (2008), this crater is centered at 30.1°S 73.4°E. It has a diameter of ~ 50 km. The inlet valley is located in the northern rim of the crater while the outlet is in the southern rim (**Fig. 4.11**).

Goudge et al. (2012) identified a layered deposit at the center of this basin and studied the superimposed CRISM observation HRL000A153, but no signs of hydrated minerals were detected. Then the CRISM spectrometer observed this basin once again taking an observation at the outlet valley (FRT00009720). We found this observation interesting because it corresponds to the valley course and because so far there are few observations of minerals in a valley course, especially in small valley like this one.

For this observation I applied the same procedure adopted for the other CRISM data here analyzed: I corrected the data and then used spectral indexes to detect the possible presence of hydrated minerals.

Also in this case the application of the PHY and PFM spectral indexes showed the possible presence of hydrated minerals (magenta pixels in **Fig. 4.11d** and cyan pixels in **Fig. 4.11e**) in the valley course (yellow boxes - **Fig. 4.11d** and **e**).

As reported in Section 4.4.1 and shown in **Table 4.3**, in the PHY parameter map, magenta pixels correspond to Fe/Mg-phyllosilicates and in the PFM map cyan pixels correspond to Fe/Mg-smectites.

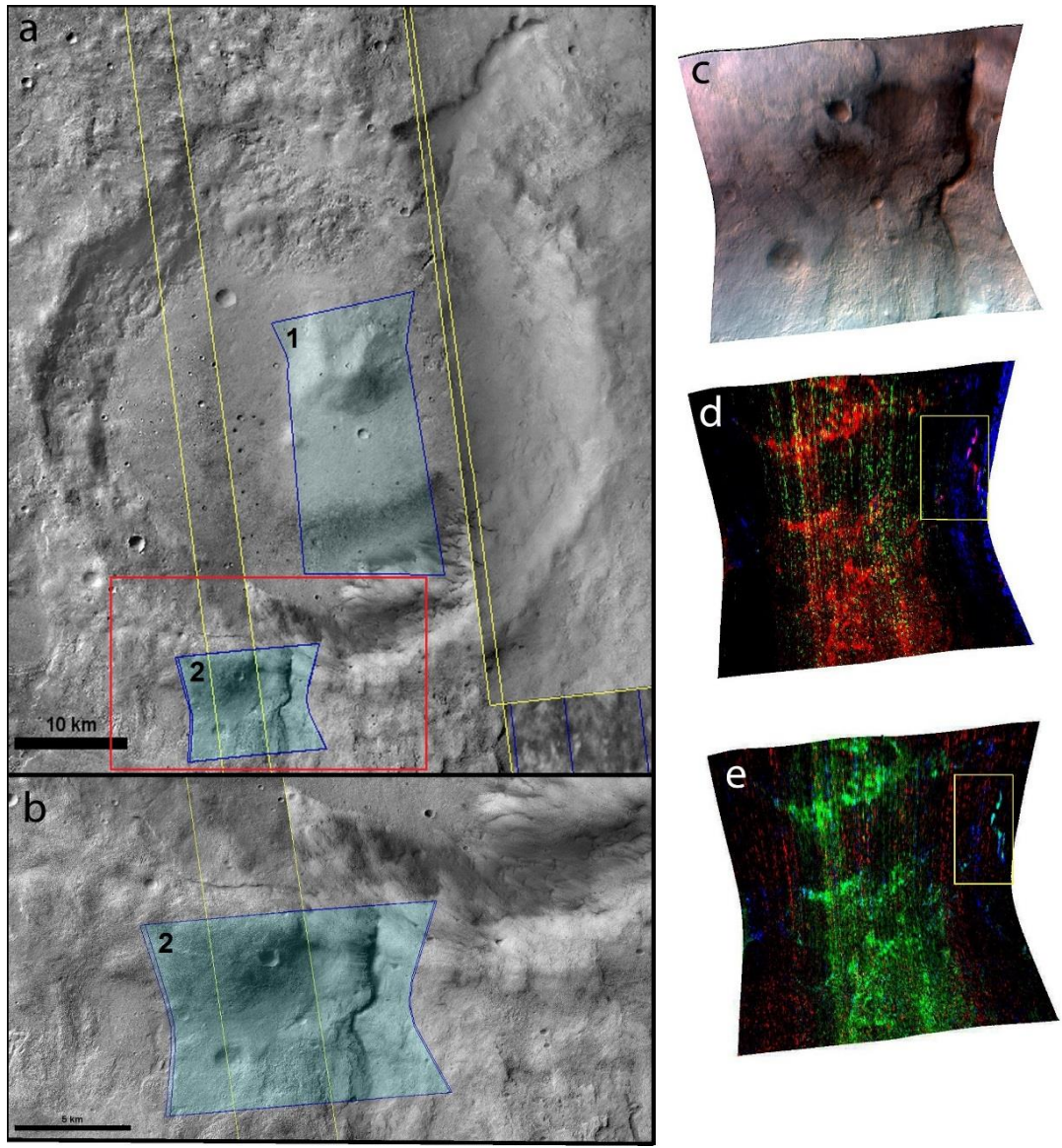

**Fig. 4.11.** a) Crater centered at 30.1°S 73.4°E shown in a CTX mosaic (P12_005618_1514_XI, P20_008690_1481_XI and P13_006185_1518_XN, whose boundaries are indicated by the yellow lines). The CRISM observation footprints HRL000A153 (1) are FRT00009720 (2) superimposed in blue swaths. b) close-up of the outlet of the basin (red box in a); c) Map-projected RGB of FRT00009720 observation showing the enhanced IR false color (red: 2.52; green: 1.50; blue: 1.08 µm). d) PHY parameter map (red: D2300; green: D2200; blue: BD1900r2); e) PFM parameter map (red: BD2355; green: D2300; blue: BD2290). Yellow boxes in c) and e) indicate the region where pixels were collected for creating the studied ROI.

For this CRISM observation I created a single ROI (R1) corresponding to the magenta pixels in **Fig. 4.11d** (yellow box) and to the cyan pixels in **Fig. 4.11e** (yellow box). I focused the attention on the pixels corresponding to the small outlet valley. In this context, this detection of hydrated minerals is particularly important because it is located in a small valley where it is usually difficult to detect such kind of materials.

The spectra obtained from these ROI were then normalized to average spectra of neutral spectral regions. This normalized spectrum is shown in the top paned of **Fig. 4.12** and the water bands centered at 1.41, 1.92 µm and 2.31 µm are visible.

To confirm the identification of hydrated minerals, also in this case I compared CRISM spectra with RELAB spectra of several materials. Among the latter, those which are more closely similar with the CRISM spectra are the ones of a olivine with hydrated alteration (C1OL09), a saponite (LASA59) and two nontronites (PHY07 and NDJB26).

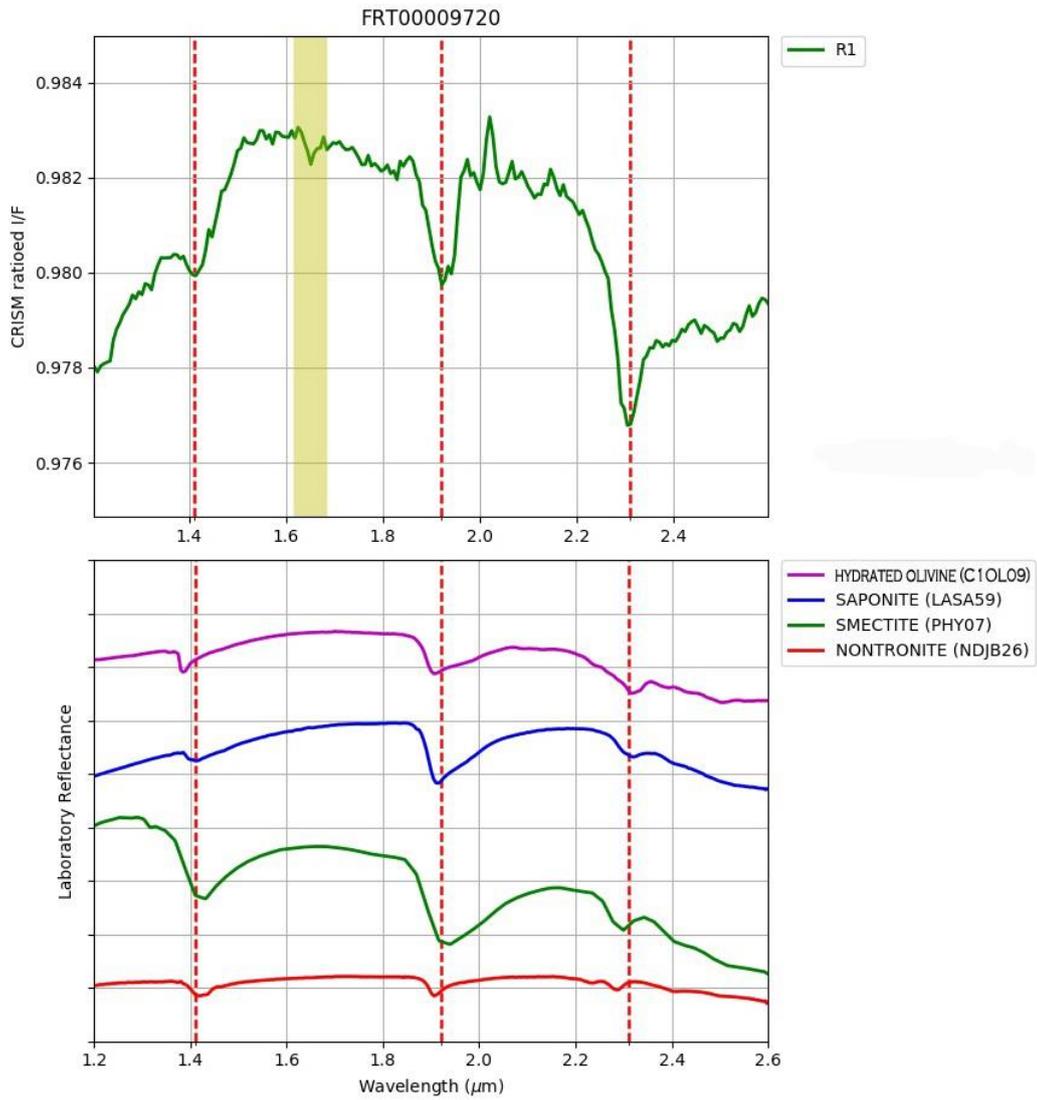

**Fig. 4.12.** Top panel: ratioed CRISM spectra obtained from R1. Bottom panel: RELAB spectra (shifted for clarity). Vertical red dashed lines at 1.41, 1.92, 2.30 µm indicate the central positions of the bands observed in the CRISM spectra. The yellow stripe indicates the spectral region around 1.65 µm where an instrumental artifact is present.

## *4.5 Discussion and conclusions*

The presence of phyllosilicate deposits on Mars is an opportunity to evaluate aqueous activity on the surface of the planet and to investigate the possibility that habitable environments may have existed during the Noachian period. In fact, some works suggest that phyllosilicates could have played an important role in the origin of life on Earth (Bishop et al., 2013) and in the preservation of biosignatures (Orofino et al., 2010).

In most environments, essential nutrients, biologically accessible energy and liquid water are necessary for the development of life (Nealson, 1997). McSween and colleagues (2009) conducted an analysis on terrestrial marine basalts with a chemical composition consistent with that of the Martian basalt. Their study showed that these rocks contain sufficient requirements to sustain life.

Other studies (e.g. Alt and Mata, 2000; Konhauser et al.,2002; Benzerara et al., 2005) have shown that microbes may facilitate palagonitization of basalt into clays and other altered phases. In addition, experiments showed that soil bacteria and viruses can survive in a variety of soil and clay environments including extreme environments replicating Martian conditions (Hawrylewicz et al.,1962; Foster et al.,1978; Moll and Vestal, 1992; Bishop et al., 2013).

Moreover, an analysis on terrestrial fossils in phyllosilicates-rich soils showed that the level of degradation may be much lower if carbonate fossils are found in clay deposits (Orofino et al., 2010). The presence of such kind of minerals, in fact, seem to enable them to retain some of their biotic features for very long times (Orofino et al., 2010).

All of these experiments and observations imply that the phyllosilicate regions on Mars may represent very interesting environments that can provide conditions favorable to preserving evidence of biomarkers (Farmer and Des Marais, 1999; Orofino et al., 2010; Bishop et al., 2013).

As discussed in Section 1.2.5 (Chapter I), analyses of spectral orbital data from CRISM and OMEGA, revealed a large variety of phyllosilicates on Mars (e.g., Bibring et al., 2005; Poulet et al., 2005; Mustard et al., 2008; Murchie et al., 2009). The most common phyllosilicate detected on the Martian surface is nontronite or Fe/Mg-smectite (e.g., Murchie et al., 2009). The largest outcrops of these minerals are found in two locations: 1) at *Mawrth Vallis* (e.g., Bishop et al., 2008; Loizeau et al., 2010; Noe Dobrea et al., 2010); 2) at *Nili Fossae* regions (e.g., Ehlmann et al., 2009).

The findings of this work are consistent with previous analyses. Our findings suggest, on the basis of the previous observations, that the areas under examination may represent very interesting environments that could have preserved evidences of biosignatures.

In particular, we found for the closed basin #16 a light-toned bedrock exposure that shows the presence of aqueous altered materials (Fe-phyllosilicates) along with unaltered mafic material (olivine) located on the rim of the crater, north of the exposure. Also for the open basin #25 we observed the simultaneously presence of altered (Fe/Mg-smectites) and unaltered (olivine) material. The observation of altered and unaltered material in the same location is important because it indicates the occurrence of the alteration process.

The other two basins that show the presence of aqueous alteration minerals are the open basins #10 and #28. Fe/Mg-smectites were detected in both locations. Particularly important is the observation in the open basin #28 because it shows the presence of aqueous alteration minerals (Fe/Mg-smectites) in the small outlet valley of the basin. To the best of our knowledge, this is the first time that this occurrence has been observed. This is also indicative of the possible presence of such kind of minerals even in small valleys. Probably so far, it was not possible to detect aqueous altered materials in such structures, owing to the low spatial resolution of the instruments now in orbit around Mars. In this view more thorough study of the spectroscopy of small Martian fluvial systems could be highly desirable.

# *Summary and conclusions*

*This work presents a study of the Martian surface and in particular of those structures produced by liquid water flow in order to further constrain habitable environments on Mars. The study of these structures from a geologic and spectroscopic point of view is in fact a key for a better understanding the presence and the duration of liquid water on the surface of the Red Planet.*

*During my PhD I updated previous manual maps of Martian valleys using a new photomosaic of the Martian surface and data at higher resolution with respect to those used in the previous works (Carr, 1995; Luo and Stepinsky, 2009; Hynek et al., 2010). I used, in fact, the so far best resolution THEMIS image mosaic (~100 m/pixel) plus CTX data (~6 m/pixel) and MOLA topographic data (463 m/pixel). Compared to previous works, these data, along with a manual approach and a thorough analysis, allowed me to better map these structures at a fine-scale and consequently to remove false positives and identify new tributaries for a considerable number of systems along with some small valleys not previously mapped. To improve the map, I associated to the mapped valleys an attribute table containing valuable information such as coordinates, total length of the systems, and an estimated maximum age. For the latter, I combined the data obtained from the valleys' map with those of the geologic map by Tanaka et al. (2014) which represents, to date, the most accurate dating of the planet surface. In this way I obtained an indication of the maximum age of each valley based on the assumption that a valley is as old as the terrain on which it has been carved (Carr, 1995; Hynek et al., 2010). Even though this assumption may seem too strong, it is, actually, a good choice to have a global idea of the age distribution of these valleys and it is also a way to assign an approximate maximum age to valleys that are too small for age determination by means of crater counting techniques (Carr, 1996).*

*To allow further analysis, the updated global map has been released to the scientific community on Zenodo (Alemanno and Orofino, 2017) and is also included in the Open Planetary Map (OPM) platform (Manaud et al., 2017).*

*Furthermore, we selected a subset of the mapped valleys which includes 63 large fluvial systems with a main branch longer than 150 km and a total length greater than*

*600 km. For each of these valleys a DEM (Digital Elevation Model) was created to determine their volume. On the basis of the volume obtained for this sample of valleys, I estimated the total eroded volume associated with whole mapped valley networks that is in good agreement, in terms of order of magnitude, with that previously obtained by Luo et al. (2017).*

*For these 63 valleys I also evaluated the duration of the water flow using a model of sediment transport and a thorough analysis of terrestrial values adapted to the Martian case. Our sample was divided into two groups: valleys with an interior channel (13 valleys) and valleys without visible interior channels (50 valleys). For the first group I estimated the formation times using a method based on the calculation of water and sediment discharges. We chose four different possible situations: continuous flow; 5% of intermittency, typical of terrestrial humid or sub-humid conditions; 1% as for semiarid or arid environments; and finally, 0.1% for hyper-arid conditions. Once we obtained the formation times for each valley we evaluated the erosion rates. Subsequently, the mean erosion rates obtained for the first group of valleys were used to calculate the formation times for the remaining 50 valleys using the ratio between the eroded volume and the erosion rate.*

*Owing to the variability in climatic phenomena, a continuous flow is unlikely on Mars, as well as on Earth. On the other hand, using an intermittency of 0.1% I obtained for eight valleys formation timescales greater than $10^9$ yr, with a maximum of 10 billion years, and these values are unreasonable or even impossible when compared with the age of the planet (4.5 billion years). So, there are two plausible scenarios for the formation of the Martian valleys: a humid and temperate environment or a semiarid/arid one. In the first case I obtained formation timescales in the range from $6 \times 10^3$ to $1 \times 10^8$ yr (median $3 \times 10^5$ yr). While in the second case, the formation timescales are between $3 \times 10^4$ and $5 \times 10^8$ yr (median $2 \times 10^6$ yr). We think that the latter is preferable according to the scenario proposed by several authors which suggests that, during the period of valleys formation, the climatic conditions on Mars were arid or semiarid (Barnhart et al., 2009; Matsubara et al., 2014) even if we cannot completely rule out the possibility of humid climatic conditions.*

*I also evaluated the temporal distribution of a subset of valleys in our sample for which the ages were obtained by other authors (Fassett and Head, 2008; Hoke and Hynek, 2009). The small number of Martian valleys that were active at the same time during the period of hydrological activity suggests that this activity should have been somewhat localized and sporadic, and that towards its late stages it would have been more intense than at the beginning.*

*Finally, the last part of my PhD thesis is dedicated to some spectroscopic analysis performed on a sample of open/closed basin lakes observed on the Martian surface and associated with the mapped valleys. During the above discussed mapping procedure of Martian valleys, I found 17 more closed basins and 41 more open basins. We then decided to look in the CRISM database for all the available observations of these basins as well as for new observations of the previously catalogued basins with sedimentary deposits (Fassett and Head, 2008; Goudge et al., 2012; 2015). I found interesting CRISM observations in one of the analyzed open-basin lakes and in three of the closed ones. The morphology of the basin lakes was analyzed using a combination of CTX data (with a resolution up to 6 m/pixel) and THEMIS data (resolution of ~100 m/pixel).*

*The CRISM data were analyzed through the application of atmospheric and photometric correction and data filtering. Once I obtained the corrected spectra, three fundamental steps have been accomplished for a good spectral analysis:*

*a) location of interesting material(s) (by means of the use of spectral parameters);*
*b) collection of best possible spectra (pixel average, Region Of Interest -ROI);*
*c) interpretation of the endmember spectra through the comparison with laboratory data.*

*Adopting this methodology, we found for the analyzed basins the possible presence of hydrated minerals, in particular (Fe/Mg-smectite), sometimes in association with mafic materials (olivine). These findings agree with previous works (Bishop et al., 2008; Ehlmann et al., 2013) that suggest a widespread presence of Fe/Mg-phyllosilicates on the surface of Mars. In addition, for one of the analyzed open-basins I found an interesting CRISM observation corresponding to the outlet*

*valley. This observation represents an important detection of hydrated minerals because it represents the first time that these kinds of minerals are detected in a relatively small valley.*

*Globally the results obtained in this PhD work suggest an ancient long-lasting fluvial activity on Mars, even though other proposed scenarios cannot be entirely excluded (Forget et al., 2013; Wordsworth et al., 2013).*

Noe Dobrea E.Z. et al. (2010). *Mineralogy and Stratigraphy of Phyllosilicates-bearing and dark mantling units in the greater Mawrth Vallis/west Arabia Terra area: constanits on geological origin.* Jou. Geophys. Res. **115**, doi: 10.1029/2009JE003351.

Nordin C.F. and Beverage J.P. (1965). in *"Sediment transport in the Rio Grande, New Mexico"*. USGS Prof. Paper 462-F, US Government printing office, Washington.

Núñez J.I. et al. (2016). *Compositional Constraints on Martian Gully Formation as Seen by CRISM on MRO.* 47th Lunar and Planetary Science Conference, TheWoodlands, Texas. LPI Contribution #1903, 3054.

Orofino V. et al. (2009). *Evaluation of carbonate abundance in putative martian paleolake basins.* Icarus **200**, 426-435, doi: 10.1016/j.icarus.2008.11.020

Orofino V et al. (2010). *Study of terrestrial fossils in phyllosilicate-rich soils: Implication in the search for biosignatures on Mars.* Icarus **208**, 202-206, doi: 10.1016/j.icarus.2010.02.028

Osterloo M.M. et al. (2010). *Geologic context of proposed chloride-bearing materials on Mars.* Jou. Geophys. Res. **115**, E10012.

Palucis M.C. et al. (2014). *The origin and evolution of the Peace Vallis fan system that drains to the Curiosity landing area, Gale Crater, Mars.* Jou. Geophys. Res., **119**, 705-728.

Paola C. and Mohrig, D. (1996). *Palaeohydraulics revisited: palaeoslope estimation in coarse-grained braided rivers.* Basin Res. **8**, 243–254.

# *List of Figures*

# *List of Tables*

# *List of publications*

*A) Publications with peer review process*

1. Mancarella F., Fonti S., **Alemanno G.**, Orofino V., Blanco A.: 2017, *Aqueous alteration detection in Tikhonravov crater, Mars.* Planetary and Space Science **152**, 165 – 175 doi: https://doi.org/10.1016/j.pss.2017.12.005.

2. Orofino V., **Alemanno G.**, Di Achille G., Mancarella F.: 2017, *Estimate of the water flow duration in large Martian fluvial systems.* Submitted to: Planetary and Space Science Journal (moderate revision in progress).

*B) Submitted publications with peer review process*

1. **Alemanno G.**, Orofino V., Mancarella F.: 2018, *Global map of Martian fluvial systems: age and total eroded volume estimations.* Submitted to: Earth and Space Science Journal.

2. Varatharajan I., Maturilli A., Helbert J., **Alemanno G.**, Hiesinger H.: 2018, *Spectral Behavior of Sulfides in Simulated Daytime Surface Conditions of Mercury: Supporting past (MESSENGER) and future missions (BepiColombo).* Submitted to: Earth and Planetary Science Letter.

*C) Published datasets*

1. **Alemanno G.**, Orofino V.: 2017, *Global map of Martian fluvial systems [Data set].* Zenodo. http://doi.org/10.5281/zenodo.1051038.

*D) Publications without peer review process*

1. **Alemanno G.**, Orofino V., Mancarella F., Fonti S.: 2017, *Global map and spectroscopic analyses of Martian fluvial systems: paleoclimatic implications.* EGU2017-504-2, Vol. 19.

2. Mancarella F., Fonti S., **Alemanno G.**, Orofino V., D'Elia M.: 2017, *Systematic study of parameters influencing spectral behavior of laboratory mixtures.* EGU2017-9090, Vol. 19.

3. **Alemanno G.**, Orofino V., Di Achille G., Mancarella F.: 2016, *A new model for evaluating the duration of water flow in the Martian fluvial systems.* Mem. S.A.It, vol. 87.

4. **Alemanno G.**, Orofino V., Di Achille G.: 2015, *Global mapping and formation timescales of Martian fluvial systems*. Annual Report, G. Co', F. Paparella and D. Dell'Anna eds.

5. **Alemanno G.**, Orofino V., Di Achille G., Mancarella F.: 2015, *Mapping and formation timescales of Martian Valley Networks*. EPSC 2015, Abstract #237, vol. 10.

*All the above-mentioned publications (except for A1, B2 and D2) have evolved from my doctoral dissertation thesis.*

# *Acknowledgements*

*"Every one of us is, in the cosmic perspective, precious. If a human disagrees with you, let him live. In a hundred billion galaxies, you will not find another."*

*Carl Sagan*

At the end of my PhD work, I would like to thank all the precious people that made this experience the biggest of my life.

First and foremost, I wish to thank my academic supervisor, **Prof. Vincenzo Orofino**, for the continuous support, motivation and knowledge with whom he guided me in these years. He has been supportive since the days I began doing research as an undergraduate, during my bachelors' and masters' degree theses. Ever since, he supported me not only by providing a research assistantship but also academically and emotionally through the rough road to finish this PhD thesis.

Similar, profound gratitude goes to **Prof. Sergio Fonti**, who has been a truly dedicated mentor. I am particularly indebted to him for his constant faith in my laboratory work, for his support and for believing in me in every occasion.

I could not have imagined having better supervisors than them for my PhD research activity and I could go on mentioning many other reasons to thank them, but I think that there is no need to say much other… they can both read all my gratitude through my eyes.

Thanks to all the members of the ***Astrophysics group*** that ended up being my second family. Thanks for every lunch together, for every activity done together always with enthusiasm and love for our work. I have very fond memories of our time together.

Let's skip to the base of everything: ***my family***! Without them nothing of this could have been possible. Their support has been unconditional all these years; they have cherished with me every great moment and supported me whenever I needed it. I have an amazing family, unique in many ways and crazy in many others.

Special thanks to my sister ***Elisa*** and my beautiful little niece ***Carlotta***, seeing their smiles helped me through the most difficult days.

Heartfelt thanks go also to my special friends. Once I read that you never feel quite complete when your other half isn't around: they are my other half, without them I don't feel complete! So, thanks to: my closest and oldest friend *Sabrina*; my soulmate *Fabiana*; my wise *Agnese*; the always smiling *Eugenia*; the very patient and always willing to help *Mosè*; the full of energy and trustful *Mattia*; and last but not least, the lovely and sweet *Marco*. Thanks for every crazy moment together, for every word, for every smile and for every magical sunset watched together with our amazing sea view.

During my PhD years I met a lot of new people from all over the world and I had many amazing experiences doing what I most love to do: Research!

So, the acknowledgements of this dissertation thesis will be not complete without mentioning all the wonderful people that this experience gave me the chance to meet.

Special thanks to *Alessandro* and *Jörn* for their support and help during my period of research abroad. They have been perfect tutors and working with them in the Planetary Spectroscopy Laboratory in Berlin has been a great experience.

Thanks also to: my special Indian friends *Indhu*, *Mansi*, *Sujatha*; my Brazilian twin *Amanda*; the London funny guy *Giambo*; my Italian friend *Barbara* that I never met in Italy, my sweet Dutch *Sophie*…and all the crazy people of the Icelandic group. I cannot mention everyone, but this thought is for each of them. Thanks because, even if we met for a brief time, all of you remain impressed in my mind and my heart. Every occasion is good to cover the distance with a simple message and the promise that one day, in one way or another, we will meet again.

Many thanks to the two patient reviewers of this PhD dissertation thesis: *Cristian Carli* and *Jules Goldspiel*. Their fruitful comments and suggestions improved the manuscript.

Finally, thanks to my lovely *Red Planet* that is the protagonist of the story of my PhD life. After all these years studying its surface I can say that I feel like a little Martian on Earth. Don't worry Mars, I'll keep studying you! It's a promise!

*"...maybe we're on Mars because of the magnificent science that can be done there - the gates of the wonder world are opening in our time. Maybe we're on Mars because we have to be, because there's a deep nomadic impulse built into us by the evolutionary process, we come after all, from hunter gatherers, and for 99.9% of our tenure on Earth we've been wanderers. And, the next place to wander to, is Mars. But whatever the reason you're on Mars is, I'm glad you're there. And I wish I was with you."*

Carl Sagan's message to future explorers of Mars